\RequirePackage{amsmath}
\documentclass{llncs}

\usepackage{amssymb}
\usepackage{graphicx}
\usepackage{bbm}
\usepackage{amsmath}
\usepackage{bibnames}
\usepackage{multirow}

\usepackage{url}
\newcommand{\keywords}[1]{\par\addvspace\baselineskip
\noindent\keywordname\enspace\ignorespaces#1}

\begin{document}

\mainmatter  

\title{Inferring Population Preferences\\ via Mixtures of Spatial Voting Models}

\titlerunning{}

%
%
\author{Alison Nahm$^*$\and Alex Pentland$^\dagger$\and Peter Krafft$^\dagger$}

\institute{$^*$Harvard University, $^\dagger$Massachusetts Institute of Technology\\
\path|anahm@post.harvard.edu, pentland@mit.edu, pkrafft@mit.edu|}

\toctitle{Lecture Notes in Computer Science}
\tocauthor{Authors' Instructions}
\maketitle

\begin{abstract}
Understanding political phenomena requires measuring the political preferences of society. We introduce a model based on mixtures of spatial voting models that infers the underlying distribution of political preferences of voters with only voting records of the population and political positions of candidates in an election. Beyond offering a cost-effective alternative to surveys, this method projects the political preferences of voters and candidates into a shared latent preference space. This projection allows us to directly compare the preferences of the two groups, which is desirable for political science but difficult with traditional survey methods. After validating the aggregated-level inferences of this model against results of related work and on simple prediction tasks, we apply the model to better understand the phenomenon of political polarization in the Texas, New York, and Ohio electorates. Taken at face value, inferences drawn from our model indicate that the electorates in these states may be less bimodal than the distribution of candidates, but that the electorates are comparatively more extreme in their variance. We conclude with a discussion of limitations of our method and potential future directions for research.

\end{abstract}

\section{Introduction}
Within a representative democracy, understanding the extent to which elected officials represent their constituencies is critical to evaluating the efficiency of the political system. Here we focus on the political system in the United States, where surveys typically evaluate the preferences of the electorate. Despite their widespread use, surveys can be costly, time consuming to execute, and often lack broad geographical coverage. Fortunately, there is an alternative source of readily available data about the preferences of the electorate---votes cast in elections. 

The key challenge of using voting data to infer localized distributions of political preferences is the coarseness of the data. Consider inferring the distribution of political preferences of voters from votes cast in a two-candidate election. Since there are only two data points from the election (and a constraint that the two points sum to the total voting population size), inferring the distribution of preferences from these vote shares appears underdetermined.

In this paper, we introduce a model-based machine learning method to measure the political preferences of voters at a fine level of geographical granularity. To solve the underdetermination problem, we introduce a Bayesian mixture model that pools data from similar election outcomes of different geographical voting units (precincts). The method connects the distribution of preferences within each precinct to voting outcomes using a spatial voting model---a standard rational voting model from the political science literature \cite{downs,hinich1984spatial}. Our model utilizes vote share data and a preprocessed form of campaign finance data to infer distributions of political preferences with potentially better coverage and lower cost than traditional survey methods \cite{abramowitz,fiorina-abrams,fiorina-abrams-pope,lee,lev-pope}.

An additional benefit of our method over surveys is that the inferred political preferences of voters are represented on the same scale as those of candidates. This is important for social science applications involving the comparison of politicians and the electorate \cite{barbera2015birds}. To demonstrate the potential utility of our method and of related future work in this area, we apply our technique to understand the extent of political polarization in the Texas, New York, and Ohio electorates in comparison to that of the political candidates. While it is well-known that elected officials are
highly polarized in their political positions, the political science community has not reached consensus as to whether the preferences of voters mirror this elite polarization or are comparatively moderate \cite{abramowitz,fiorina-abrams}. Using congressional election data for the states from the 2006, 2008, and 2010 election cycles,
we find varying answers to the question depending on the polarization metric we use. We find that the distribution of the political preferences of voters is likely more extreme than that of candidates in terms of variance, while less extreme than that of candidates in terms of bimodality.

In the remainder of this paper, we begin with a discussion of
related works. We then provide an overview of our novel probabilistic generative model of voting behavior. We validate this model with comparisons to results of related work and with simple prediction tasks. We then apply our model to better understand political polarization in Texas, New York, and Ohio.  Finally, we conclude with a discussion of the limitations of this method and suggestions for future work.

\subsection{Related Work}

There has been recent work in quantitative political science that is closely related to our work. For instance,
researchers recently developed a technique for estimating the
preferences of the electorate and elected officials from Twitter data using a
probabilistic generative network model related to the spatial voting
model we use \cite{barbera2015birds}.  Some political scientists have used ideal point
models, which are closely related to spatial voting models,
to infer distributions of voter preferences from fine-grained voter
data \cite{lewis2001estimating,gerber2004beyond}. Unlike our work, these previous works using voting data relied on individual-level voting data, which is difficult to obtain. Other political scientists have developed meta-analysis-like methods for aggregating survey results to improve
accuracy and representativeness \cite{warshaw}, but this work still
suffers from the limitation of low coverage of survey data due to collection difficulties. Thus, the methods can only consider a coarser level of geographical granularity.

Within the computer science field, our work falls closest to a
growing line of research dedicated to developing novel machine
learning models for computational social science.  Machine
learning researchers in this area have
not yet addressed the exact problem we study in our work, to the best of our knowledge. However, they have
been interested in similar problems and related classes of models
(e.g. \cite{gerrish,flaxman2015supported,krafft2012topic,airoldi2009mixed}). More
tangentially, a large body of work in computer science has been
dedicated to drawing inferences from public observational data.  Some researchers have suggested using social media data to better
understand public opinion \cite{derek-ruths}, while others have developed models based on inconsistent user behavior to infer their implicit preferences \cite{ding2015learning}.

\section{Model}
\label{sec:model}

We first discuss mathematical theories of voting behavior that inform our novel model. Then, we describe our model for inferring political preferences of voters.

\subsection{Spatial Voting Models}

Our model generalizes the widely used ``spatial'' or ``Downsian'' voting model, which is a standard model in political science of rational voting and turnout behavior in two-candidate majority vote elections \cite{downs,hinich1984spatial}. The spatial voting model defines each voter and each political candidate as points in a one-dimensional policy space. The model defines the utility to a voter of a specific candidate winning as the Euclidean distance between their two points. Assuming the election involves exactly two candidates, the spatial voting model predicts that voters will select for their votes the candidates closest to them according to Euclidean distance in the single-dimensional policy space.

\subsection{Mixtures of Spatial Voting Models}

Our statistical model consists of a generative process for the vote shares of candidates in an election. Each precinct
$i$ with $N_i$ total voters is associated with an election of exactly two candidates, $c_{0i}$ and $c_{1i}$. In line with the spatial voting model, we assume that voters $v_{ji}, j \in \{1,\ldots,N_i\}$ and candidates $c_{ki}, k \in \{0, 1\}$, have positions in the same one-dimensional latent space, which we consider their {\em political preferences}.  We assume the candidate positions are known to all election participants. Like other spatial voting models, we assume that each voter $j$ of precinct $i$ votes for the closest candidate in the one-dimensional latent policy space. In other words, voter $j$ in precinct $i$ votes for candidate 0 in precinct $i$ if $|v_{ji} - c_{0i}| < |v_{ji} - c_{1i}|$.

We assume that each precinct is associated with a particular distribution over political preferences, which determines the preferences of the voters in that precinct. However, it is problematic to assume that these distributions are all distinct from and independent of each other. In this case we are limited to using
only one data point per precinct to infer a distinct distribution. To solve this issue we use a mixture model to pool
data across precincts. We assume that certain subsets, or \textit{clusters}, of precincts share the same distribution of
preferences. These assignment of precincts to clusters is determined dynamically during inference according to similarity in observed voting patterns.  This modeling assumption seems reasonable given that it is likely neighboring precincts will have similar distributions of preferences.

The expected proportion of precincts that will be assigned to each of $K$ clusters is given a Dirichlet prior, $\vec{\theta} \sim Dirichlet(\vec{1})$ and $|\vec{\theta}|=K$. The assignment of each precinct $i$ to a particular cluster is then drawn as in a standard mixture model, $x_i \sim \vec{\theta}$. The position of each voter $j$ in each precinct $i$ is drawn according to a component distribution associated with the cluster assignment of that precinct.  The parameters of these component distributions are given weakly informative priors. The number of votes $n_{ki}$ received by candidate $k$ in precinct $i$ are then given
deterministically by the spatial voting model specified above. Mathematically, $n_{0i} = \sum_{j=1}^{N_i} \mathbbm{1}(|v_{ji} - c_{0i}| < |v_{ji} - c_{1i}|)$, where $\mathbbm{1}$ is an indicator function, and $n_{1i} = N_i - n_{0i}$.

We treat candidate positions as fixed and given since we have data on these values, but we treat voter positions, precinct assignments, and cluster distribution parameters as unknown. After marginalizing over voter positions, then conditioning on direct estimates of candidate positions and on observed vote shares per candidate, we can use Bayesian inference to arrive at likely values for the remaining unknown parameters, thus estimating the distributions of voter preferences within each precinct.

\section{Data}
\label{sec:data}
For our empirical analysis, we use three main sources of data.

\paragraph{Precinct-Level Voting Results.}

Precincts are the finest granularity of publicly accessible aggregated vote shares. We examine congressional elections of Texas, New York, and Ohio \cite{heda}. In these cycles, Republicans won 65\% of the Texas elections, and Democrats won 80\% of the New York elections. We also analyze Ohio to test the ability of our method to generalize to more extreme voter distributions, as Ohio is commonly labeled by political scientists as a ``swing state''. We consider the election cycles 2006, 2008, and 2010 because they all depend on the same district geographic boundaries set by the 2000 U.S. Census. We omit the 2002 and 2004 election cycles to focus on recent elections. Future work could analyze longer periods.

\paragraph{Candidate CFscores.}

We incorporate quantitative estimates of the political preferences of candidates called campaign-finance scores (CFscores) \cite{dime}. CFscores are one-dimensional quantitative estimates of the political ideology of political candidates, with lower values indicating more liberal ideologies and higher values more conservative. CFscores are recognized as effective estimates of candidate ideology when estimates for unelected candidates are needed (in contrast to DW-NOMINATE scores which only exist for winning candidates \cite{dw-nominate}).

\paragraph{Geographic Precinct Boundaries.}

We link the above-mentioned data sets with the congressional candidates running in each precinct election. We assign any precincts to the district whose geographic center fall within the specific congressional district boundary lines using the Geospatial Data Abstraction Library (GDAL/OGR) package \cite{cdmaps,tigerweb}.

\section{Inference}
\label{sec:inference}

The goal of our inference procedure is to determine likely precinct assignments $(\vec{x})$ and likely parameters of the $K$ cluster distributions in the model described in Sect. \ref{sec:model}. These estimates allow us to characterize the distribution of voter preferences of each precinct.

For more efficient inference, we first integrate out the voter positions $(\vec{v})$ and the precinct assignments $(\vec{x})$. All component distributions we consider allow the voter positions to be integrated out analytically. This yields the following posterior distribution:
    \begin{align}
      &P(\vec{\eta}, \vec{\theta} \,|\, \vec{v}) \\
      &\propto P(\vec{\theta}) P( \vec{\eta})
      \sum_{\vec{x}} P(\vec{x} \,|\, \vec{\theta})
      \int_{\vec{y}} P(\vec{y} \,|\, \vec{x}, \vec{\eta}) P(\vec{v} \,|\, \vec{\eta}, \vec{y})   \\
      &\propto P(\vec{\eta})
      \prod_{i=1}^M
      \left[
        \sum_{x_i = 1}^K
        \theta_{x_i}
        (\Phi_{i,x_i})^{n_{0i}}
        (1 - \Phi_{i,x_i})^{n_{1i}}
        \right] \label{eq:asst-simple}
    \end{align}
where $\vec{\eta}$ is the distribution parameters of each cluster, and $\Phi_{i,x_i}$ is the cumulative distribution of the component distribution of precinct $i$ after integrating out $\vec{y}$, in other words $\Phi_{i, x_i}  = P\left(y < \frac{c_{1i} - c_{0i}}{2} \,|\, \eta_{x_i}\right)$. For prior distributions over the cluster parameters, when using Normal component distributions, we use a Normal prior for $\vec{\mu}$ with a mean of 0 and a variance of 100, and for $\vec{\sigma}$ we use an Inverse Gamma distribution with scale and shape parameters both set to 1. 

We then use a Metropolis-Hasting Markov Chain Monte Carlo (MCMC) algorithm to arrive at likely values for the component distribution parameters and the mixture proportion $(\vec{\theta})$\footnote{Code and data for analyses are available at \url{https://github.com/anahm/inferring-population-preferences}}. To generate the results presented in this paper, we ran four independent MCMC chains from randomly generated initialized values. We then selected the set of parameter values from all generated sets that yield the highest posterior. We infer parameter values separately for the data of each state and election cycle combination described in Sect. \ref{sec:data}. After inferring the parameters of the mixture distribution, we infer all precinct assignments to clusters, $\vec{x}$, by selecting the maximum a posteriori (MAP) assignment variable for each precinct.

Our model-based method provides a better approach than analyzing vote shares because we can derive an overall distribution of preferences per precinct rather than a single point. This approach also has benefits even looking at coarser geographic granularity as well. We ultimately aggregate our inferences to district or state level for validation purposes, and inferring precinct-level distributions opens the possibility for the distribution at a less-granular level to be a complicated combination of precinct-level distributions.

\section{Validation}
\label{sec:validation}

To assess validity, we compare summary statistics of the distributions we infer with corresponding values from related works. In addition, we compare the predictive performance of our model with prediction methods based on empirical data and results of related works. In this section, we present results assuming the number of clusters, $K$, is 4 and underlying Normal component distributions, but we reach similar results when we vary the value of $K$ and the component distribution type, which can be seen in Sect. \ref{sec:addl-results}. These validation methods are meant to qualitatively and quantitatively assess the ``face validity'' of our proposed model.

\subsection{Comparison with Related Works}
\label{sec:dist-agg-desc}
We compare our results with the results of two survey-based methods for estimating district-level political preferences. This comparison would ensure that our inferred distributions are qualitatively reasonable from the perspective of the prior related work. A district is a coarser granularity geographical unit than a precinct, but precinct-level surveys are rarely implemented due to high costs. To compare, we obtain a single-point estimate of each congressional district preference in Texas, New York, and Ohio from our inferred precinct-level distributions. 

To compute single-point estimates of district-level voter preferences, we use our model's assumption that each precinct has the same parameters as the inferred parameters of its assigned cluster. The district-level estimates are averages of the precinct-level inferences of precincts in the same district weighted by the population of those precincts.

\subsubsection{Comparison with Raw Survey Results.}
We first compare with the Cooperative Congressional Election Study (CCES) \cite{cces}. The CCES surveys over 50,000 Americans every election cycle. Many political scientists use the CCES to understand the American public opinion. Moreover, the CCES respondents report their congressional districts, which yields more fine-grained data than most other national surveys \cite{cces}. We compare our results with the survey responses to two questions. The first question asks:

    \begin{quote}
    Thinking about politics these days, how would you describe your own political viewpoint? (Very Liberal, Liberal, Moderate, Conservative, Very Conservative, Not sure)
    \end{quote}

\begin{figure}
        \centering
        \includegraphics[width=0.32\linewidth]{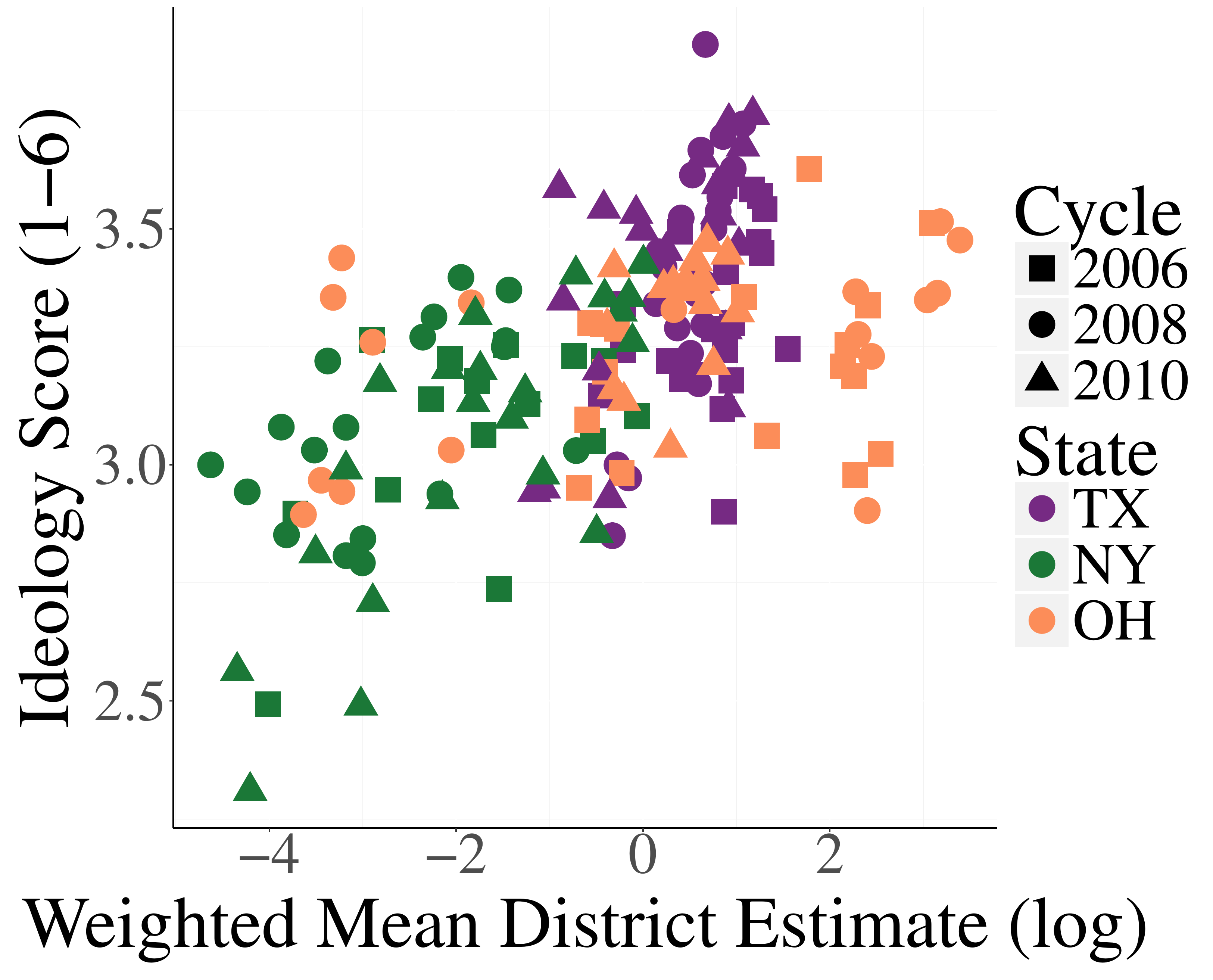}
        \includegraphics[width=0.32\linewidth]{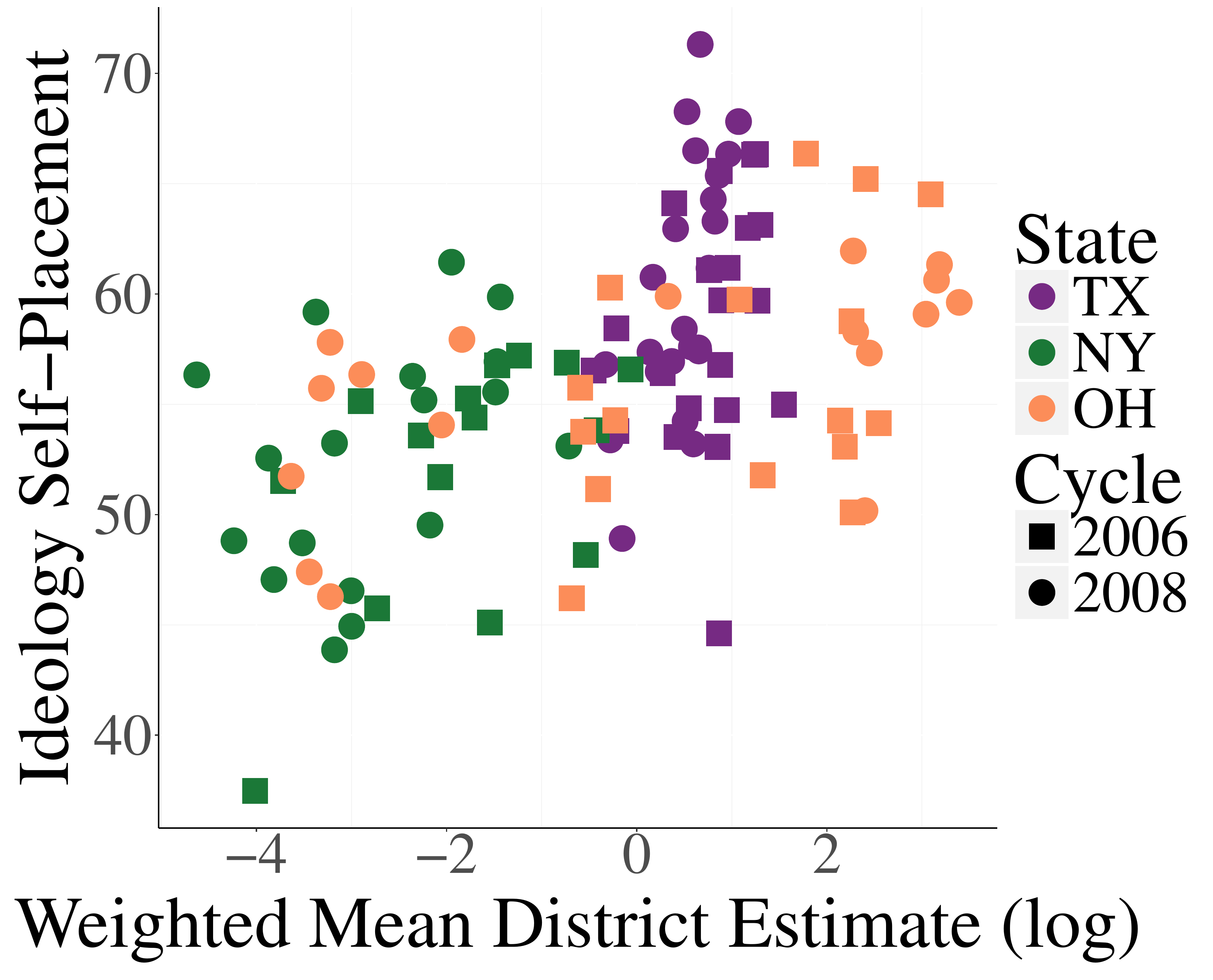}
        \includegraphics[width=0.32\linewidth]{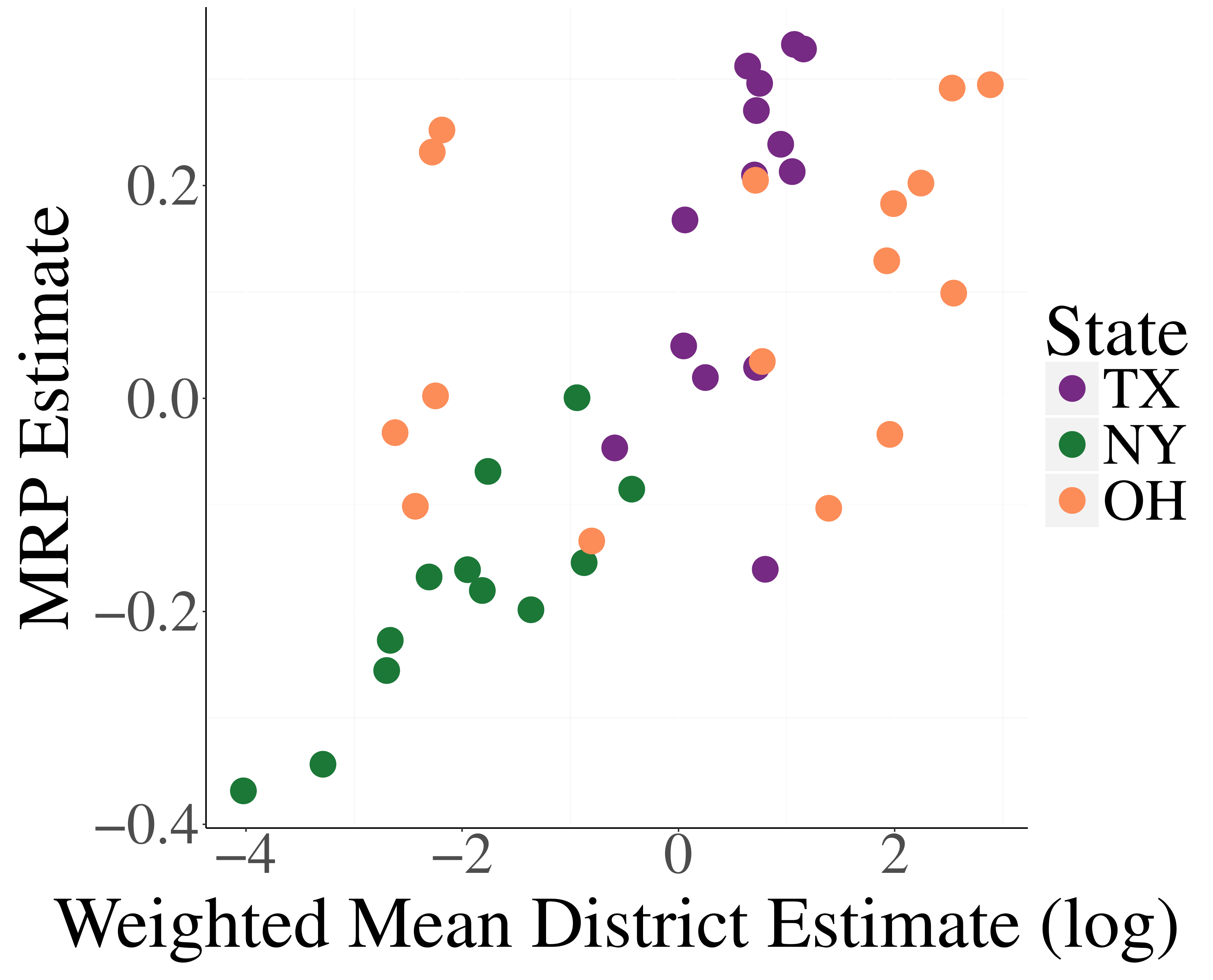}
        \caption{(Left) Inferred district-level voter preferences compared with CCES question of self-reported ideologies on a discrete scale. (Center) Inferred preferences compared with CCES self-reported ideologies on a continuous scale from 0 to 100. (Right) District-level inferences of a  decade compared with MRP estimates \cite{warshaw}. In all cases, the inferred district-level voter preferences are weighted mean district estimates transformed from $x$ to $sign(x) \, log( |x| + 1 )$.}
        \label{fig:cces-comp}
    \end{figure}

As shown in Fig. \ref{fig:cces-comp}, we find a significant positive correlation between our district-level point estimates and the responses to this question. The correlation level of the results of our method and reported survey values is 0.3127 with a p-value less than 0.01.
    
The second CCES question asks survey respondents to score their political ideology on a continuous scale. The question is phrased:
    
    \begin{quote}
    One way that people talk about politics in the United States is in terms of left, right, and center, or liberal, conservative, and moderate. We would like to know how you view the parties and candidates using these terms. The scale below represents the ideological spectrum from very liberal (0) to very conservative (100). The most centrist American is exactly at the middle (50). Where would you place yourself?
    \end{quote}
    
This question was only used in the 2006 and 2008 CCES surveys \cite{cces}. As shown in Fig. \ref{fig:cces-comp}, we find a significant positive correlation between our estimates and the responses to this CCES question. The correlation level of the log of the estimates of our method and reported survey values is 0.2535 with a p-value less than 0.01. Using a monotonic transformation is acceptable since the answers to survey questions and our inferred preferences are not necessarily on comparable scales.

\subsubsection{Comparison with Aggregated Survey Results.}

In addition to CCES results, we validate our results against district-level ideological scores developed by two political scientists, Chris Tausanovitch and Christopher Warshaw. They use disaggregation and multilevel regression with post-stratification (MRP) on survey data from 2000 to 2010 to estimate mean policy preferences of congressional districts \cite{warshaw}. Their work is one of the recent related works understanding preferences, and they analyze election cycles in a similar time frame to this paper \cite{warshaw}. Further, these ideological scores might be more representative of the U.S. population because the work's inference methods account for possible sampling bias.

One caveat of the work by Tausanovitch and Warshaw is that their estimates span a decade of voter behavior rather than a single election. To ensure comparison across consistent measures, we aggregate the district-level results of our model across the three election cycles into one district-level estimate spanning 2006-2010. As shown in Fig. \ref{fig:cces-comp}, we find a significant positive correlation of 0.4149 between the log of the results of our method and their results with a p-value less than 0.01. By contrast, the MRP compared to the two CCES responses has correlations of 0.6543 and 0.7189.

The correlations between our results and the results of the survey-based methods suggest that our method can infer similar qualitative distributions to those of prior works. Further, our method can not only recover district-level mean political preferences, but also examine more granularity, precinct-level preferences in a shorter time period.

\subsection{Predictive Power}
As a second method of validation, we analyze the predictive power of our model. Specifically, we derive values from a \textit{comparison election} to predict the Democratic vote share of a separate \textit{target election}. Here a target election is either an election occurring in a later cycle or for a different government position occurring in the same cycle.

While prediction tasks are sometimes used to argue that a model has the best predictive value compared to alternative models, that is not the goal of this section. We are aware that the predictive power could be improved by incorporating more types of data. Rather, the purpose of these prediction tasks is to demonstrate that our model achieves predictive performance comparable to reasonable alternatives. Our method has additional benefits in terms over the comparison methods, so these prediction tasks are meant to lend quantitative face validity to our model.

\subsubsection{Methods of Prediction.}
We compute the expected vote share of the Democratic candidate (assigned to be candidate 0) of the target election by assuming voters follow the spatial voting model and that voter preferences are identical in the comparison and target elections. The predicted vote share of each precinct $i$ is given by $\Phi_{i, x_i}$, the cumulative of the inferred distribution of voter preferences at the midpoint between the CFscores of the two candidates running in the target election. We aggregate vote shares of the candidate in all precincts of the same district to facilitate comparison with less fine-grained data sources.

We compare the predictive power of values yielded by our model with three alternatives: raw vote shares of previous elections, survey data, and MRP estimates. The method using raw vote shares assumes the candidates of the same political party receive the same proportion of votes in the comparison and the target election. For instance, this na\"ive baseline predicts the vote share of the Democratic candidate in 2010 is equivalent to the vote share of the Democratic candidate in the previous election in 2008. We consider this the baseline prediction model.

The survey prediction method uses CCES responses to a question on political party affiliation as a proxy for votes for the Democratic candidate in the target election \cite{cces}. We approximate the percentage of Democrats as the number of reported Democrats and half the reported Independent or Other divided by the total number of responses. We assume the respondents who state Independent or Other divide equally between Democrat or Republican when faced with only those options.

We also develop a prediction method based on the MRP estimates developed by Tausanovitch and Warshaw \cite{warshaw}. 
This method, labeled in Fig. \ref{fig:error-comp} as MRP Cross-Val, is a simple cross-validation leave-one-out prediction method using MRP ideological scores to predict vote share. For the Cross-Val method, we obtain predictions for each district given the remaining districts and combine error terms into one mean squared error. The previously described prediction methods predict all district preferences at once, so the methods only yield one error term, which is the mean squared error term.

\subsubsection{Prediction Tasks.}

We examine two prediction tasks: next cycle and same-year. The next cycle prediction task defines target elections as congressional elections of the same district one election cycle (two years) after the comparison election cycle. In other words, we use point estimates of voter preferences from election cycle $t-1$ to predict the results of election $t$.

The second prediction task, the same-year prediction task, defines target elections as elections for a different government position, a Senate seat, in the same election cycle as the comparison election. Although our method infers estimates using results of the same year as the election we are trying to predict, the target and comparison elections are for unrelated positions. We assume that voters consider their votes for different ballot items as in independent elections. 

\subsubsection{Prediction Task Results.}
\begin{figure}
        \centering
        \includegraphics[width=0.75 \linewidth]{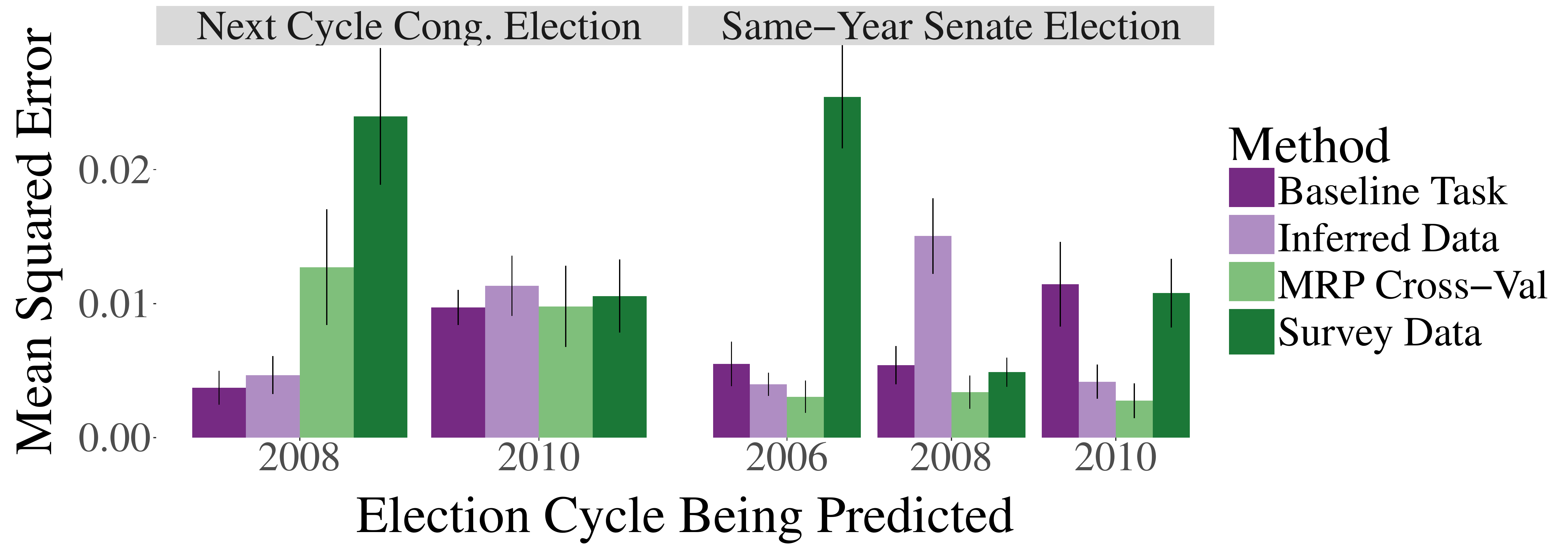}
        \caption{Plot of the mean squared error of the actual and predicted vote share yielded by various prediction methods. Error bars are standard error of the mean.}
        \label{fig:error-comp}
    \end{figure}

As we can see in Fig. \ref{fig:error-comp}, in all but one case our model tends to do as well or better than the alternatives, which further validates our model. The MRP Cross-Val method is the most competitive alternative. However, the high performance of MRP Cross-Val is likely because the cross-validation method is optimized for prediction, whereas our model and the other alternatives are not.

\section{Application}

The results of our method distinguish themselves from the baseline methods discussed in Sect. \ref{sec:validation} because they not only shed light on voter preferences, but also represent those preferences in the same latent policy space as known positions of political candidates. We can leverage the latter capability to assess the degree of political polarization in the electorate compared to candidates' political positions.

\subsection{Background of Polarization}
\label{sec:pol-background}

\subsubsection{Previous Work.}
Popular media and political science communities have observed that the American political elite is becoming increasingly polarized over time, but much less work has drawn conclusions on \textit{mass polarization} \cite{poole-rosenthal,fowler}. Some political scientists hypothesize that the distribution of the political preferences of the U.S. electorate is unimodal and moderate compared to that of the political elite \cite{fiorina-abrams-pope,fiorina-abrams}. Yet other evidence suggests increasing polarization in the American population \cite{abramowitz}. The results of our method could add a valuable new perspective since we draw on data sources separate from the survey methods many of the opposing arguments used \cite{abramowitz,fiorina-abrams,fiorina-abrams-pope,lee,lev-pope}.

Among the related works that do not base their conclusions on survey data, Fiorina and Abrams find that most observations of mass polarization trends, such as differences in sociocultural attributes and world views, are not strong enough to make definitive claims about overall trends of mass polarization \cite{fiorina-abrams}. Their main critique of utilizing voting behavior as a proxy for political polarization is that the 
past actions and political preferences of candidates are not factored into the model. Our work addresses this by factoring in candidate CFscores in addition to voter behavior. Further, most related works only use one numeric metric, the difference of the means of subsets of the population, to measure mass polarization \cite{lev-pope}. However, DiMaggio, et al. define mass polarization in terms of discrepancies between distributions, which can be described by more than a single number \cite{dimaggio}. Some have analyzed polarization in a more holistic way, but their methods depend on survey data 
 \cite{lee,lev-pope}.
 
\subsubsection{Defining Political Polarization.}

Mass political polarization refers to polarization in the electorate, while elite polarization refers to polarization among elected officials. One way to directly measure polarization in either case is to fit a mixture of two Normal distributions to the distribution of preferences of a population, then take the standardized difference of the component means of that mixture. For this procedure to be interpretable, we assume each mixture component has equal weight $(0.5)$ and the same variance. We standardize the absolute difference of the two means by dividing by the inferred standard deviation of each component. This unique ``difference-of-means'' metric is intended as a rough proxy for the probability mass missing from the center of a distribution of preferences. Related works often use a similar metric, but they compute means by aggregating survey responses or other point estimates of subsets of the population rather than fitting a mixture model to the population distribution.

We also measure political polarization in two  ways previously used in the political science literature: dispersion and bimodality \cite{dimaggio}. Dispersion represents the extent to which more varied opinions in the population increase the difficulty for a ``centrist political consensus'' to exist in the population \cite{dimaggio}. DiMaggio, et al. suggest measuring dispersion with the standard deviation of the distribution of political preferences. An increase in the standard deviation signifies that voters have more extreme political preferences and less 
moderate preferences in the middle of the distribution. Bimodality represents the level of separate opinions of different groups that can lead to a higher chance of social conflict. Bimodality can be measured with the kurtosis of the distribution. The formal definitions of these quantities are given in the appendix in Sect. \ref{sec:math-pol}.

\subsection{Analyzing Polarization}

\begin{figure}[ht]
        \begin{centering}
        \begin{minipage}{\dimexpr\linewidth-1cm\relax}%
            \raisebox{\dimexpr-.5\height-1em}{
                \includegraphics[width=0.37\linewidth]{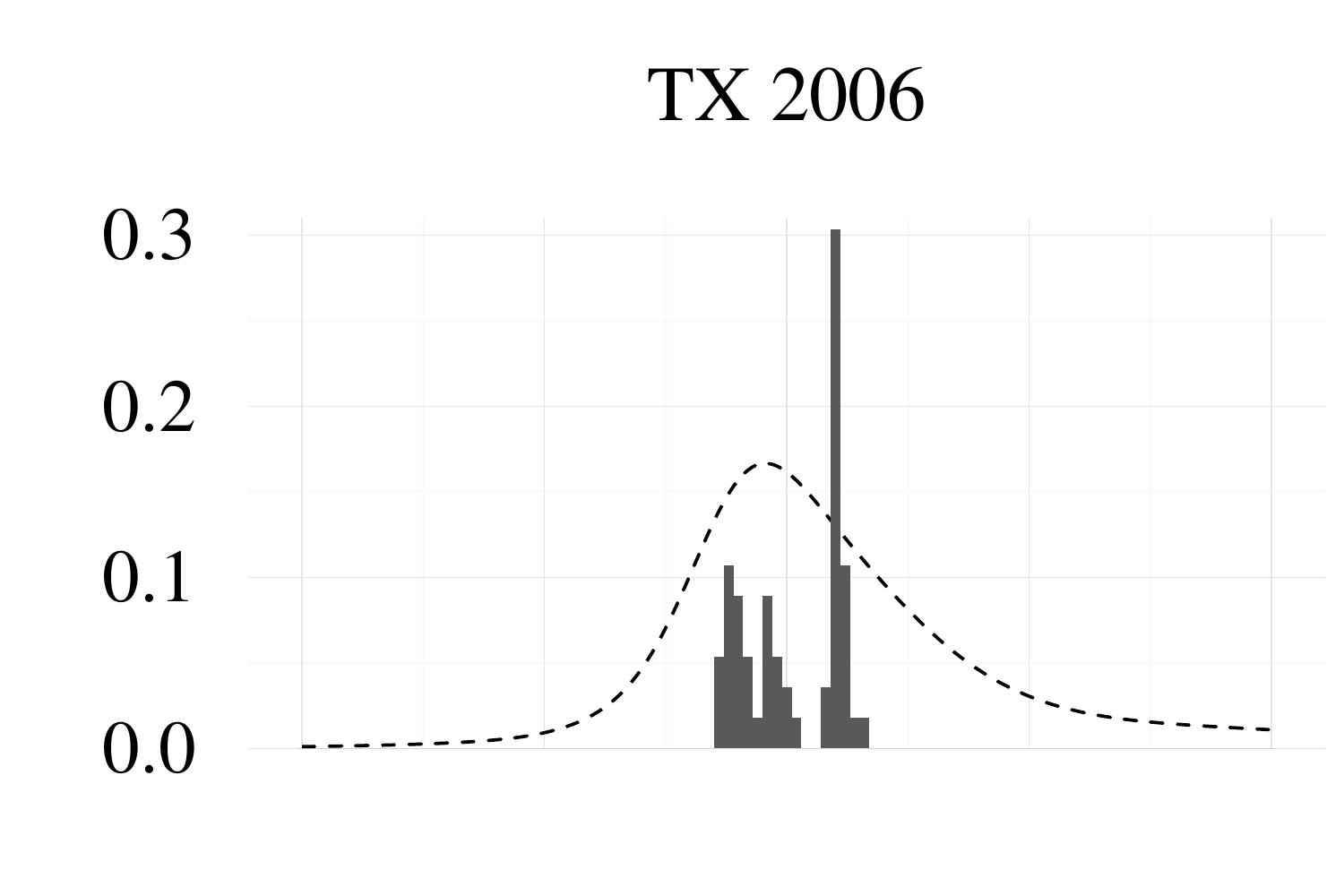}
                \hspace{-0.5cm}
                \includegraphics[width=0.37\linewidth]{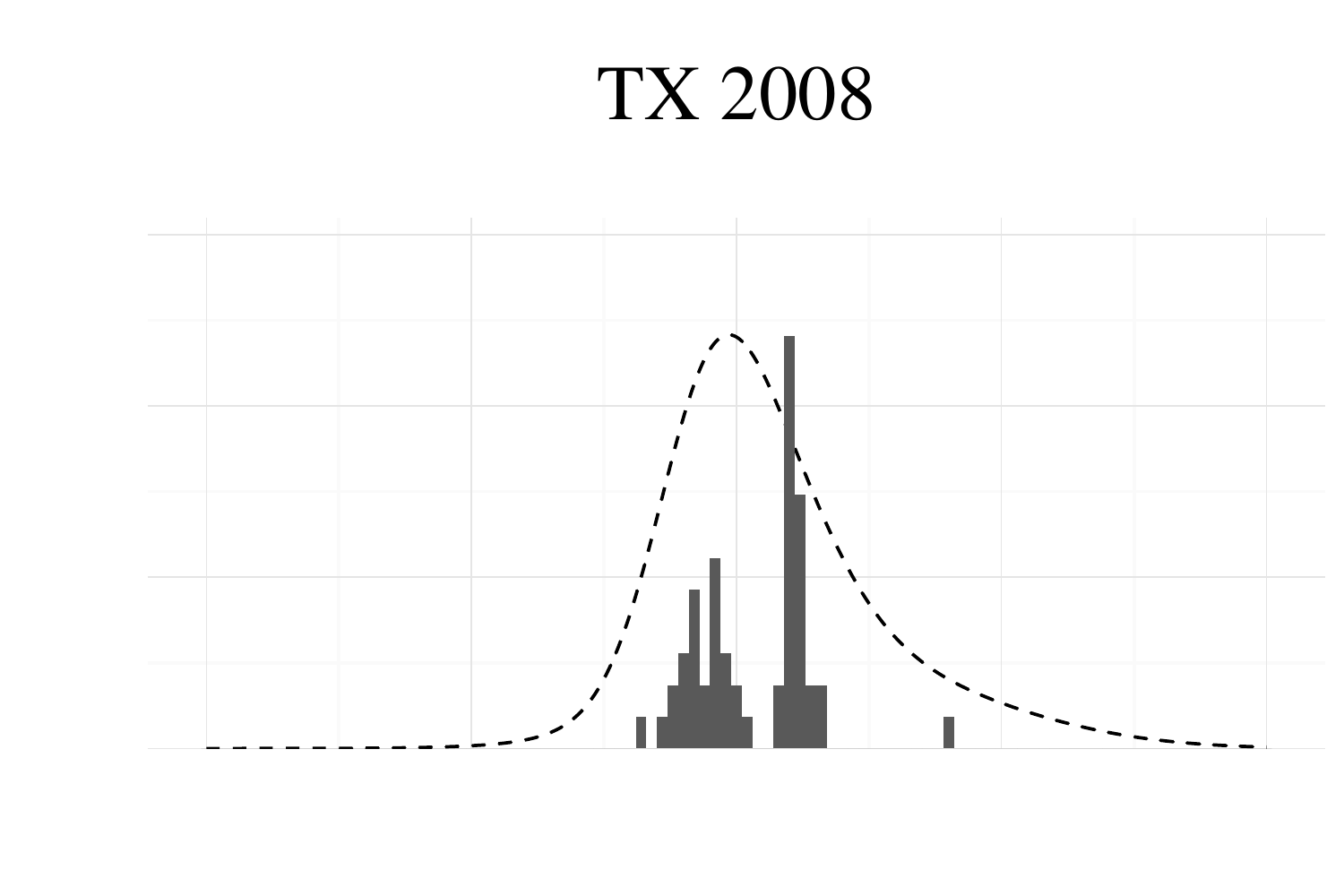}
                \hspace{-0.5cm}
                \includegraphics[width=0.37\linewidth]{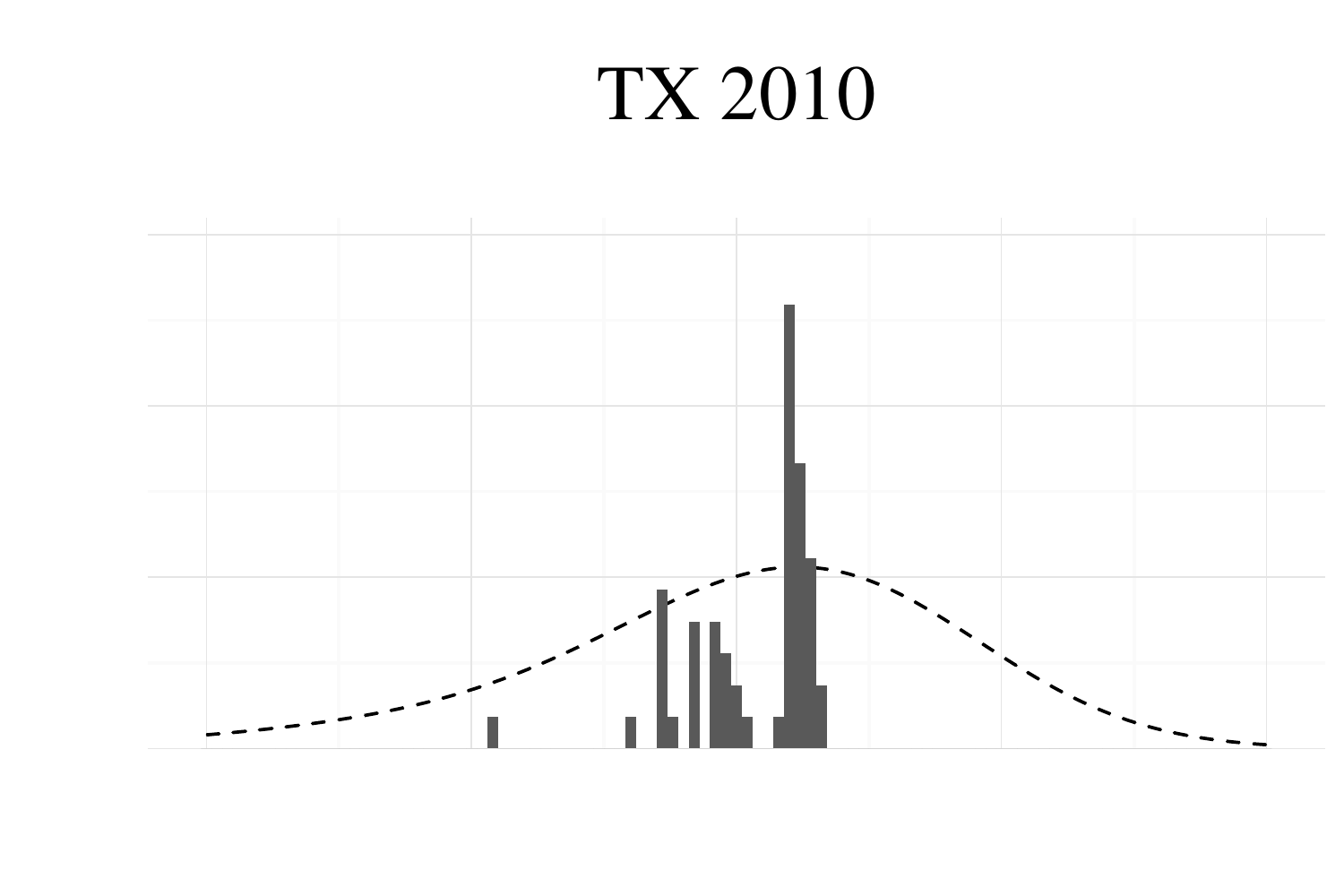}
                \par
            } \\[-5pt]
            \raisebox{\dimexpr-.5\height-1em}{
                \includegraphics[width=0.37\linewidth]{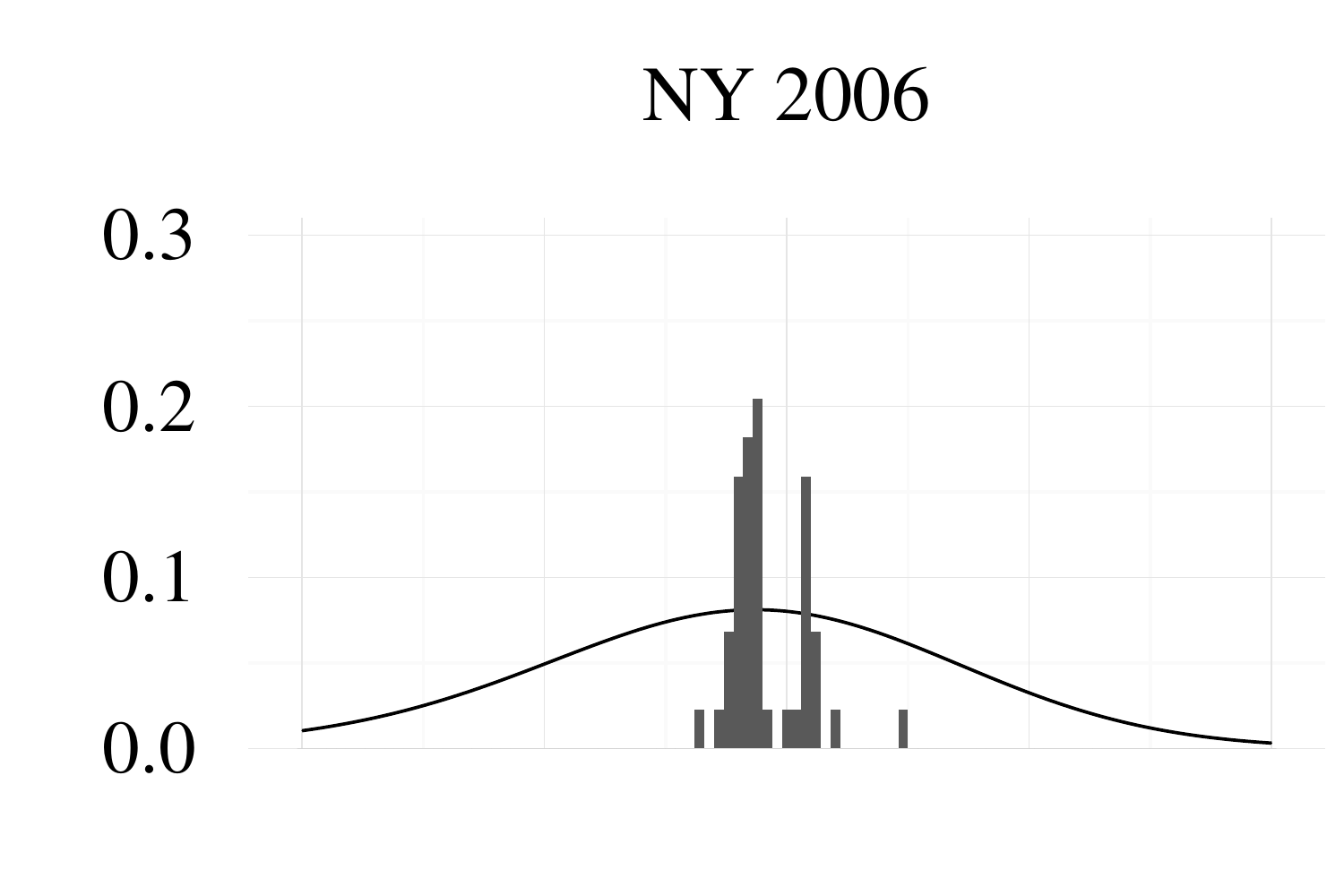}
                \hspace{-0.5cm}
                \includegraphics[width=0.37\linewidth]{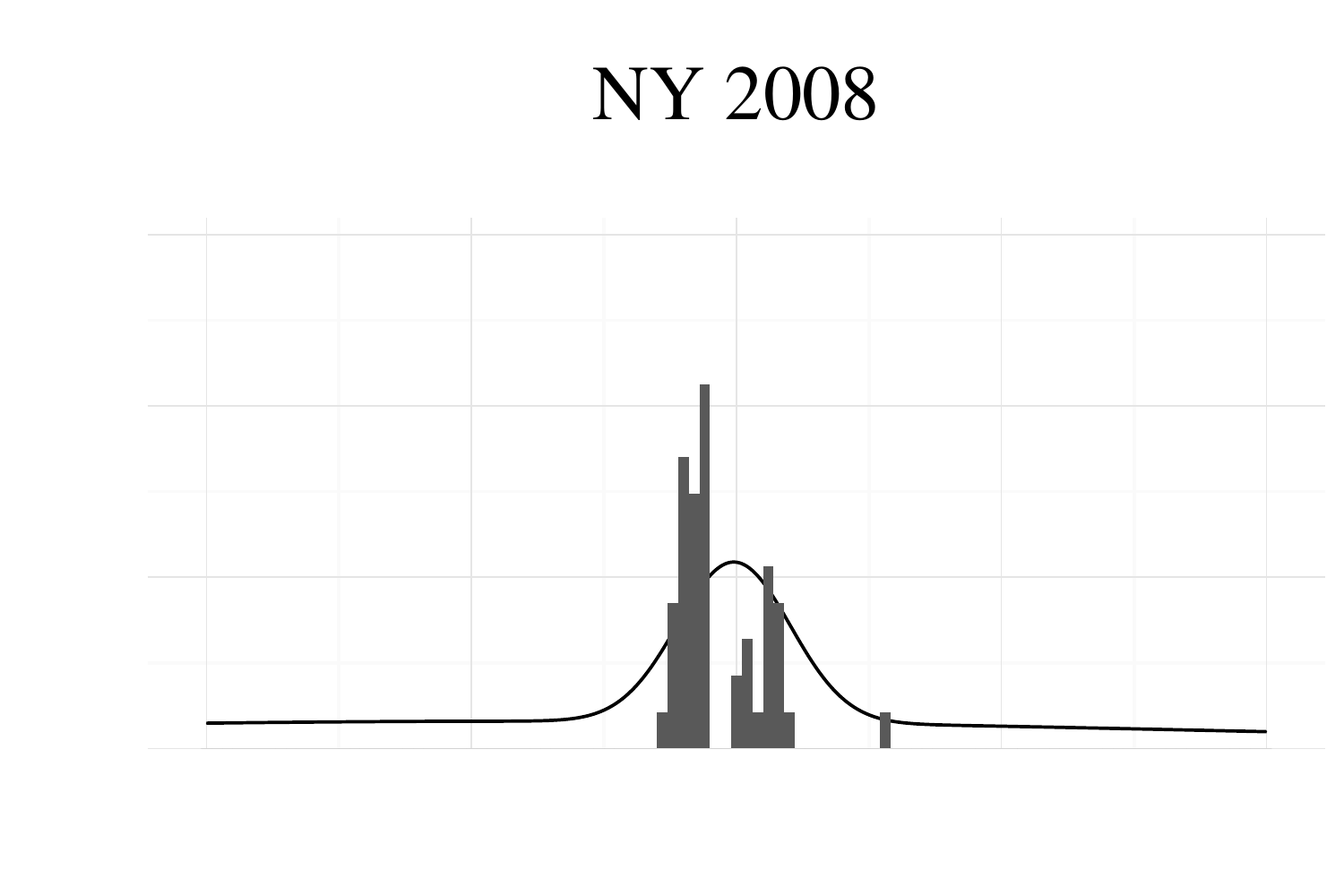}
                \hspace{-0.5cm}
                \includegraphics[width=0.37\linewidth]{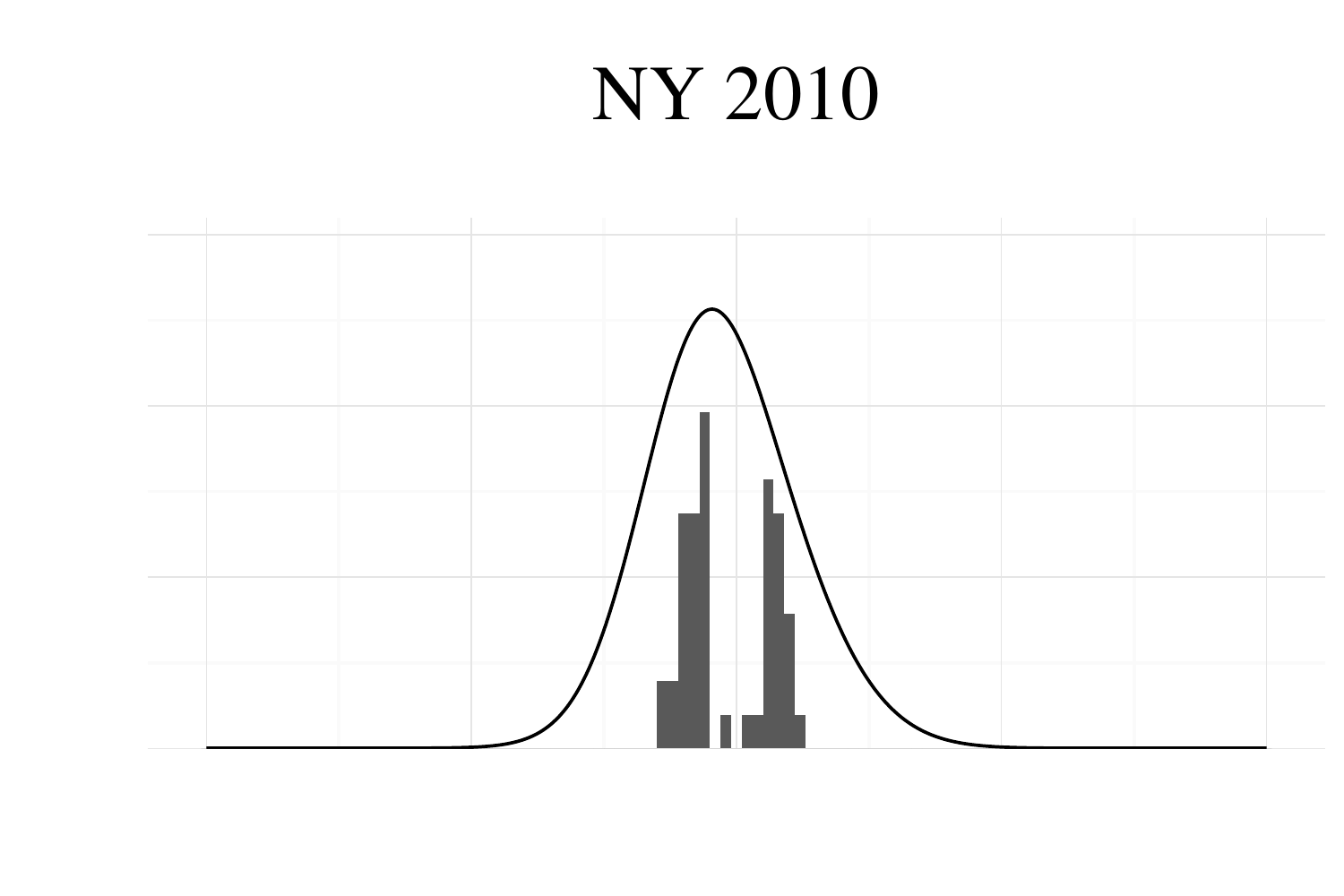}
                \par
            } \\[-5pt]
            \raisebox{\dimexpr-.5\height-1em}{
                \includegraphics[width=0.37\linewidth]{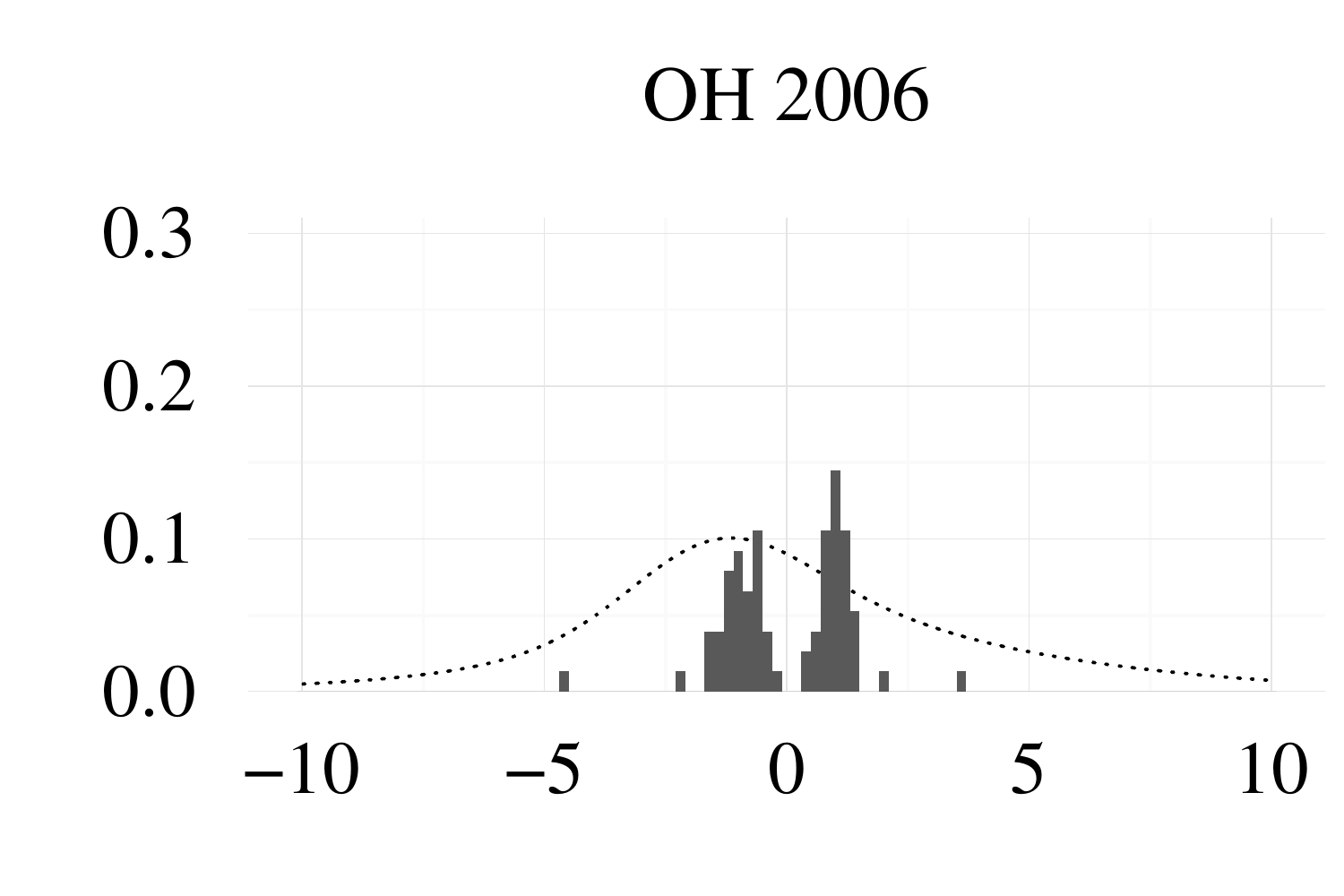}
                \hspace{-0.5cm}
                \includegraphics[width=0.37\linewidth]{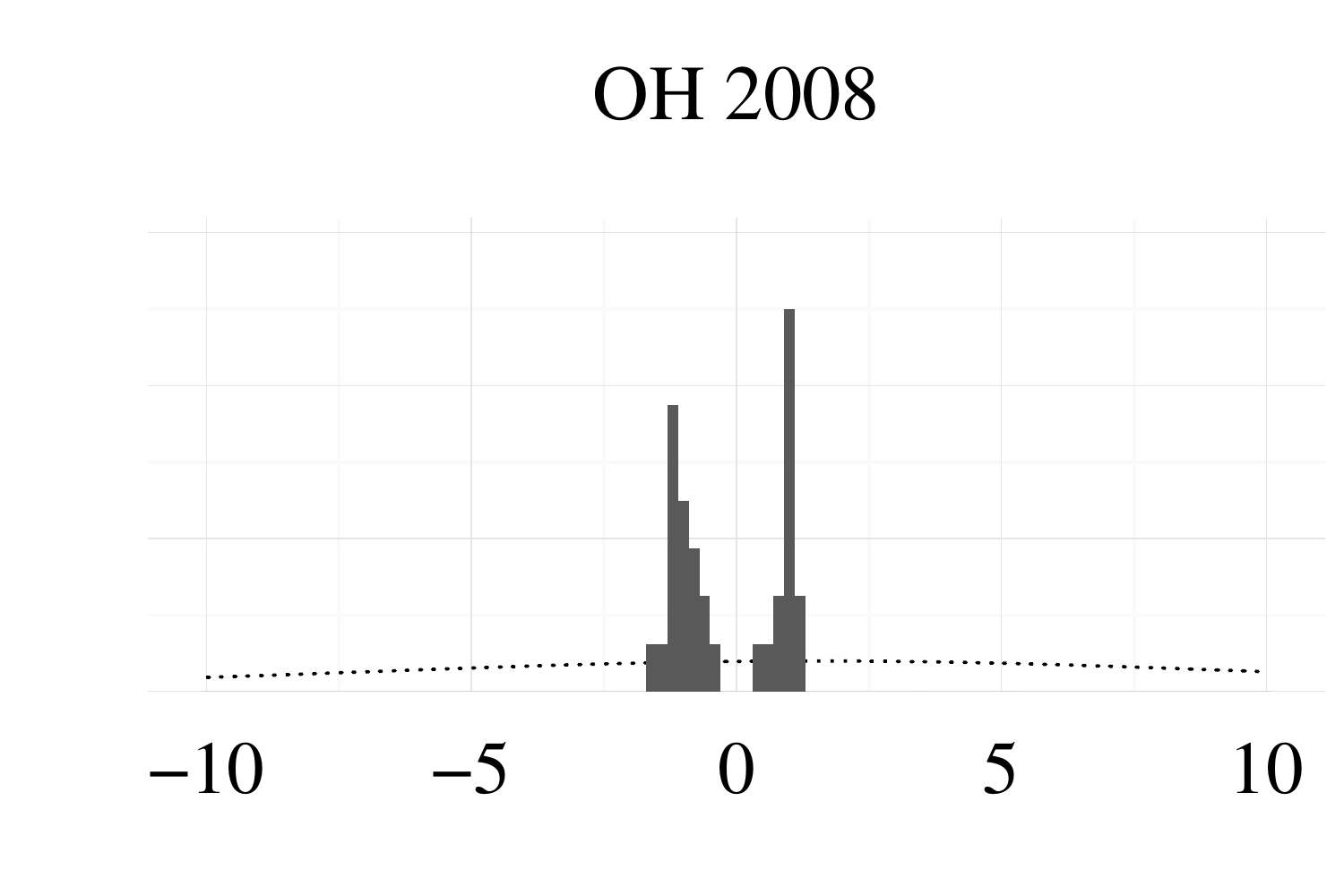}
                \hspace{-0.5cm}
                \includegraphics[width=0.37\linewidth]{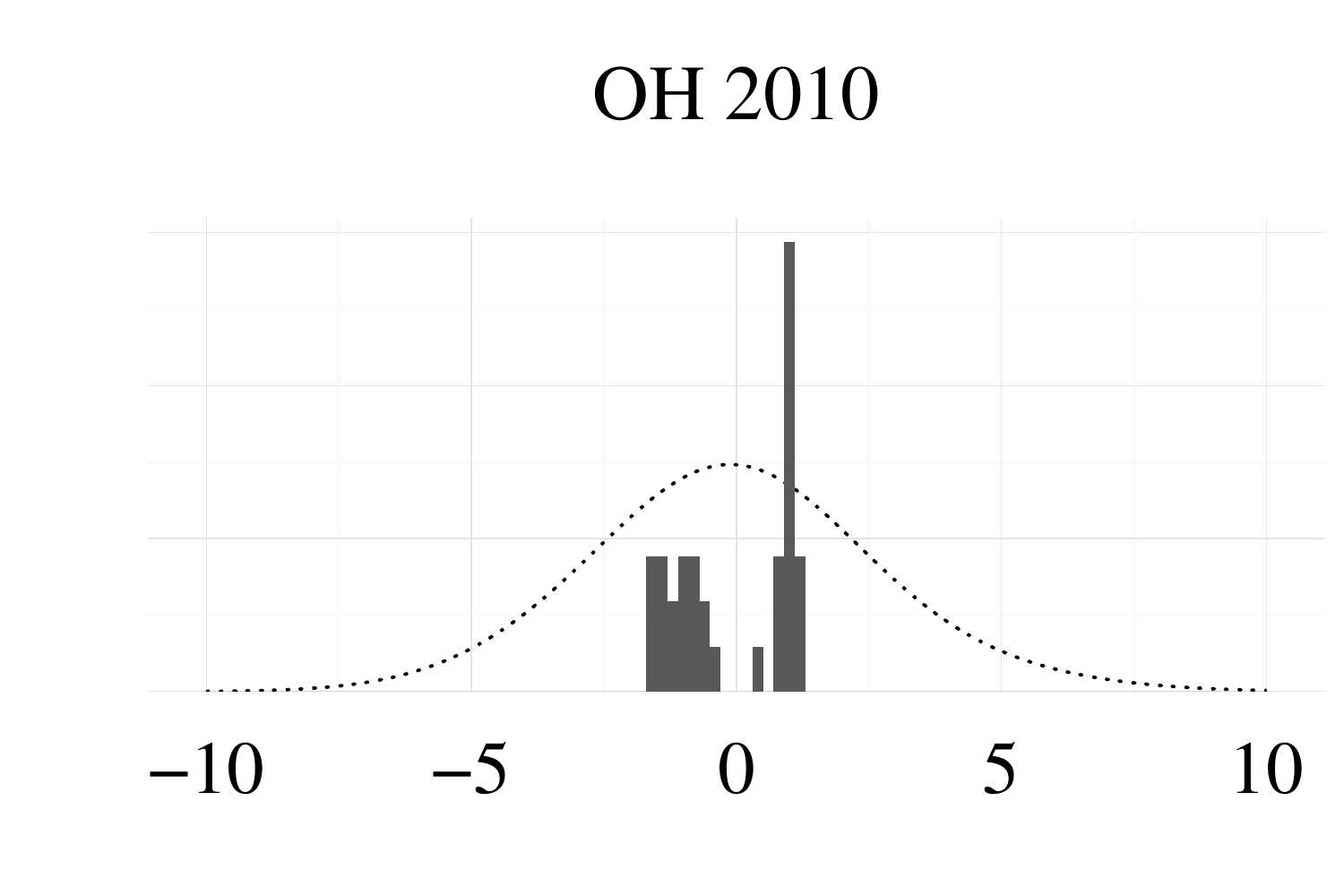}
                \par
            }
        \end{minipage}
        \end{centering}
        \par
    
      \caption{Distributions of inferred voter preferences (represented by lines) and candidate preferences (histograms) based on data of the 2006, 2008 and 2010 congressional elections in Texas, New York, and Ohio.}
      \label{fig:post-pred}
    \end{figure}

We apply our results to the question of trends in mass polarization for Texas, New York, and Ohio. We generate state-level distributions of voter preferences similar to the method described in \ref{sec:dist-agg-desc}. The resulting inferred distributions are shown in Fig. \ref{fig:post-pred}.

\subsubsection{Mass versus Elite Polarization.}
Using the state-level estimates, we compute the polarization metrics as described in Sect. \ref{sec:pol-background} for the inferred voter distributions and the corresponding candidate distributions.

First, as seen in Fig. \ref{fig:post-pred}, the inferred distributions of voter preferences visually appear unimodal, even though our model does not make this assumption at the state level. This is supported quantitatively in Table \ref{fig:pol-metrics}, where the electorate has relatively consistently smaller difference-of-means polarization metric than the candidates. This suggests there is no mass polarization in terms of bimodality within each state. Second, we find in Table \ref{fig:pol-metrics} that the standard deviation of the distribution of voter preferences is generally larger than that of the distribution of candidates. This consistent difference in dispersion suggests that voters often have more extreme preferences than the candidates running for office. Third, the kurtosis of the distribution of voter preferences is generally larger than that of the distribution of candidate preferences. Larger values of kurtosis indicate lower values of bimodality, so the consistent positive difference in kurtosis suggests that the voter distributions are unimodal. The results on kurtosis corroborate qualitative observation of our plots and the difference-of-means statistic. 

Together these findings suggest that the distribution of political candidates may not be representative of the distribution of preferences of the electorate in the years and states we study because the extremes and the center of the voter populations are underrepresented. Further, these findings also suggest that the elites are more polarized than the electorate in terms of bimodality, but the electorate is more polarized than the elites in terms of dispersion. Inferred distributions under alternative assumed numbers of cluster components and under alternative mixture component distributions are shown in the appendix in Fig.  \ref{fig:precdist-post-pred} and Fig. \ref{fig:clust-post-pred}. These robustness checks further support our dispersion finding but yield somewhat more mixed results on bimodality.

    \begin{table}
        \centering
        \caption{Polarization metrics of the voters given inferred voter preferences and of the candidates based on Bonica's CFscores \cite{bonica}. The ``difference'' of each polarization metric row is the difference between the voter metric and the candidate metric.}
        \vspace{1mm}

        \def\arraystretch{1.1}
        \setlength\tabcolsep{1mm}
        \begin{tabular}{|cl|lll|lll|lll|}
        \hline
        ~                     & ~          & \multicolumn{3}{c|}{\centering{2006}} & \multicolumn{3}{c|}{2008} & \multicolumn{3}{c|}{2010} \\
        ~                     & ~          & TX & NY & OH & TX & NY & OH & TX & NY & OH \\ \hline
        Difference- & Voters & 0.02 & 0.19 & 5.25 & 0.18 & 0.16 & 12.60 & 0.27 & 0.03 & 0.86 \\
        of-Means & Cand. & 2.00 & 1.33 & 2.11 & 0.36 & 0.62 & 1.88 & 0.13 & 1.96 & 2.16 \\
        ~ & \textbf{Diff} & \textbf{-1.98} & \textbf{-1.14} & \textbf{3.14} & \textbf{-0.17} & \textbf{-0.46} & \textbf{10.73} & \textbf{0.14} & \textbf{-1.94} & \textbf{-1.30} \\ \hline
        Standard                     & Voters &  4.36&39.54&68.20&2.14&78.46&74.23&3.88&39.30&2.75 \\
        Deviation & Cand. &
         0.97&0.77&1.22&1.07&0.83&0.99&1.22&0.81&1.07 \\
        
        ~                     & \textbf{Diff} & \textbf{3.38} & \textbf{38.77} & \textbf{66.97} & \textbf{1.07} & \textbf{77.64} & \textbf{73.24} & \textbf{2.66} & \textbf{38.49} & \textbf{1.68} \\ \hline
        ~                     & Voters     & 3.04  & 20.87 & 2.78 & 1.47 & 19.54 & 0.56  & 0.85  & 16.15 & 0.22 \\
        Kurtosis & Cand.  & -1.62 & 1.67 & 1.23 & 0.49 & 1.79 & -1.77 & 3.24  & -1.61 & -1.73   \\
        ~                     & \textbf{Diff} & \textbf{4.66}  & \textbf{19.20}   & \textbf{1.55} & \textbf{0.97} & \textbf{17.75} & \textbf{2.33} & \textbf{-2.39} & \textbf{17.76} & \textbf{1.95} \\ \hline
        \end{tabular}
        \label{fig:pol-metrics}
    \end{table}

\section{Discussion}

Our methods have some limitations that could be addressed in future work.
Firstly, biases inherent in our model could undermine our qualitative conclusions about polarization.
Because the tails of the component distributions in our model are nearly non-identifiable, the posterior distributions in our model end up placing a large mass on preference distributions with high variances. The inference procedure could be improved to yield more reliable estimates. Similarly, our inference procedure assumes unimodal component distributions, which could bias the aggregate state level output towards unimodal distributions. As a result, we may infer high-variance unimodal precinct distributions because of the biases in the component distributions. To address these concerns, we test the ability of our procedure to recover bimodal distributions at the state level with synthetic precinct-level data. We explicitly define bimodal precinct distributions in the synthetic data, and our inference procedure was still able to correctly recover bimodal district and state-level distributions under the assumption of unimodal component distributions. This indicates that our model is able to represent and recover bimodal preference distributions even if the model is making an incorrect assumption that individual precincts are unimodal. This result on synthetic data lends some confidence that our application results on real data are not artefactual. 
However, our inferences on real data still have some quirks. For example, the inferred distribution of the Ohio election in 2008 is implausibly wide. In explorations of ways to address this issue, we found that adding an informative prior on the variance helped to an extent.

Another limitation of our method is that we assume the observed vote shares of elections is tied to the underlying preference distributions. This allows the model to aggregate ideological preferences of individual voters within precincts. However, some political scientists believe voters tend to vote for candidates according to party affiliation rather than ideology \cite{bartels}. While this is a large limitation to our approach, future work could extend our model by estimating the extent to which people vote along party lines as opposed to according to policy preferences. As it stands, the distributions of preferences we infer can be interpreted as the distributions of \textit{expressed} preferences of voters, if not the distribution of actual preferences.

Finally, a more fundamental limitation is that we cannot validate the shape of the inferred precinct distributions with existing survey data, since we only have observations at the district level. Future work could create a test set through an in-depth survey of the distributions of preferences of particular precincts. Validating the distributions we infer is critical to bringing our work to a level that would be useful to practitioners. At the moment we cannot tell to what extent the shape of the individual precinct distributions we infer is determined by the data versus biases from our model.

\section{Conclusion}

In this work, we present a model to address the problem of understanding public opinion in the context of voter preferences and political polarization. Our model builds on a long literature of spatial voting models. We use mixtures of spatial voting models to infer clusters of U.S. precincts that display similar voting patterns.  Our model simultaneously infers preference distributions associated with those clusters utilizing data not only about voters, but also of candidates. These features allow us to analyze the distributions of voter preferences on their own and relative to distributions of candidate preferences. We infer voter preferences given precinct-level election results of three election cycles and three different states. We validate our inferences to the extent that we can using existing data.  We validate against alternative measures of public opinion, as well as by comparing the predictive power of our model and alternative methods.

One extension of this work could adapt the model to account for elections with an uncontested candidate using similar ideas tried by Levendusky, et al. \cite{lev}. Our model could also be updated to include an offset accounting for the number of political parties in the election, which can address the bias of results based on a two-party system. Another direction is to explore other applications of our inference methods in the field of political science. For example, inferred precinct-level preference distributions could predict the effects of congressional redistricting, the process of assigning geographic boundary lines to congressional districts. Precincts are the building blocks for districts, so we could use precinct preference estimates of our model to predict the effects of redistricting proposals on the makeup of Congress.

Variations of our model could be applied more broadly beyond the scope of political science to understand distributions of preferences on other topics. For example, surveys are also used to understand consumer preferences on certain consumer products. A variation of our model could avoid the need for surveys or supplement surveys in this and other areas.

\section{Acknowledgements}

Special thanks to David Lazer for bringing our attention to Adam
Bonica's work, to David Parkes for suggesting a reformulation of our
model that enabled integrating out voter positions, and to Matt
Blackwell for encouraging us to think more about validity and
identifiability.  This work was supported in part by the NSF GRFP
under grant \#1122374. Any opinions, findings, and conclusions or
recommendations expressed in this material are those of the authors
and do not necessarily reflect those of the sponsors.

\bibliographystyle{splncs03}
\bibliography{sigproc}

\begin{thebibliography}{10}
\providecommand{\url}[1]{\texttt{#1}}
\providecommand{\urlprefix}{URL }

\bibitem{abramowitz}
Abramowitz, A.I., Saunders, K.L.: Is polarization a myth? The Journal of
  Politics  70(2),  542--555 (2008)

\bibitem{airoldi2009mixed}
Airoldi, E.M., Blei, D.M., Fienberg, S.E., Xing, E.P.: Mixed membership
  stochastic blockmodels. In: Advances in Neural Information Processing
  Systems. pp. 33--40 (2009)

\bibitem{heda}
Ansolabehere, S., Palmer, M., Lee, A.: Precinct-level election data.
  http://hdl.handle.net/1902.1/21919 (2014)

\bibitem{cces}
Ansolabehere, S., Pettigrew, S.: Cumulative {CCES} {C}ommon {C}ontent
  (2006-2012). http://dx.doi.org/10.7910/DVN/26451 (2014)

\bibitem{barbera2015birds}
Barber{\'a}, P.: Birds of the same feather tweet together: Bayesian ideal point
  estimation using twitter data. Political Analysis  23(1),  76--91 (2015)

\bibitem{bartels}
Bartels, L.M.: Beyond the running tally: Partisan bias in political
  perceptions. Political Behavior  24(2),  117--150 (2002)

\bibitem{dime}
Bonica, A.: Database on ideology, money in politics, and elections: Public
  version 1.0. http://data.stanford.edu/dime (2013)

\bibitem{bonica}
Bonica, A.: Mapping the ideological marketplace. American Journal of Political
  Science  58(2),  367--386 (2014)

\bibitem{dimaggio}
DiMaggio, P., Evans, J., Bryson, B.: Have {A}merican's social attitudes become
  more polarized? American Journal of Sociology pp. 690--755 (1996)

\bibitem{ding2015learning}
Ding, W., Ishwar, P., Saligrama, V.: Learning mixed membership mallows models
  from pairwise comparisons. arXiv:1504.00757  (2015)

\bibitem{downs}
Downs, A.: An economic theory of political action in a democracy. The Journal
  of Political Economy pp. 135--150 (1957)

\bibitem{hinich1984spatial}
Enelow, J.M., Hinich, M.J.: The Spatial Theory of Voting: An Introduction.
  Cambridge Univ Press (1984)

\bibitem{fiorina-abrams}
Fiorina, M.P., Abrams, S.J.: Political polarization in the {A}merican public.
  Annual Review of Political Science  11,  563--588 (2008)

\bibitem{fiorina-abrams-pope}
Fiorina, M.P., Abrams, S.J., Pope, J.C.: Polarization in the {A}merican public:
  Misconceptions and misreadings. The Journal of Politics  70(2),  556--560
  (2008)

\bibitem{flaxman2015supported}
Flaxman, S.R., Wang, Y.X., Smola, A.J.: Who supported {O}bama in 2012?:
  Ecological inference through distribution regression. In: Proceedings of the
  21th ACM SIGKDD International Conference on Knowledge Discovery and Data
  Mining. pp. 289--298. ACM (2015)

\bibitem{gerber2004beyond}
Gerber, E.R., Lewis, J.B.: Beyond the median: Voter preferences, district
  heterogeneity, and political representation. Journal of Political Economy
  112(6),  1364--1383 (2004)

\bibitem{gerrish}
Gerrish, S., Blei, D.M.: Predicting legislative roll calls from text. In:
  Proceedings of the 28th International Conference on Machine Learning. pp.
  489--496 (2011)

\bibitem{krafft2012topic}
Krafft, P., Moore, J., Desmarais, B., Wallach, H.M.: Topic-partitioned
  multinetwork embeddings. In: Advances in Neural Information Processing
  Systems. pp. 2807--2815 (2012)

\bibitem{lee}
Lee, J.M.: Assessing mass opinion polarization in the {U.S.} using relative
  distribution method. Social Indicators Research pp. 1--28 (2014)

\bibitem{lev}
Levendusky, M.S., Pope, J.C., Jackman, S.D.: Measuring district-level
  partisanship with implications for the analysis of {U.S.} elections. The
  Journal of Politics  70(3),  736--753 (2008)

\bibitem{lev-pope}
Levendusky, M.S., Pope, J.C.: Red states vs. blue states going beyond the mean.
  Public Opinion Quarterly  75(2),  227--248 (2011)

\bibitem{lewis2001estimating}
Lewis, J.B.: Estimating voter preference distributions from individual-level
  voting data. Political Analysis  9(3),  275--297 (2001)

\bibitem{cdmaps}
Lewis, J.B., DeVine, B., Pitcher, L., Martis, K.C.: Digital boundary
  definitions of {U}nited {S}tates congressional districts, 1789-2012.
  http://cdmaps.polisci.ucla.edu (2013)

\bibitem{poole-rosenthal}
McCarty, N., Poole, K.T., Rosenthal, H.: Polarized America: The Dance of
  Ideology and Unequal Riches, vol.~5. MIT Press (2006)

\bibitem{dw-nominate}
Poole, K.T., Rosenthal, H.: A spatial model for legislative roll call analysis.
  American Journal of Political Science pp. 357--384 (1985)

\bibitem{derek-ruths}
Ruths, D., Pfeffer, J.: Social media for large studies of behavior. Science
  346(6213),  1063--1064 (2014)

\bibitem{warshaw}
Tausanovitch, C., Warshaw, C.: Measuring constituent policy preferences in
  congress, state legislatures, and cities. The Journal of Politics  75(2),
  330--342 (2013)

\bibitem{tigerweb}
{United States Census Bureau}: Tigerweb state-based data files: Voting
  districts - {C}ensus 2010. http://tigerweb.geo.census.gov (2010)

\bibitem{fowler}
Zhang, Y., Friend, A., Traud, A.L., Porter, M.A., Fowler, J.H., Mucha, P.J.:
  Community structure in congressional cosponsorship networks. Physica A:
  Statistical Mechanics and its Applications  387(7),  1705--1712 (2008)

\end{thebibliography}

\appendix

\renewcommand{\thefigure}{S\arabic{figure}}
\renewcommand{\thetable}{S\arabic{table}}

\setcounter{figure}{0}
\setcounter{table}{0}

\section{Mathematical Definitions of DiMaggio's Polarization Metrics}
\label{sec:math-pol}

Given the estimates of our model, we use the following analytical form of the standard deviation of a mixture model to measure political polarization in terms of dispersion:
    \begin{equation} \label{eqn:weighted-variance}
        M_{\sigma} = \sqrt{\left(\sum_{i=1}^K \frac{n_i}{\sum_{j=1}^K n_j} ({\mu_i}^2 + {\sigma_i}^2) \right) - M_\mu}
    \end{equation}
where $n_i$ is the total number of voters assigned to component $i$ and $M_{\mu}$ is the weighted mean of the mixture distribution of voter preferences.

To measure political polarization in terms of bimodality, we use kurtosis. Kurtosis is the fourth central moment of the mixture distribution divided by the square of the variance of the mixture distribution. We use the following analytical form:
    \begin{equation}
        M_k = \frac{E[(X - M_\mu)]^4}{M_{\sigma}^4}
    \end{equation}
where $X$ is a random variable drawn from the mixture distribution and hence the numerator is the fourth central moment of the mixture distribution. The analytical form to compute the z-th central moment of the mixture distribution is below. 
    \begin{equation}
        E[(X - M_\mu)^z] = \sum_{i=1}^K \sum_{j=1}^z {z \choose j}  (\mu_i - M_\mu)^{z - j} w_i \, E[(Y_i - \mu_i)^z]
    \end{equation}
where $Y_i$ is a random variable drawn from component $i$ of the mixture distribution, $w_i$ is the weight of each component, and $E[(Y_i - \mu_i)^z]$ is the z-th central moment of the $i$th component distribution. In our analysis, we weight each component in the mixture distribution by the proportion of the population assigned that component.

\section{Additional Results}
\label{sec:addl-results}

In Sect. \ref{sec:validation}, we presented the results of our method assuming the underlying component distribution is Normal and the number of clusters $(K)$ is 4. This section tests the robustness of these assumptions and presents our results when varying the underlying component distribution and the number of clusters.

\subsection{Varying the Underlying Component Distribution}
\begin{figure}
    \begin{minipage}{0.2cm}
        \rotatebox{90}{\textbf{Uniform distribution}}
    \end{minipage}
    \begin{minipage}{\dimexpr\linewidth-1cm\relax}%
        \raisebox{\dimexpr-.5\height-1em}{
            \includegraphics[width=0.35\linewidth]{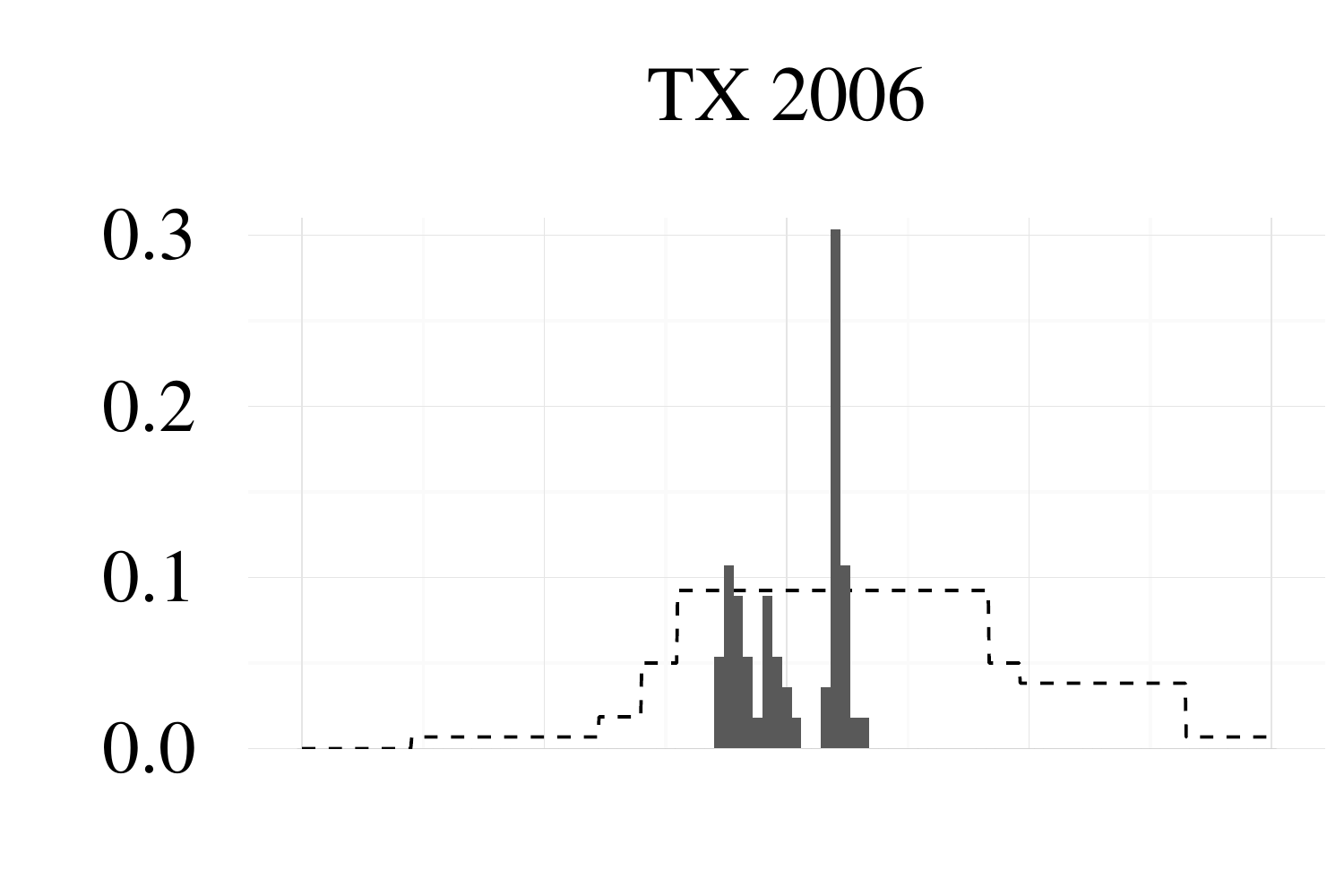}
            \hspace{-0.5cm}
            \includegraphics[width=0.35\linewidth]{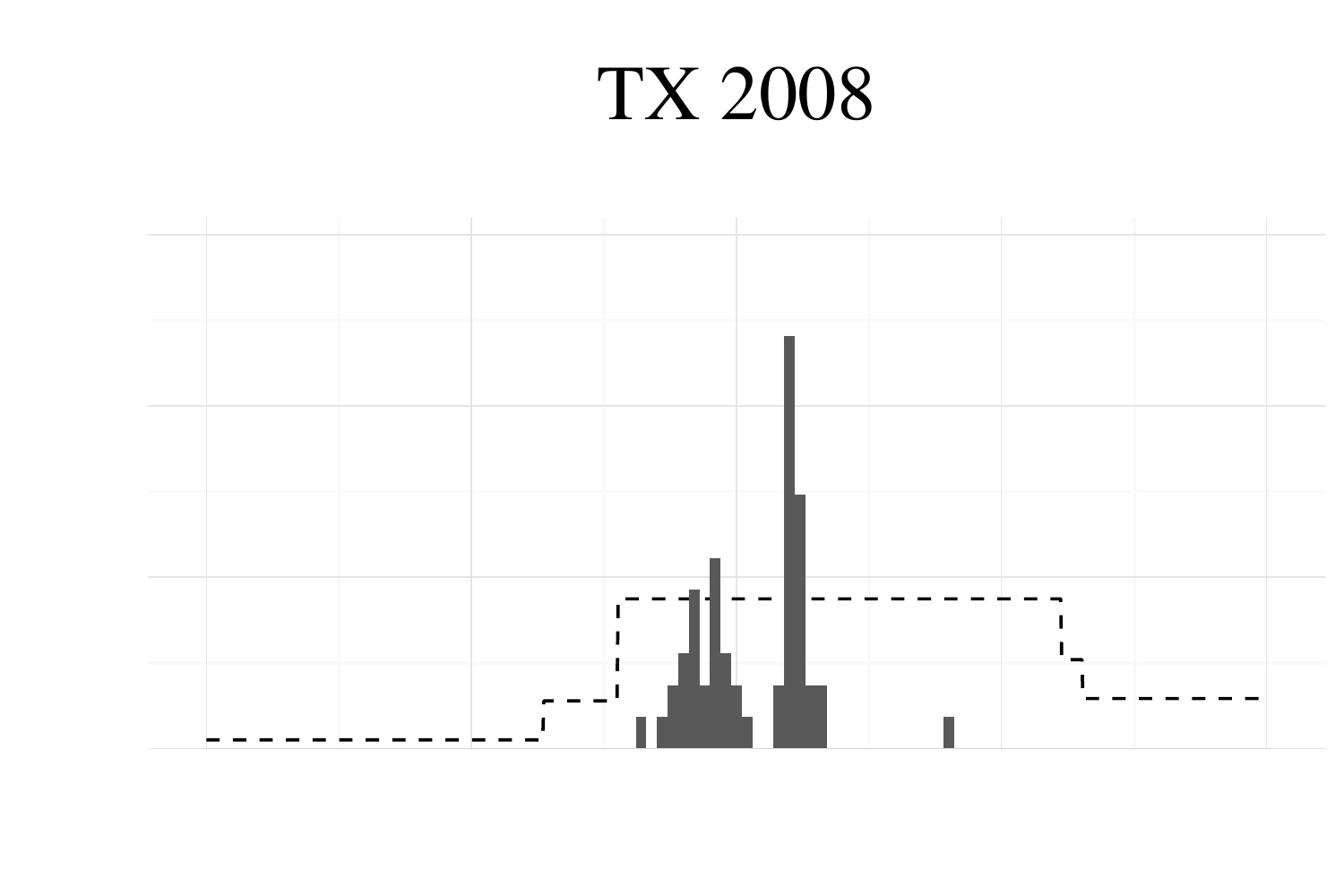}
            \hspace{-0.5cm}
            \includegraphics[width=0.35\linewidth]{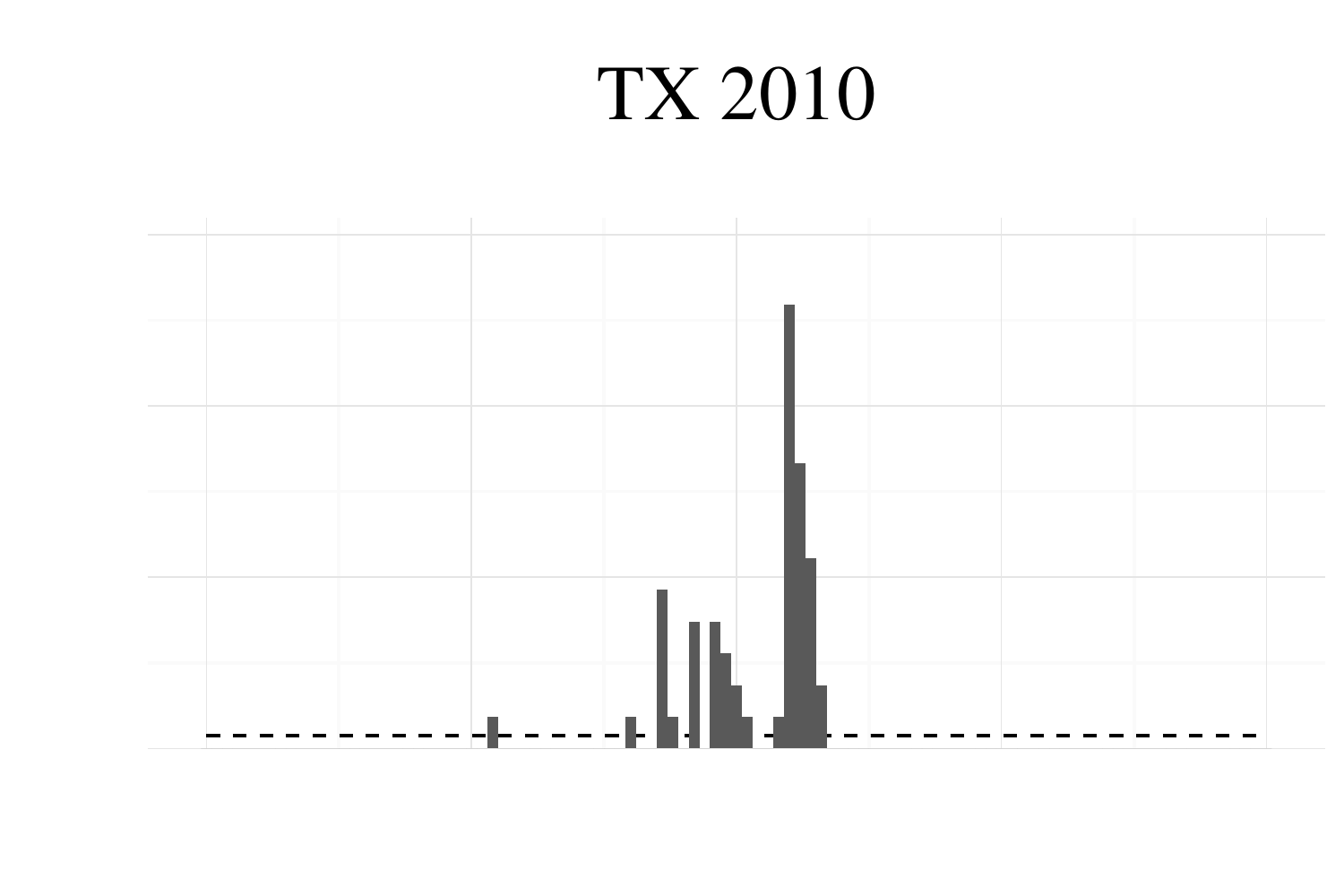}
            \par
        } \\[-15pt]
        \raisebox{\dimexpr-.5\height-1em}{
            \includegraphics[width=0.35\linewidth]{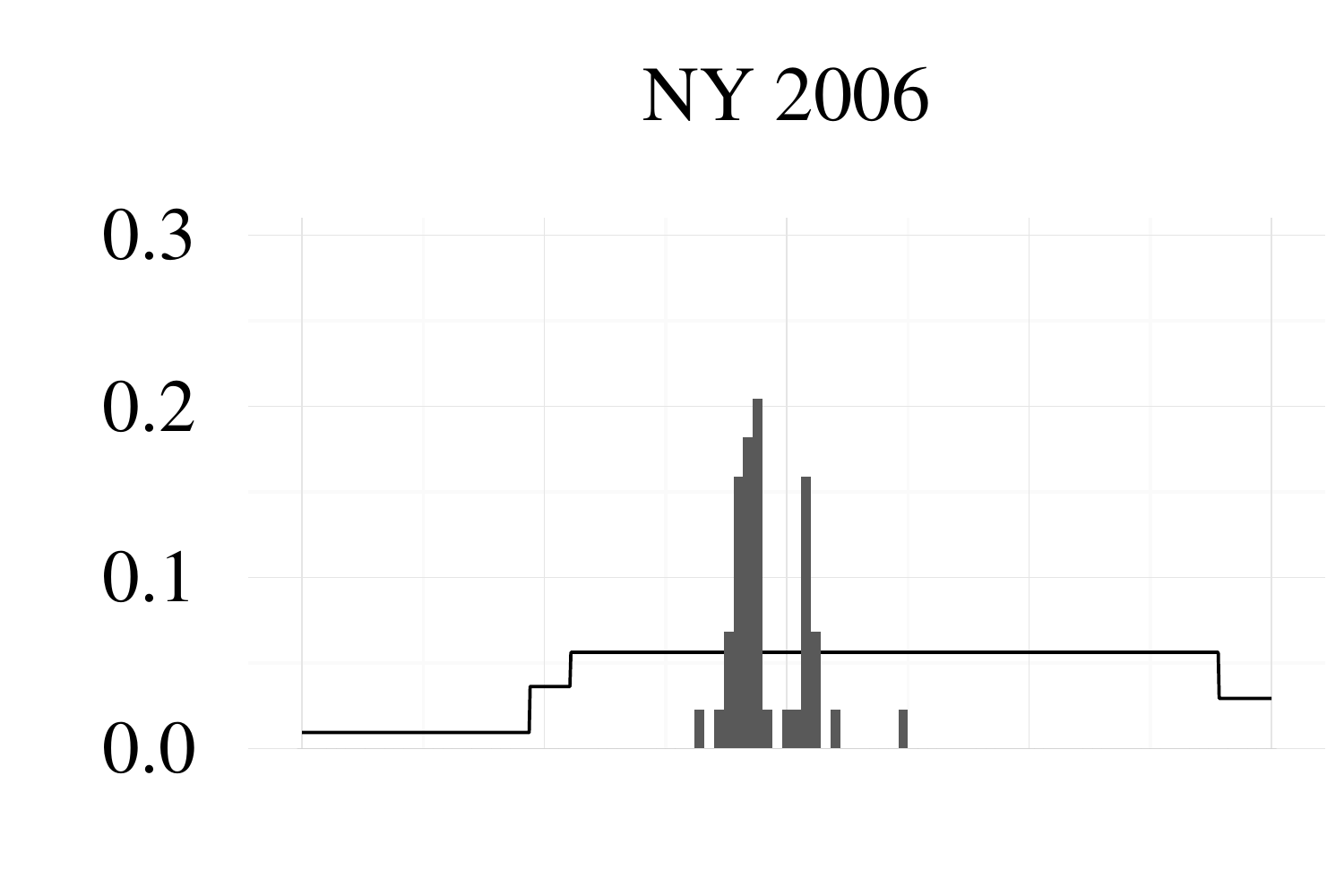}
            \hspace{-0.5cm}
            \includegraphics[width=0.35\linewidth]{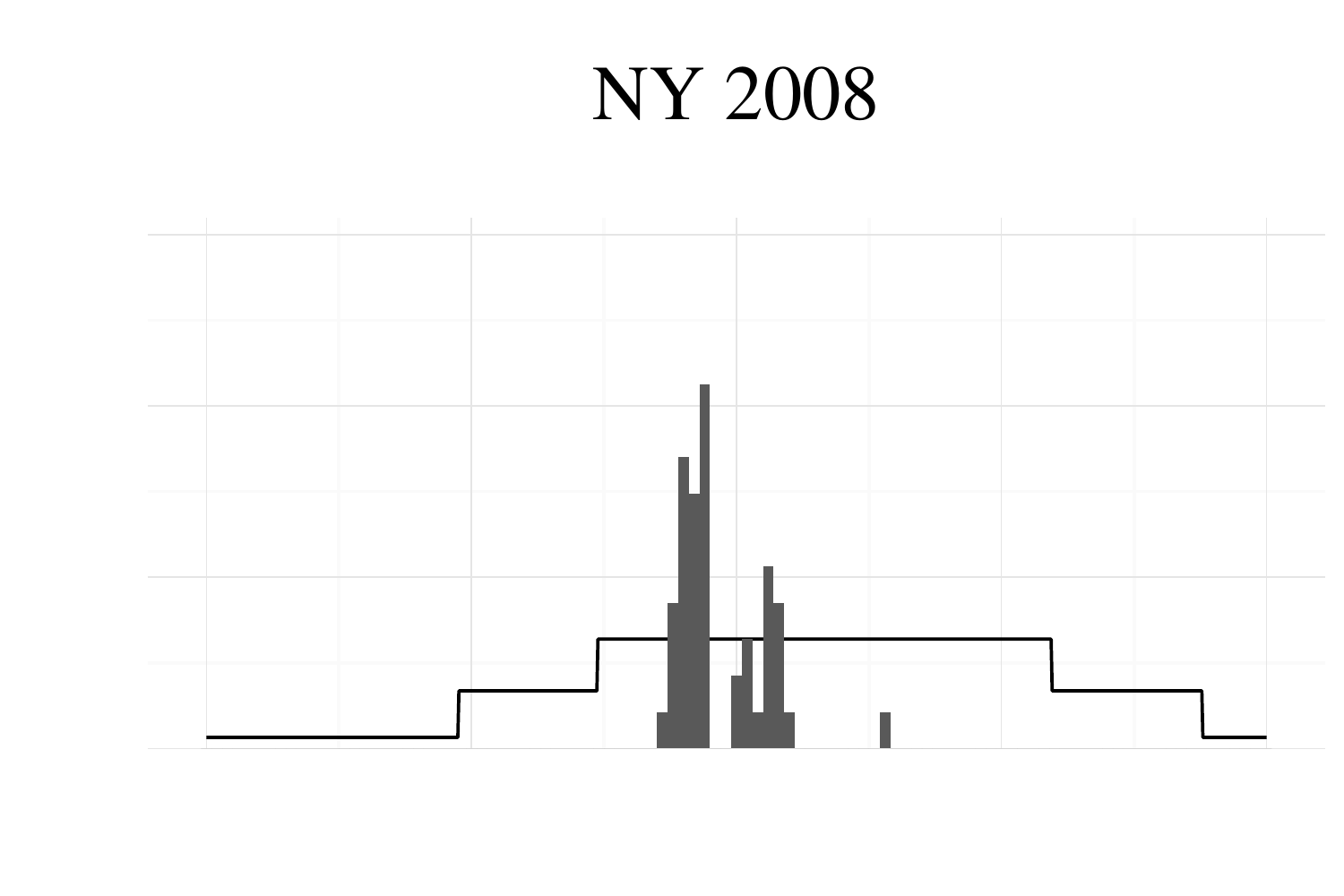}
            \hspace{-0.5cm}
            \includegraphics[width=0.35\linewidth]{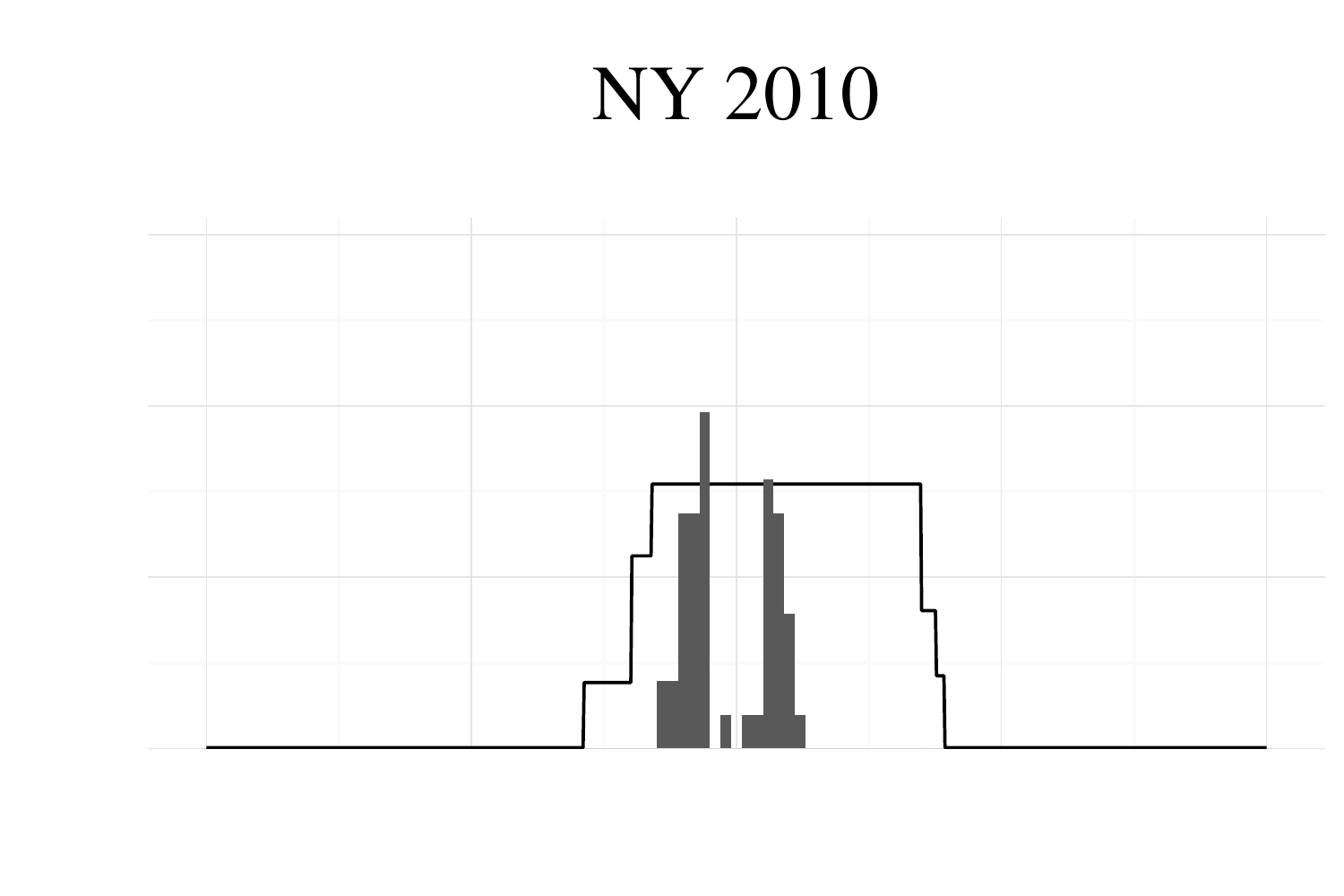}
            \par
        } \\[-15pt]
        \raisebox{\dimexpr-.5\height-1em}{
            \includegraphics[width=0.35\linewidth]{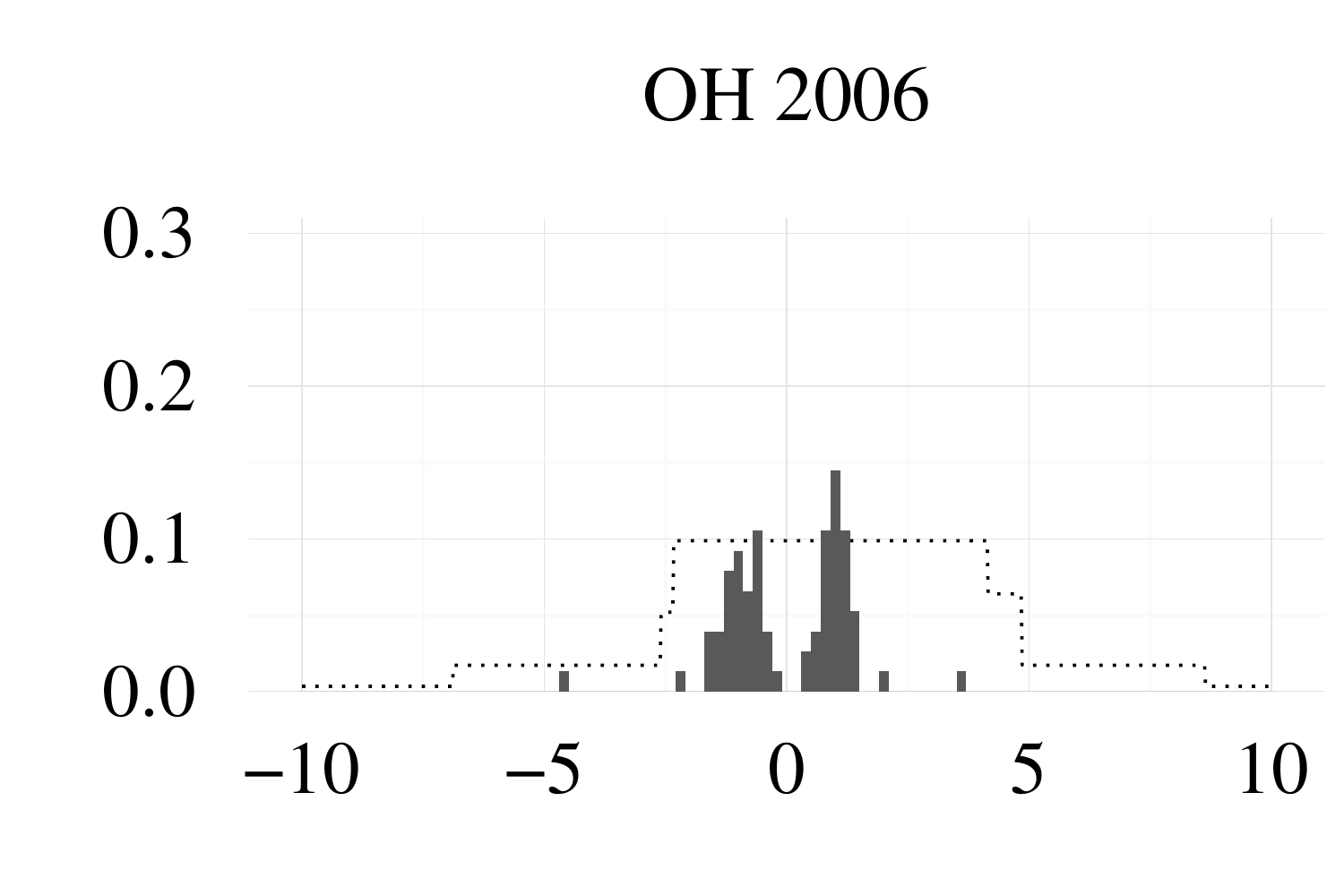}
            \hspace{-0.5cm}
            \includegraphics[width=0.35\linewidth]{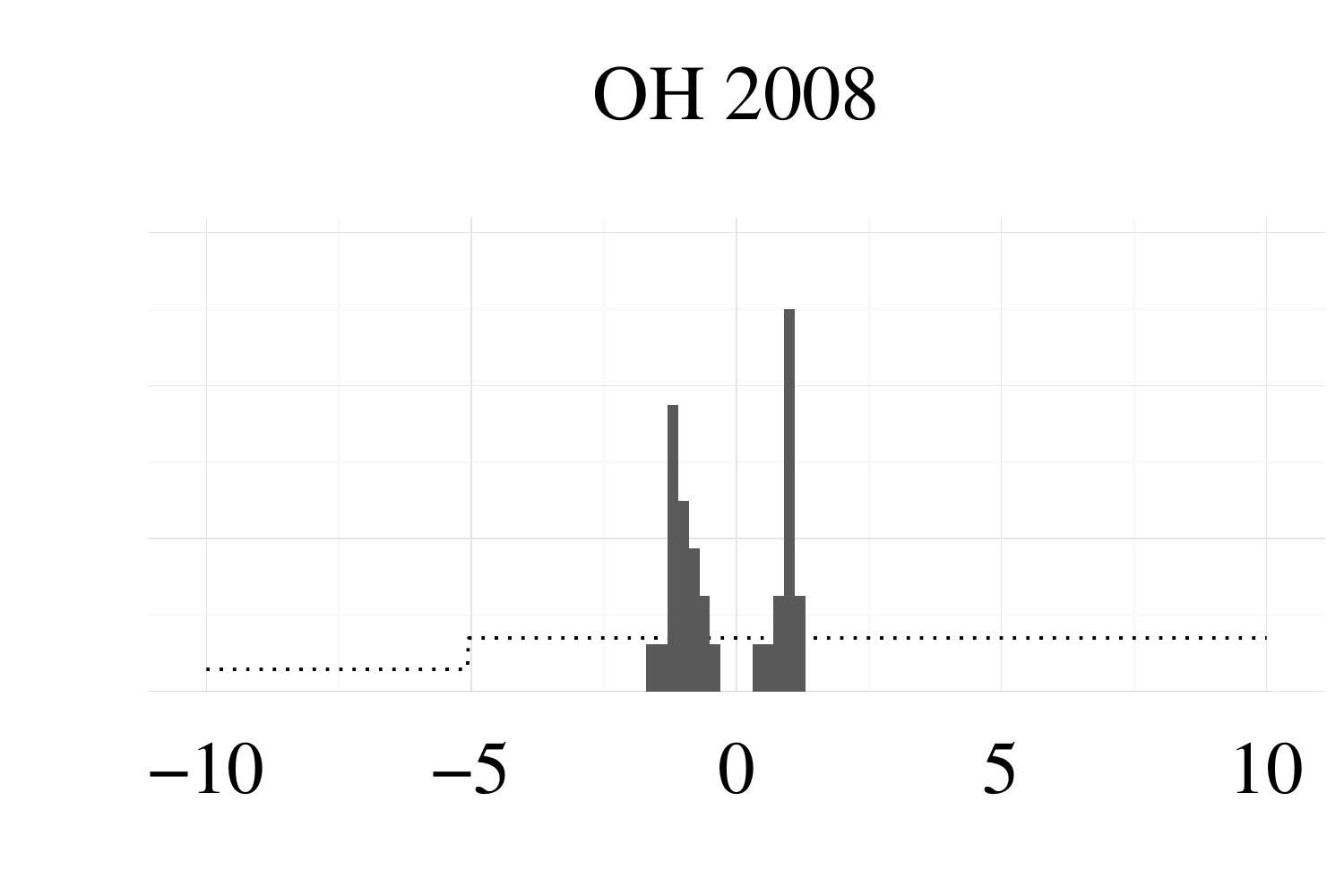}
            \hspace{-0.5cm}
            \includegraphics[width=0.35\linewidth]{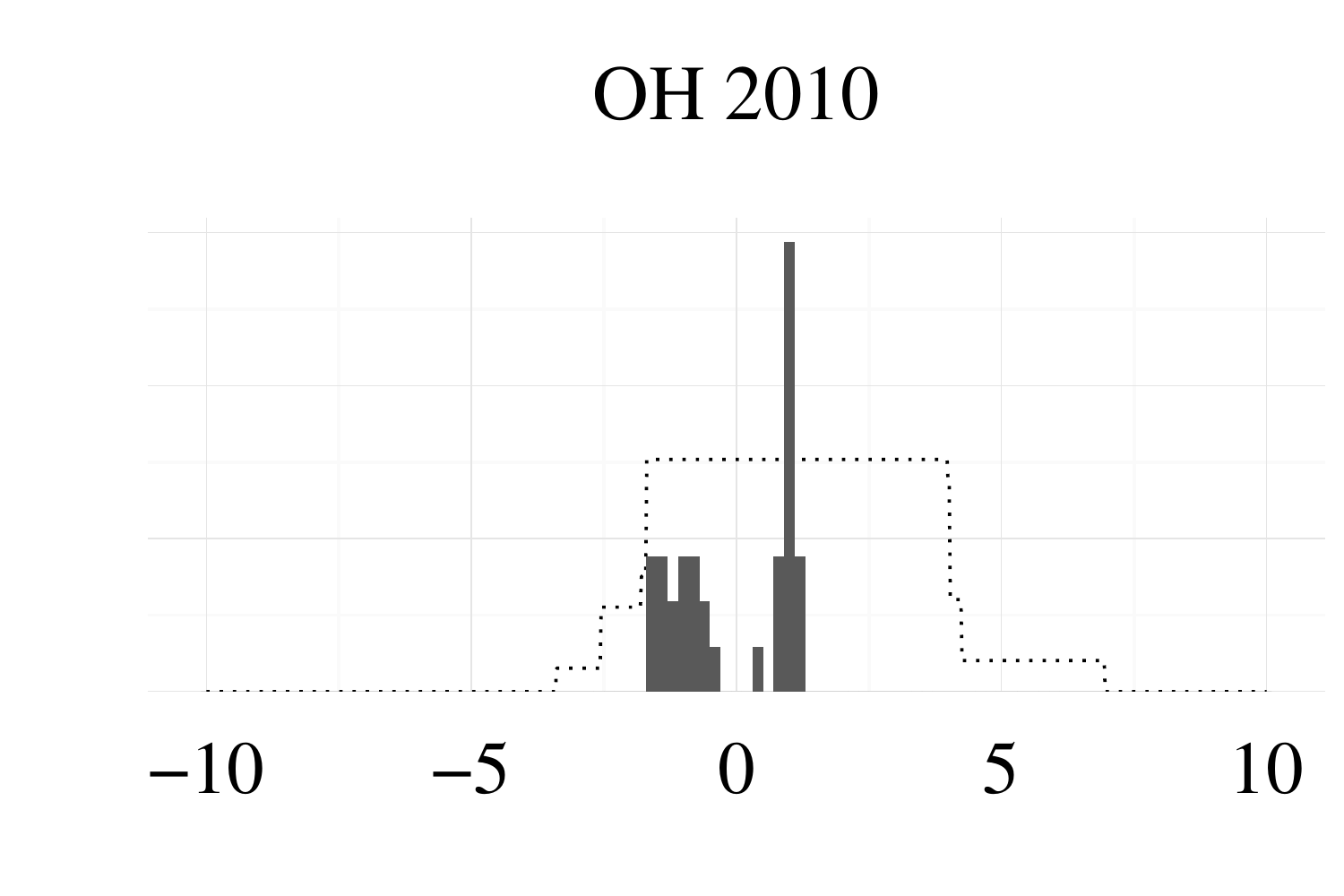}
            \par
        }
    \end{minipage}
    \par

    \begin{minipage}{1cm}
        \rotatebox{90}{\textbf{Laplace distribution}}
    \end{minipage}
    \begin{minipage}{\dimexpr\linewidth-1cm\relax}%
        \raisebox{\dimexpr-.5\height-1em}{
            \includegraphics[width=0.35\linewidth]{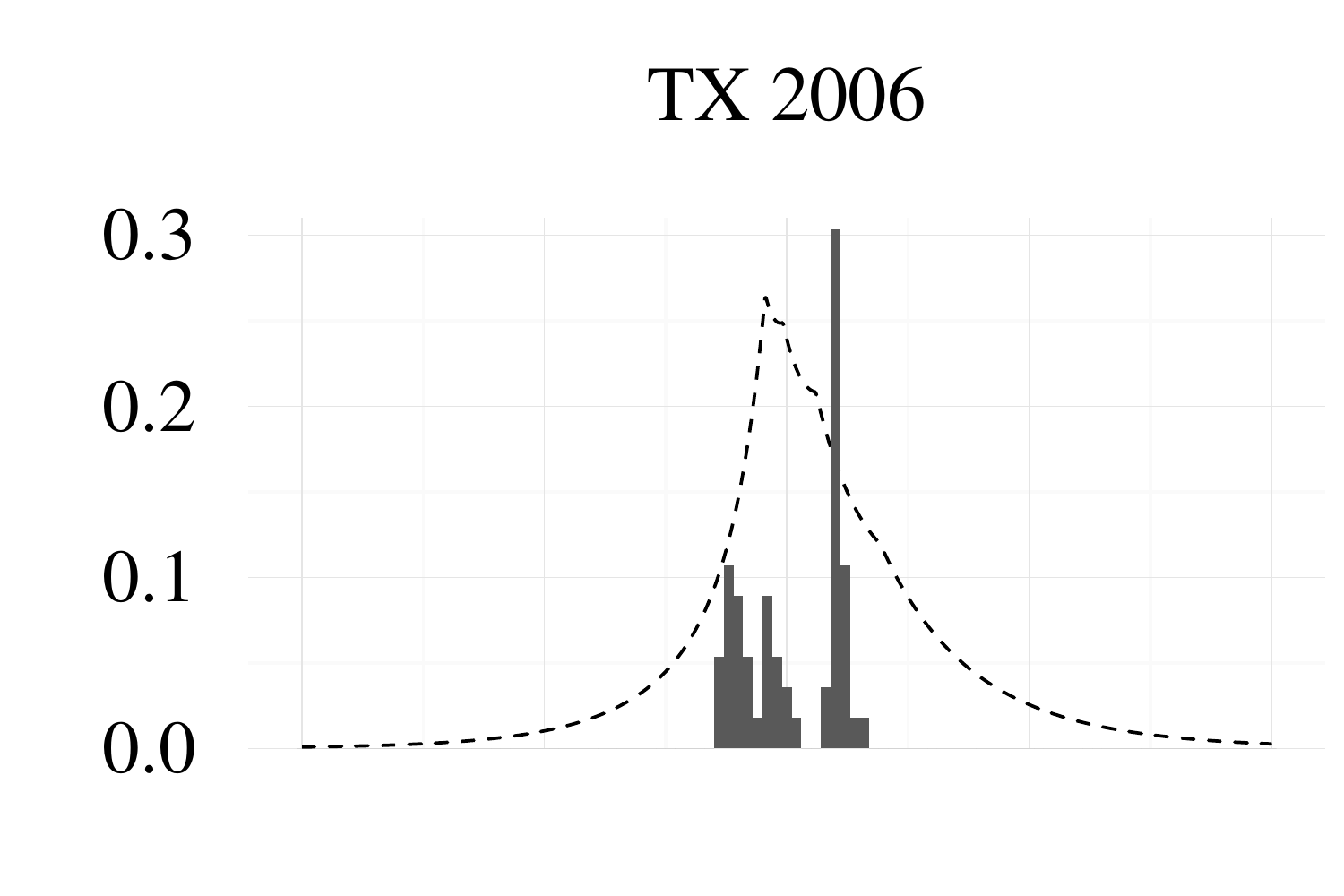}
            \hspace{-0.5cm}
            \includegraphics[width=0.35\linewidth]{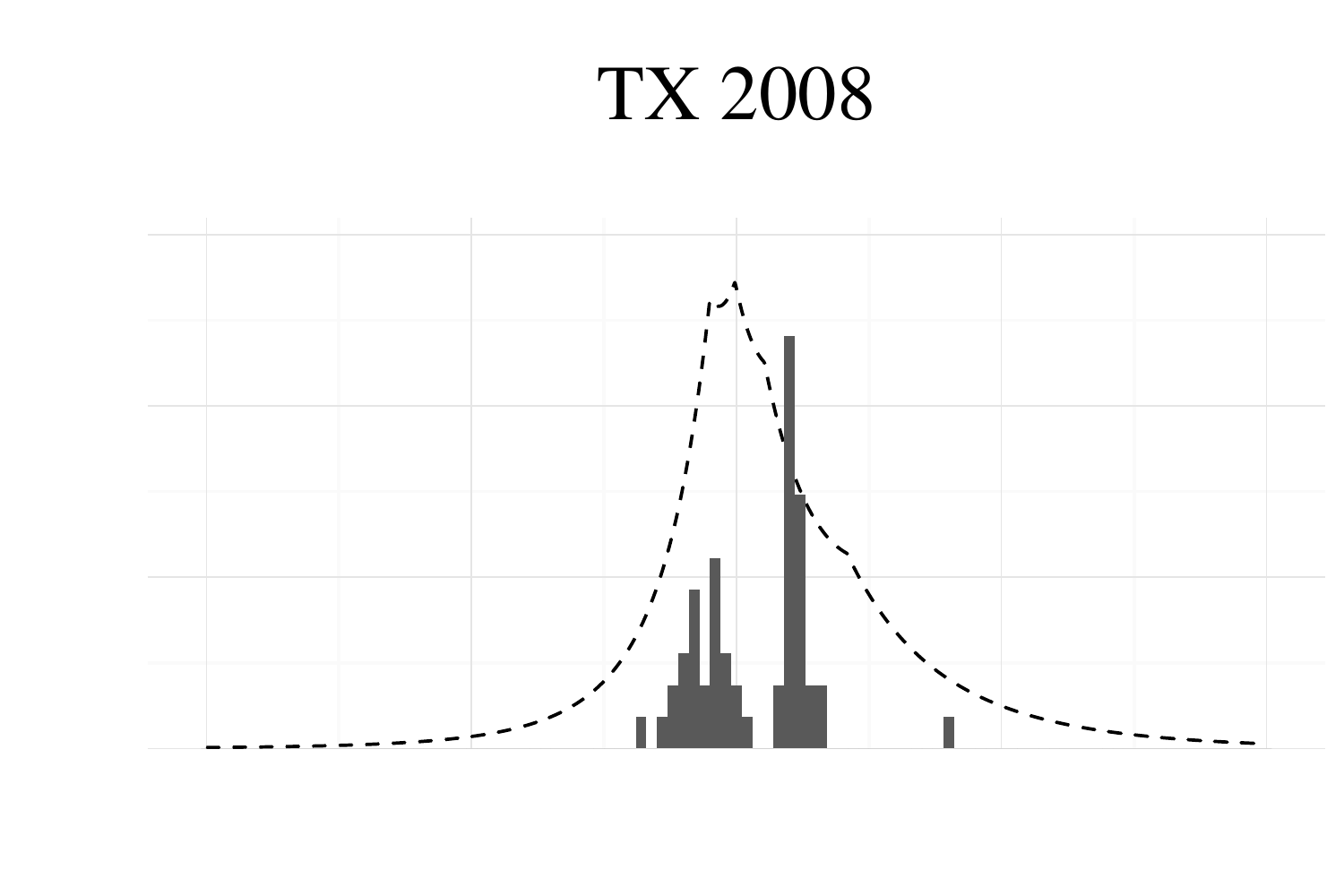}
            \hspace{-0.5cm}
            \includegraphics[width=0.35\linewidth]{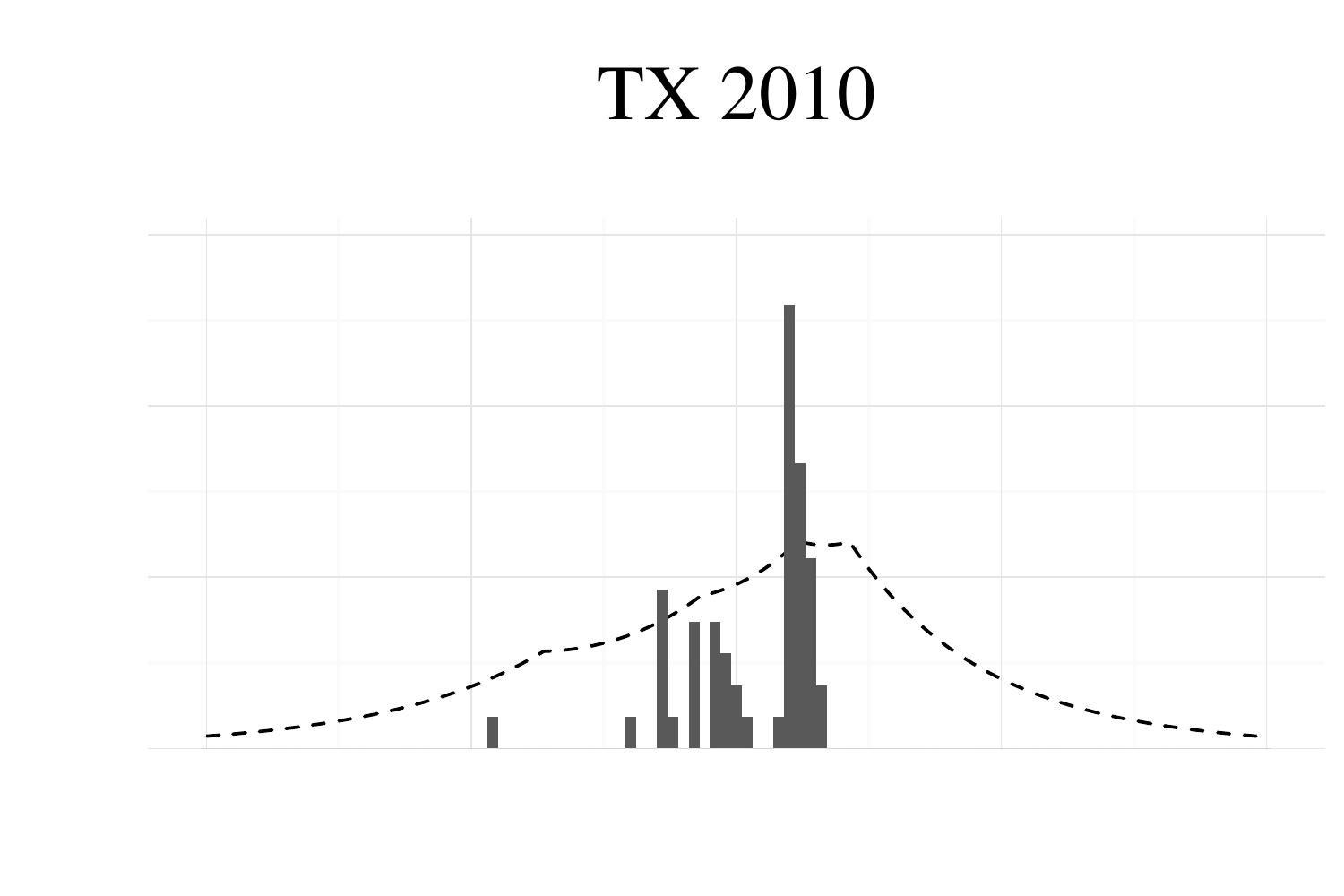}
            \par
        } \\[-15pt]
        \raisebox{\dimexpr-.5\height-1em}{
            \includegraphics[width=0.35\linewidth]{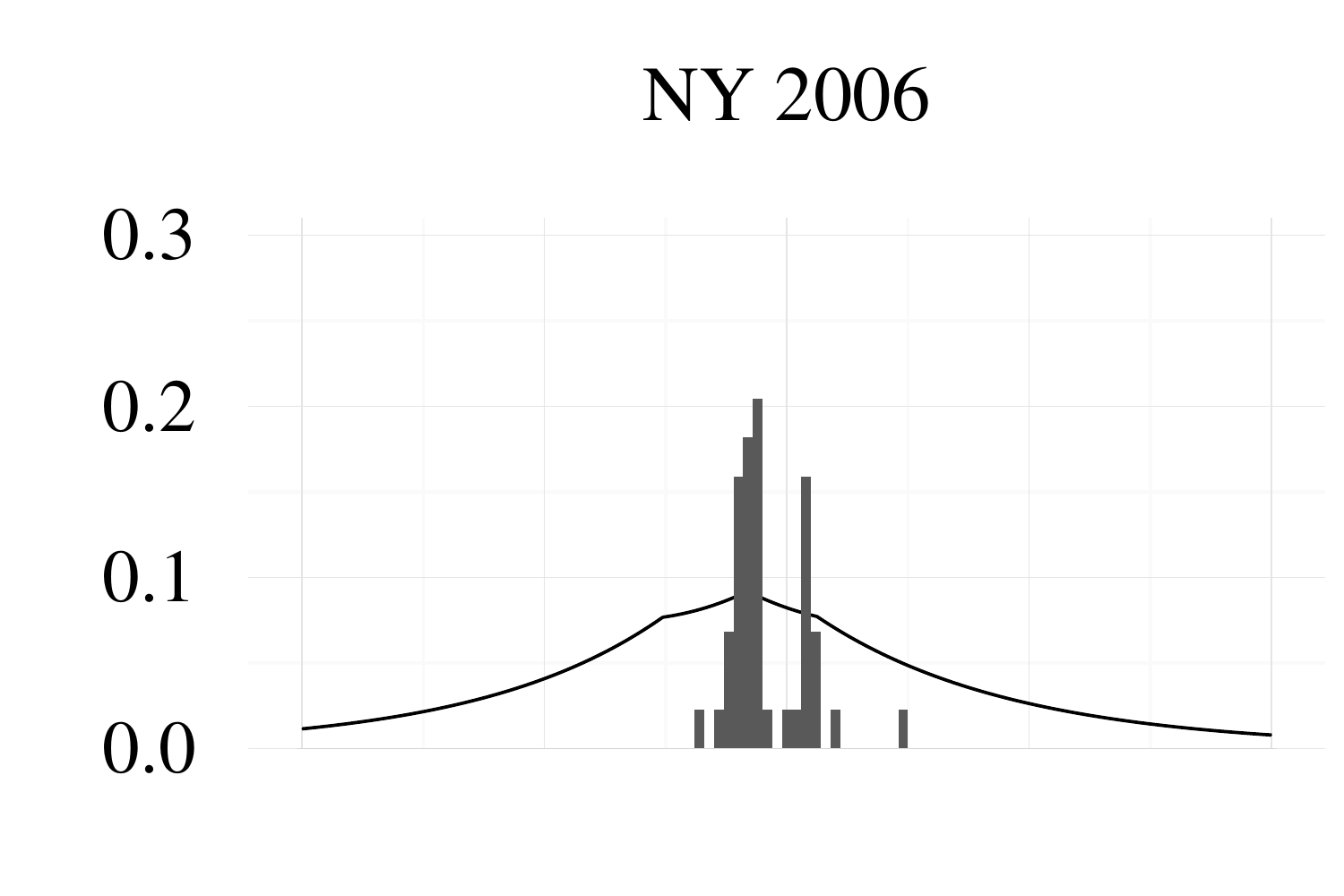}
            \hspace{-0.5cm}
            \includegraphics[width=0.35\linewidth]{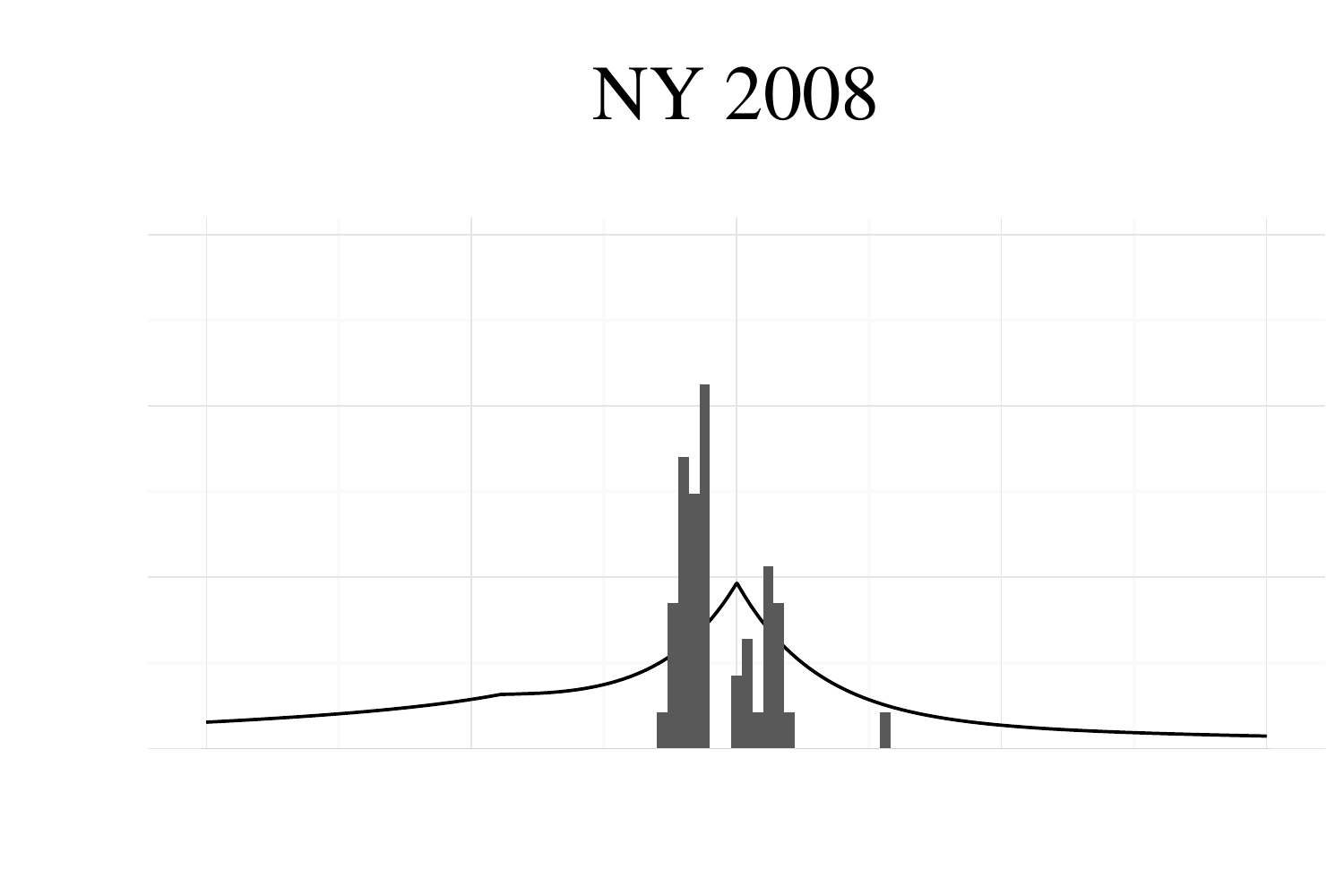}
            \hspace{-0.5cm}
            \includegraphics[width=0.35\linewidth]{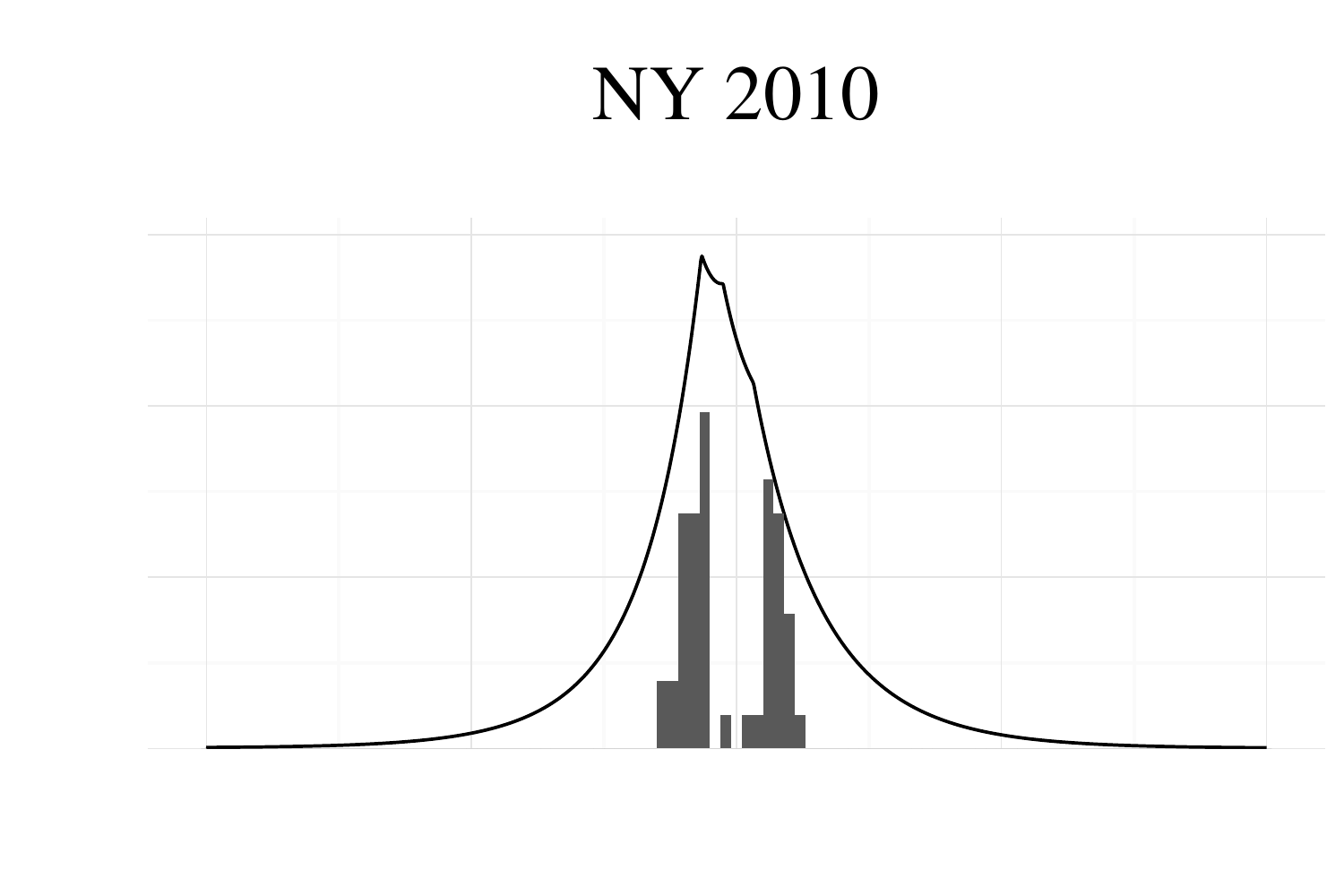}
            \par
        } \\[-15pt]
        \raisebox{\dimexpr-.5\height-1em}{
            \includegraphics[width=0.35\linewidth]{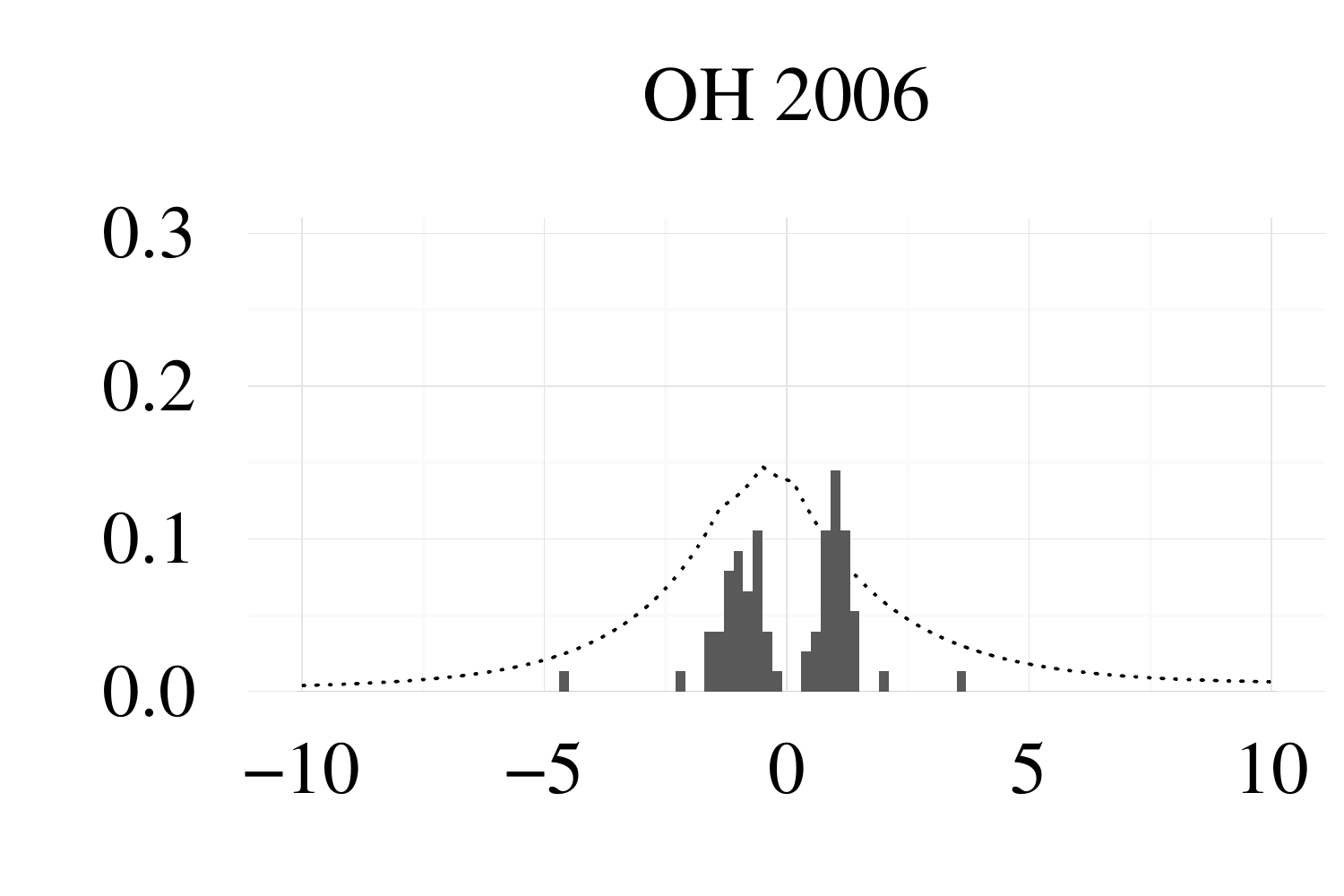}
            \hspace{-0.5cm}
            \includegraphics[width=0.35\linewidth]{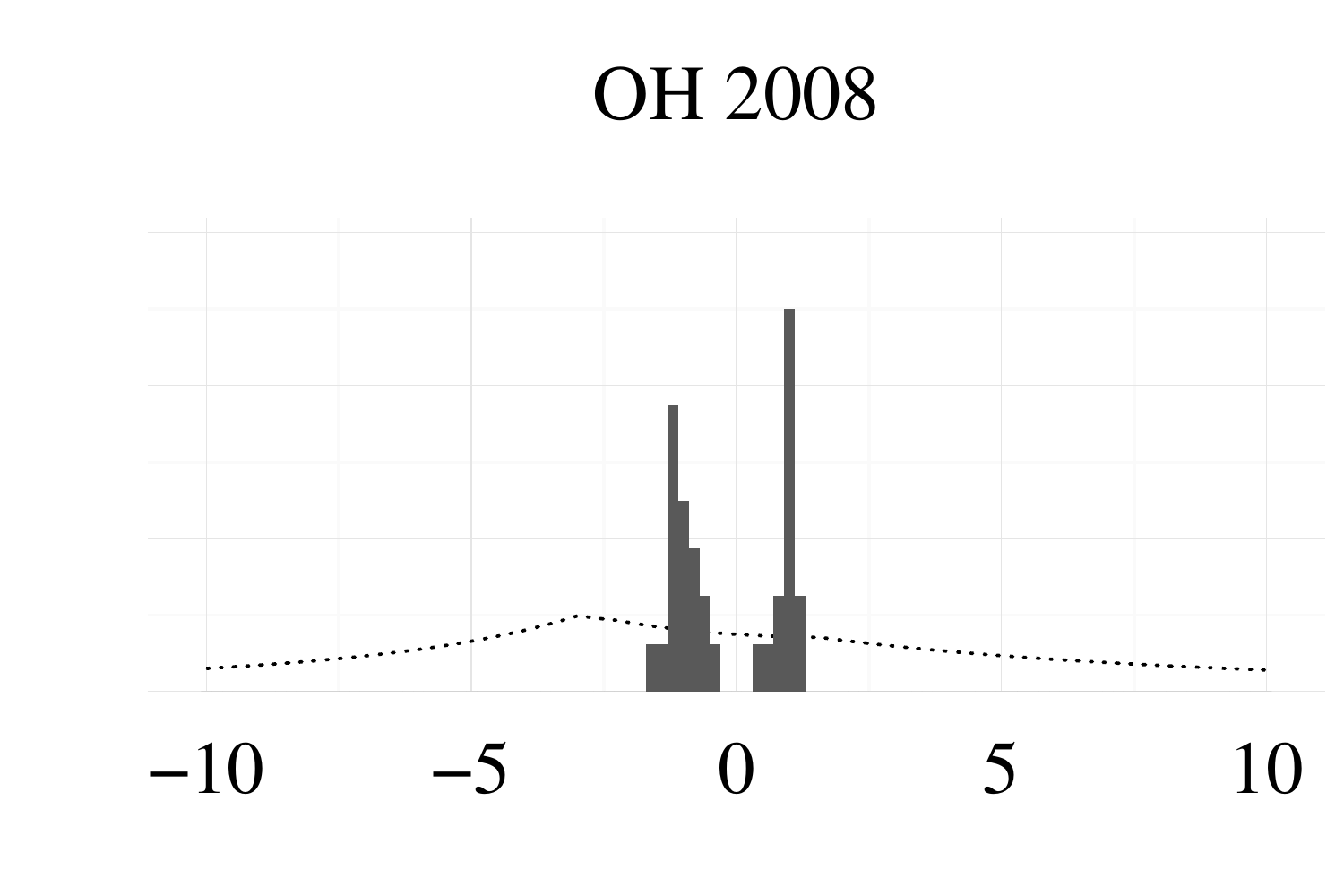}
            \hspace{-0.5cm}
            \includegraphics[width=0.35\linewidth]{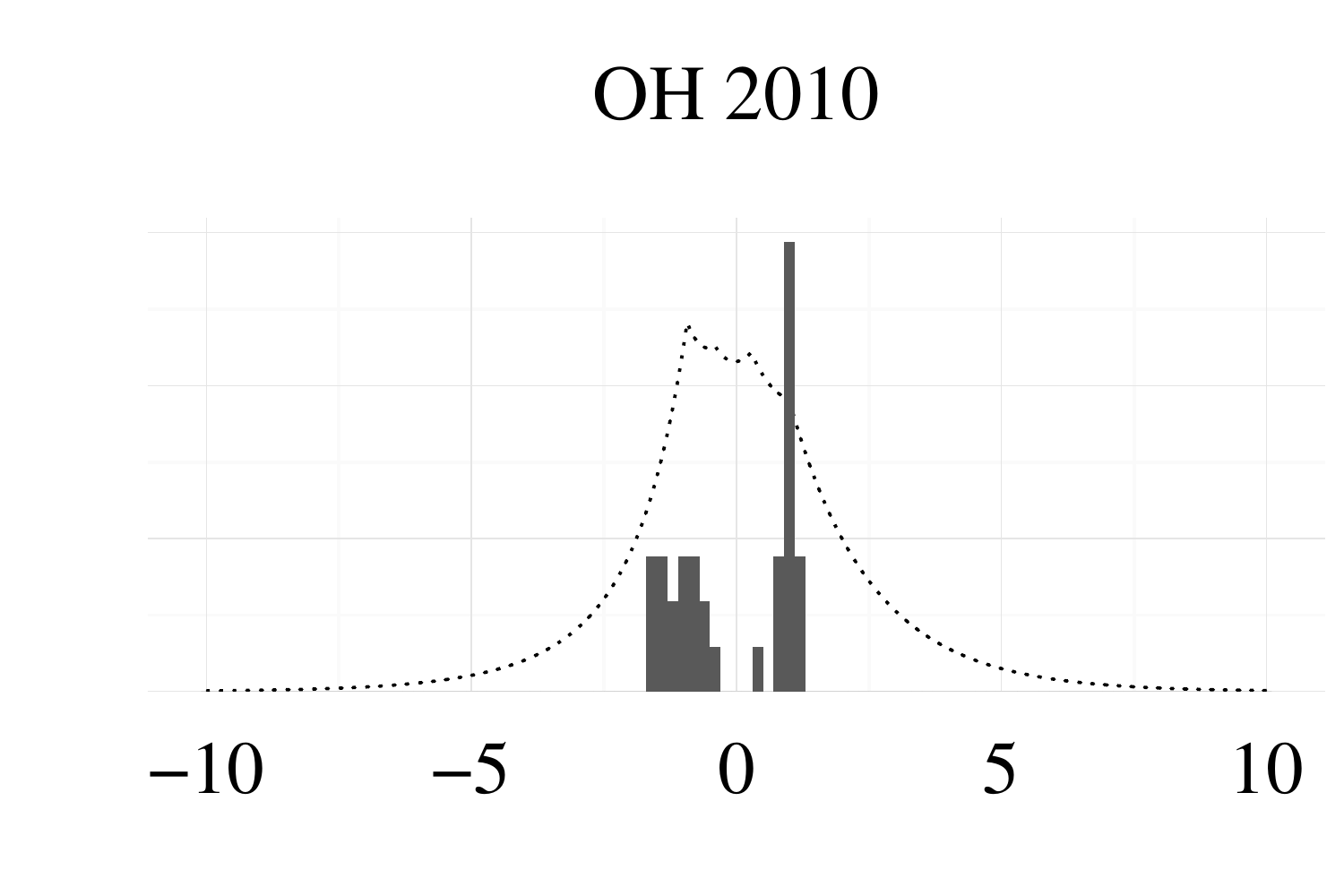}
            \par
        }

    \end{minipage}
    \par

  \caption{Distributions of inferred voter preferences (represented by lines) under alternative assumed mixture component distributions, and candidate preferences (histograms) based on data of the 2006, 2008 and 2010 congressional elections in Texas, New York, and Ohio.}
  \label{fig:precdist-post-pred}
\end{figure}

\begin{figure}
    \begin{centering}
        \begin{minipage}{0.75cm}
            \rotatebox{90}{\textbf{Uniform distribution}}
        \end{minipage}
        \begin{minipage}{\dimexpr\linewidth-3cm\relax}%
            \centering
            \includegraphics[width=\linewidth]{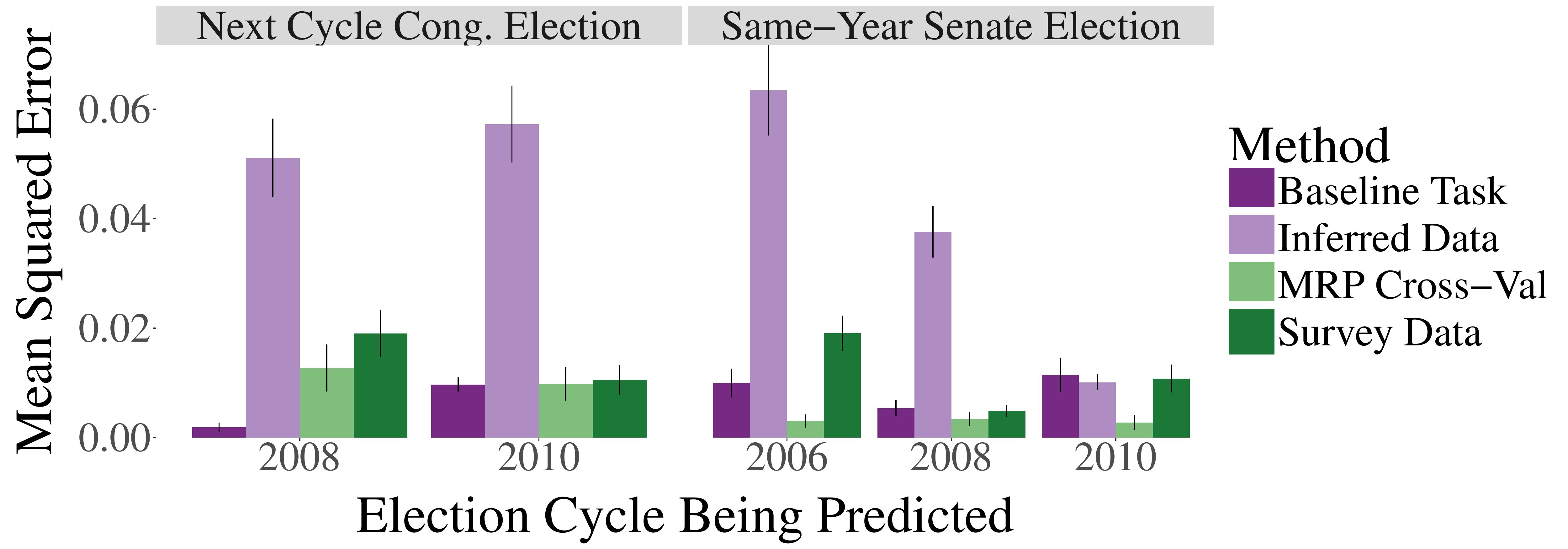}
        \end{minipage}
        \par \vspace{0.5cm}
    
        \begin{minipage}{0.75cm}
            \rotatebox{90}{\textbf{Laplace distribution}}
        \end{minipage}
        \begin{minipage}{\dimexpr\linewidth-3cm\relax}%
            \centering
            \includegraphics[width=\linewidth]{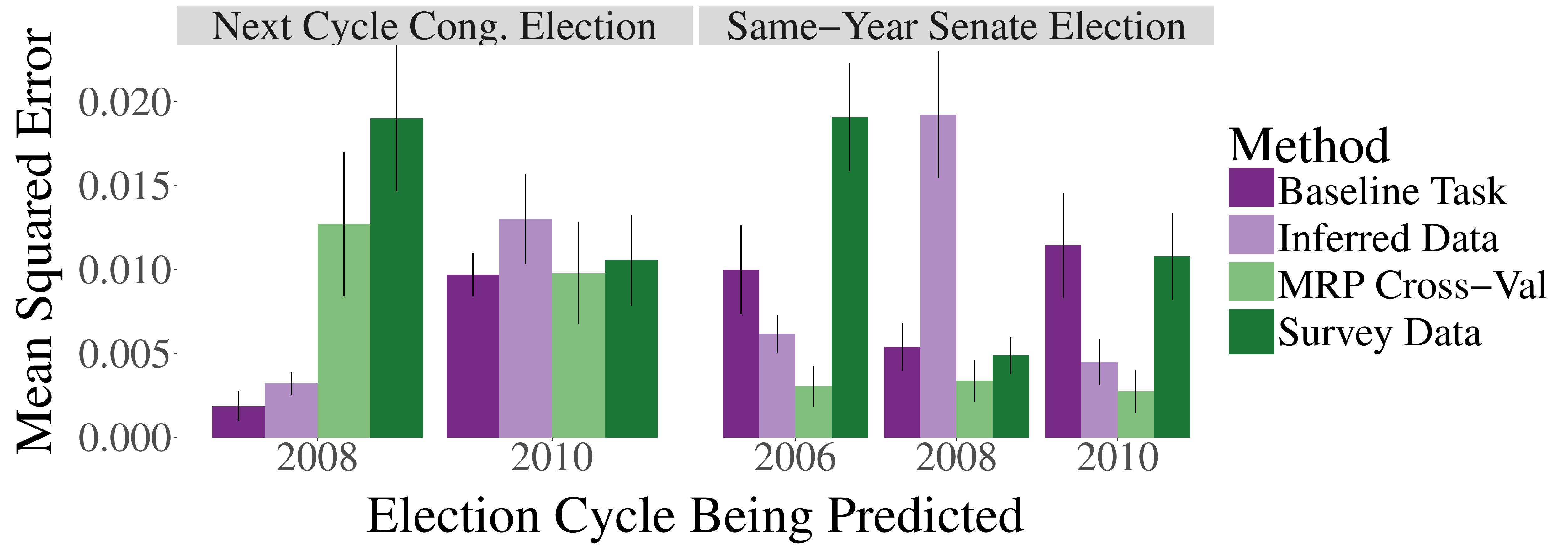}
        \end{minipage}
        \par
    \end{centering}
            
        \caption{Mean squared error of the actual and predicted vote share yielded by various prediction methods. Inferred data prediction method is based on our model assuming alternative underlying component distributions.}
        \label{fig:precdist-error-comp}
    \end{figure}

We test the inference procedure of our model not only assuming Normal component distributions, but also Uniform and Laplace component distributions. When assuming the distributions of voters follow a Laplace distribution, we use the same Normal prior defined for the mean of the Normal component for the location parameter and the same Inverse Gamma prior defined for the standard deviation of the Normal component for the scale parameter. When we use Uniform component distributions, we use the same Normal prior defined for the mean of the Normal component for both the minimum and the distance between the minimum and maximum parameters. The priors defined for the Normal component parameters can be found in Sect. \ref{sec:inference}. For each alternative underlying component distribution, the inferred distributions can be seen in Fig. \ref{fig:precdist-post-pred}, derived polarization metrics can be seen in Table \ref{tbl:precdist-pol-metrics}, and prediction comparisons can be seen in Fig. \ref{fig:precdist-error-comp}. 

    \begin{table}
        \centering
        \caption{Polarization metrics computed similarly to Tab. \ref{fig:pol-metrics} of voters and candidates assuming different underlying component distributions.}

        \def\arraystretch{1.1}%
        \setlength\tabcolsep{0.7mm}
        \begin{tabular}{|c|cl|lll|lll|lll|}
        \hline
        ~ & ~ & ~ & \multicolumn{3}{c|}{\centering{2006}} & \multicolumn{3}{c|}{2008} & \multicolumn{3}{c|}{2010} \\
        ~ & ~ & ~ & TX & NY & OH & TX & NY & OH & TX & NY & OH  \\ \hline
        \multirow{9}{*}{\rotatebox[origin=c]{90}{\textbf{Uniform distribution}}} & Difference- & Voters & 0.02 & 5.73 & -0.26 & 0.47 & 6.69 & 3.27 & 0.26 & 3.01 & 0.02 \\
        ~ & of-Means & Cand. & 1.99 & 0.62 & 1.08 & 1.40 & 1.52 & 1.85 & 0.60 & 1.65 & 2.09 \\
        ~ & ~ & \textbf{Diff} & \textbf{-1.97} & \textbf{5.11} & \textbf{-0.25} & \textbf{-0.93} &	\textbf{5.17} & \textbf{1.42} & \textbf{-0.34} & \textbf{1.37} & \textbf{-2.06} \\ \cline{2-12}

        ~ &Standard & Voters &  2.11 & 27.51 & 24.55 & 1.94	& 86.17 & 27.71 & 3.42 & 21.46 & 5.12\\
        ~ &Deviation & Cand. & 0.97 & 0.77 & 1.22 & 1.07 & 0.83 & 0.99 & 1.22 & 0.81 & 1.07 \\
        ~ &~ & \textbf{Diff} & \textbf{1.13} & \textbf{26.73} & \textbf{23.33} & \textbf{0.87} & \textbf{85.34} & \textbf{26.72} & \textbf{2.20} & \textbf{20.65} & \textbf{4.05} \\ \cline{2-12}

        ~ &~ & Voters & -0.77	& 42.17 & -2.96 & 0.72 & 42.50 & -2.99 & 0.38 & 39.39 & -3.00 \\
        ~ &Kurtosis & Cand.  & -1.62	& 1.67 & 1.23	& 0.49	& 1.79 & -1.77	& 3.24	& -1.61 & -1.73 \\
        ~ & ~ & \textbf{Diff} & \textbf{0.85} & \textbf{40.50} & \textbf{-4.19} & \textbf{0.23} & \textbf{40.71} & \textbf{-1.23} & \textbf{-2.86} & \textbf{41.00} & \textbf{-1.27} \\ \hline
        
        \multirow{9}{*}{\rotatebox[origin=c]{90}{\textbf{Laplace distribution}}} & Difference- & Voters & 1.76 & 5.13 & 6.01 & 1.47 & 3.21 & 4.07 & 5.60 & 4.17 & 0.56 \\
        ~ & of-Means & Cand. & 1.80 & 0.41 & 1.84 & 1.73 & 1.15 & 2.00 & 1.79 & 1.64 & 2.06 \\
        ~ & ~ & \textbf{Diff} & \textbf{-0.04} & \textbf{4.72} & \textbf{4.17} & \textbf{-0.26} & \textbf{2.07} & \textbf{2.07} & \textbf{3.81} & \textbf{2.54} & \textbf{-1.50} \\ \cline{2-12}

        ~ & Standard & Voters & 19.25 & 61.64 & 14.51	& 10.64 & 86.42 & 26.79 & 193.24	& 53.56 & 1.67 \\
        ~ & Deviation & Cand. & 0.97 & 0.77 & 1.22 & 1.07 & 0.83 & 0.99 & 1.22 & 0.81 & 1.07 \\
        ~ & ~ & \textbf{Diff} & \textbf{18.28} & \textbf{60.87} & \textbf{13.29} & \textbf{9.57} & \textbf{85.60} & \textbf{25.80} & \textbf{192.03} & \textbf{52.74} & \textbf{0.60} \\ \cline{2-12}
        ~ & ~ & Voters & -2.99	& 1.78 & 1.84 & -2.48 & 1.49 & 2.34 & -2.90 & 1.47 & -1.42 \\
        ~ & Kurtosis & Cand.  & -1.62	& 1.67 & 1.23	& 0.49	& 1.79 & -1.77	& 3.24	& -1.61 & -1.73\\
        ~ & ~ & \textbf{Diff} & \textbf{-1.37} & \textbf{0.11} & \textbf{0.61} & \textbf{-2.97} & \textbf{-0.30} & \textbf{4.10} & \textbf{-6.14} & \textbf{3.08} & \textbf{0.30} \\ \hline

        \end{tabular}
        \label{tbl:precdist-pol-metrics}
    \end{table}

\begin{figure}
    \begin{minipage}{0.2cm}
        \rotatebox{90}{\textbf{Uniform}}
    \end{minipage}
    \begin{minipage}{\dimexpr\linewidth-1cm\relax}%
        \raisebox{\dimexpr-.5\height-1em}{
        \includegraphics[width=0.32\linewidth]{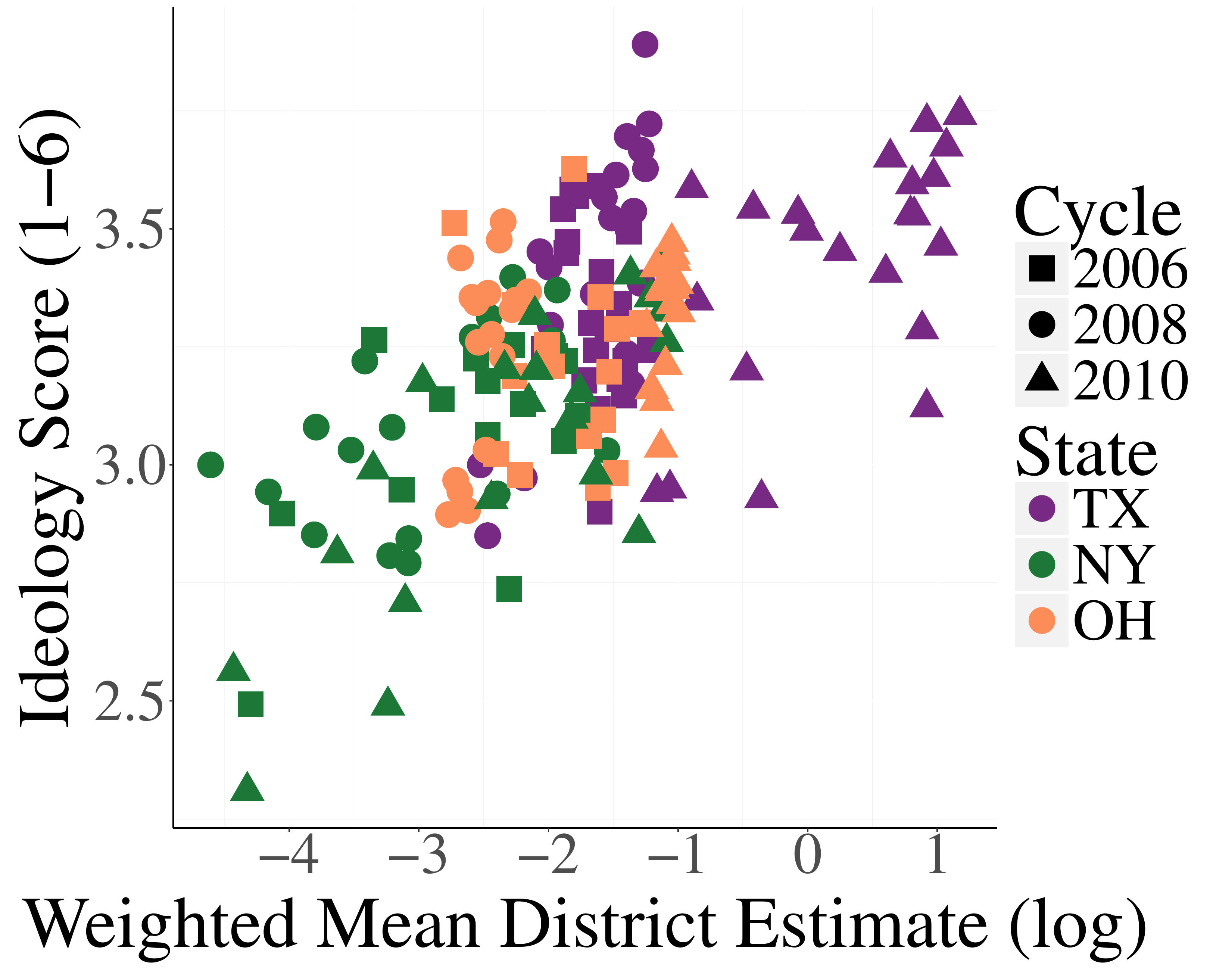}
        \includegraphics[width=0.32\linewidth]{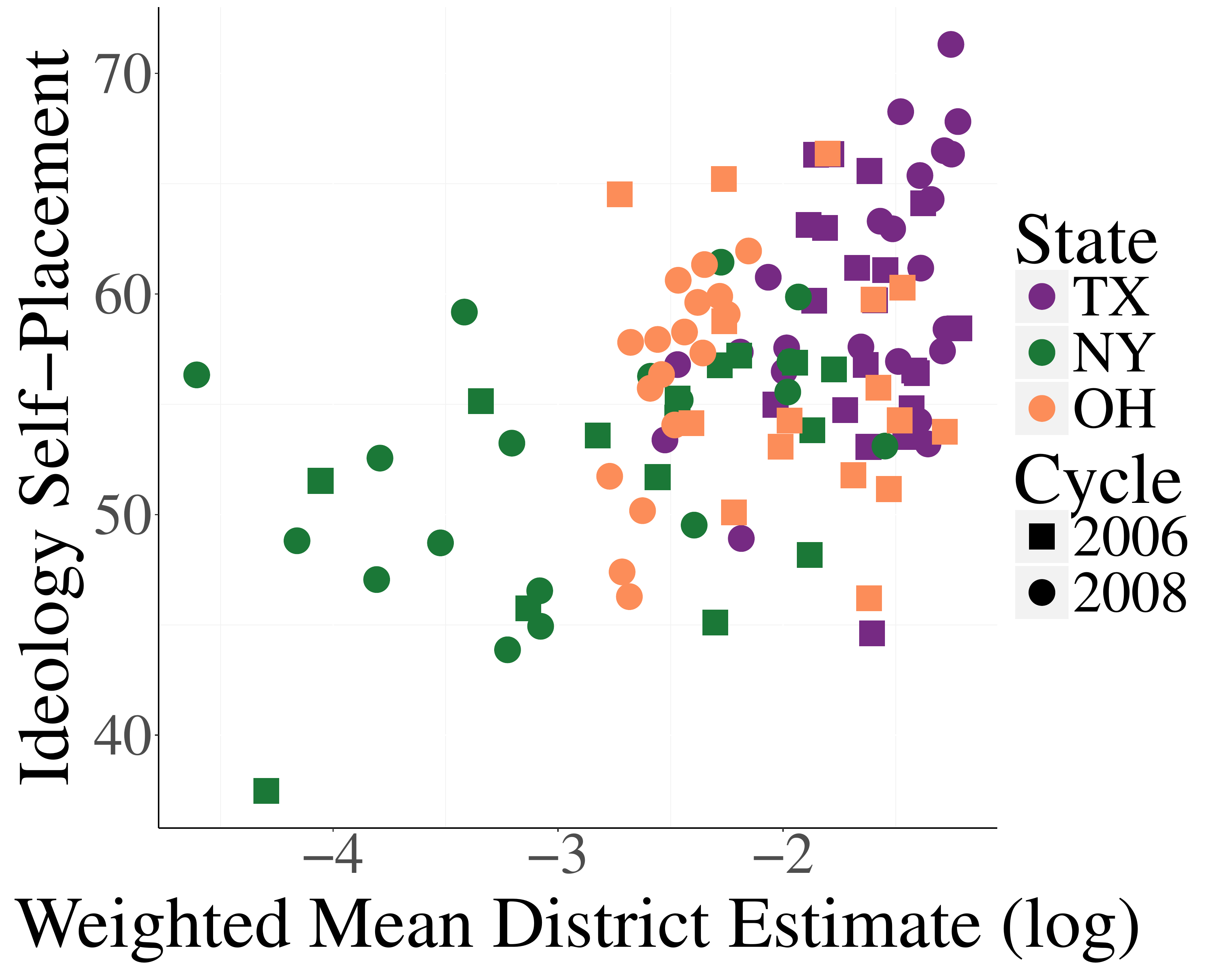}
        \includegraphics[width=0.32\linewidth]{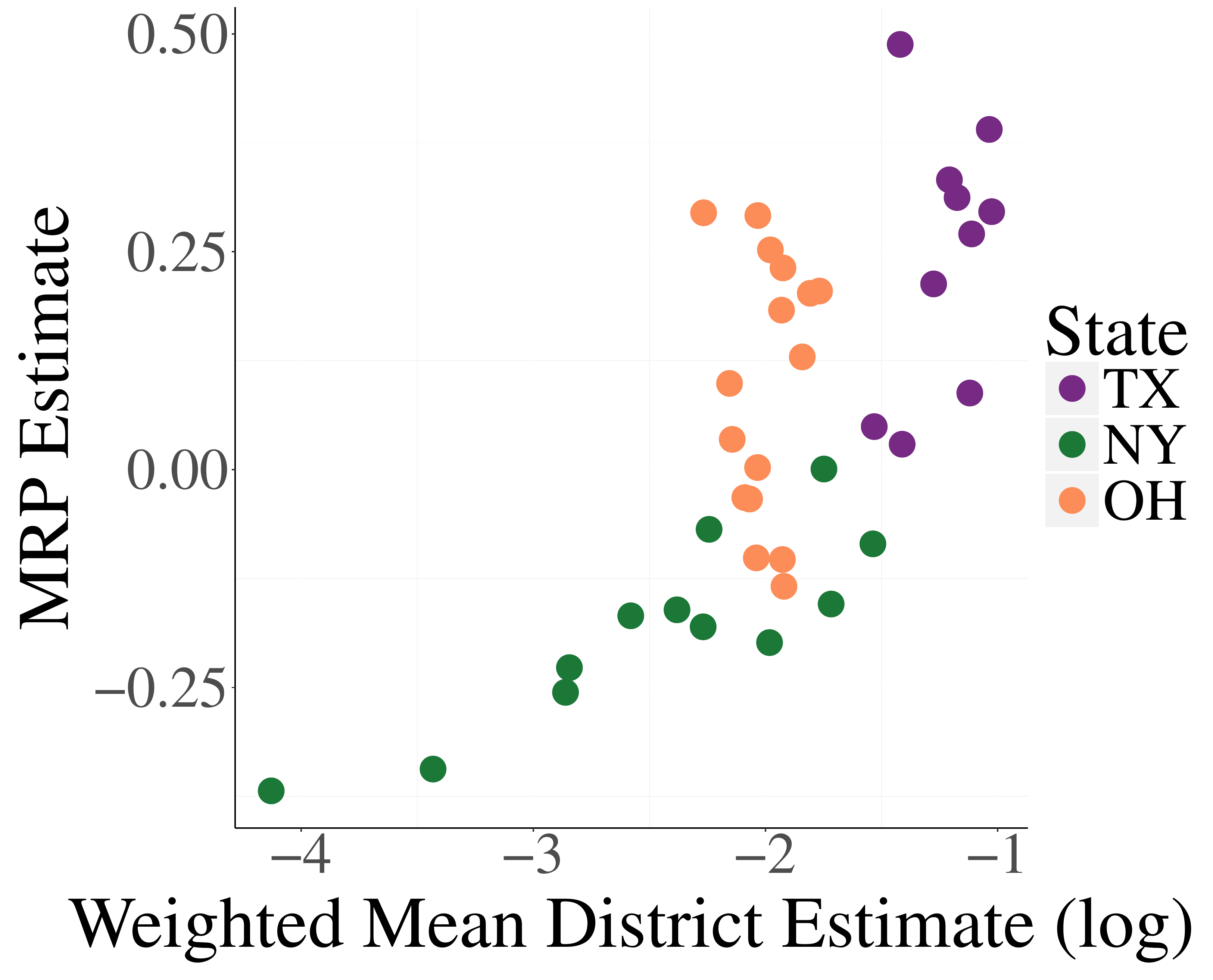}
            \par
        }
    \end{minipage}
    \par

    \begin{minipage}{0.2cm}
        \rotatebox{90}{\textbf{Laplace}}
    \end{minipage}
    \begin{minipage}{\dimexpr\linewidth-1cm\relax}%
        \raisebox{\dimexpr-.5\height-1em}{
        \includegraphics[width=0.32\linewidth]{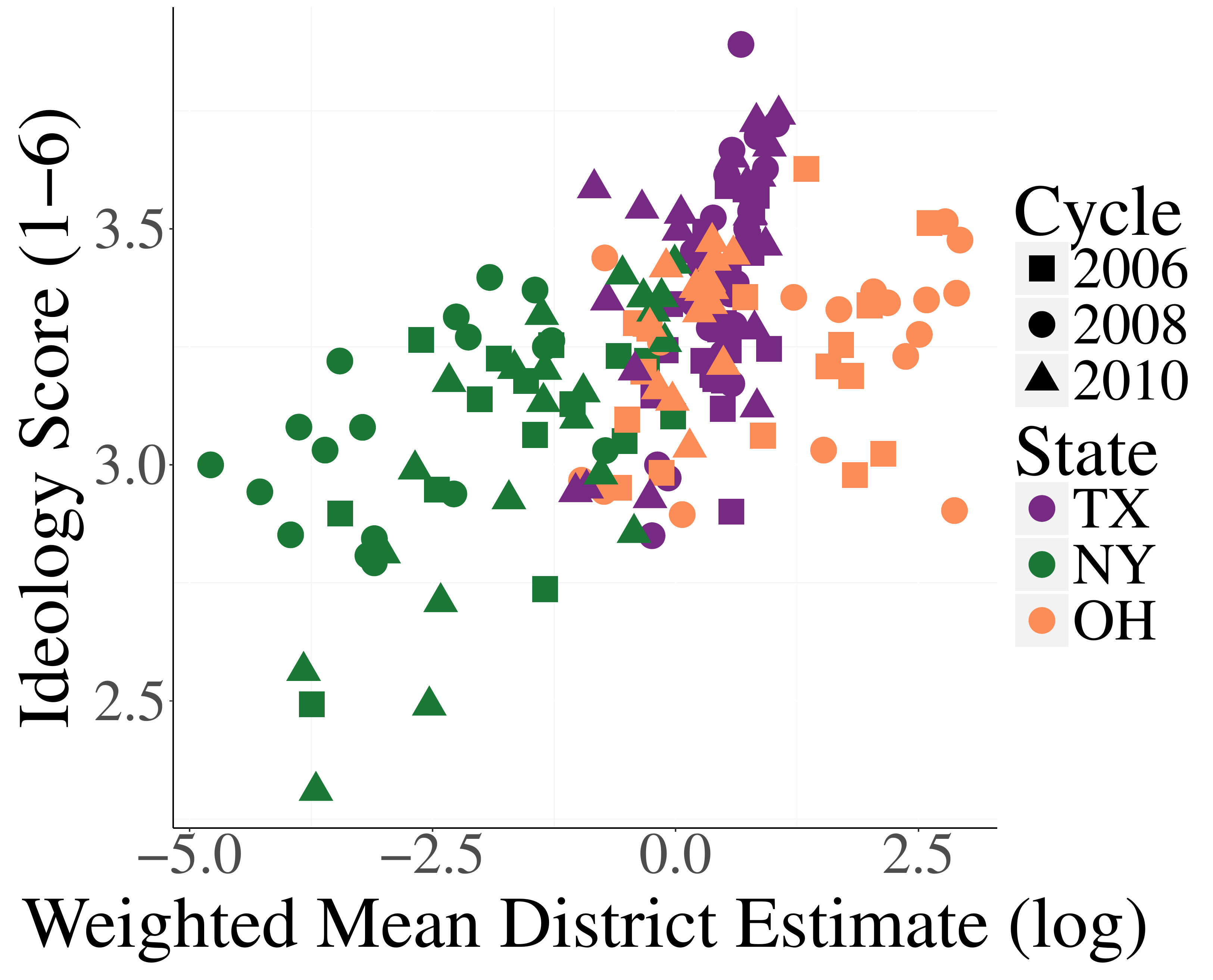}
        \includegraphics[width=0.32\linewidth]{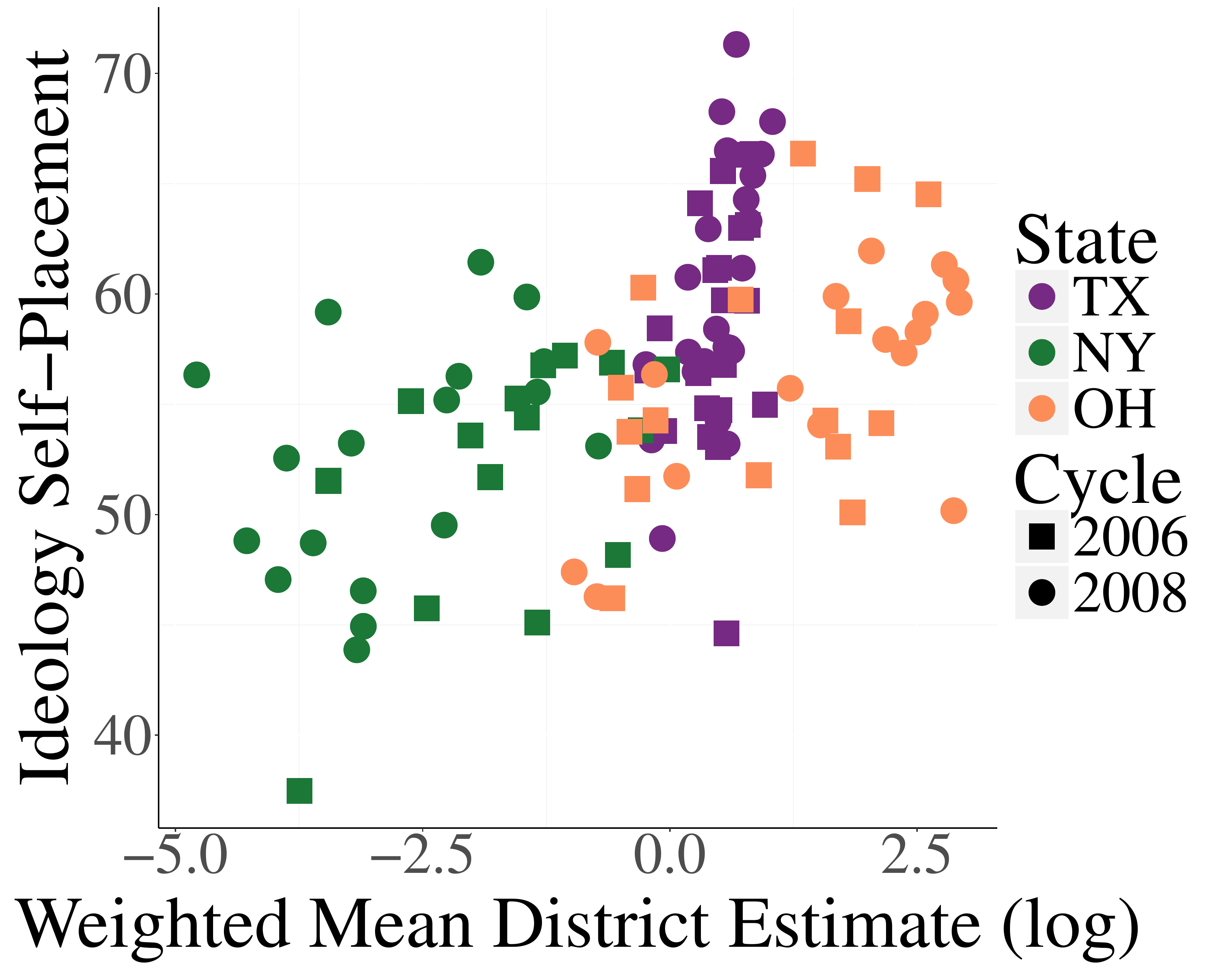}
        \includegraphics[width=0.32\linewidth]{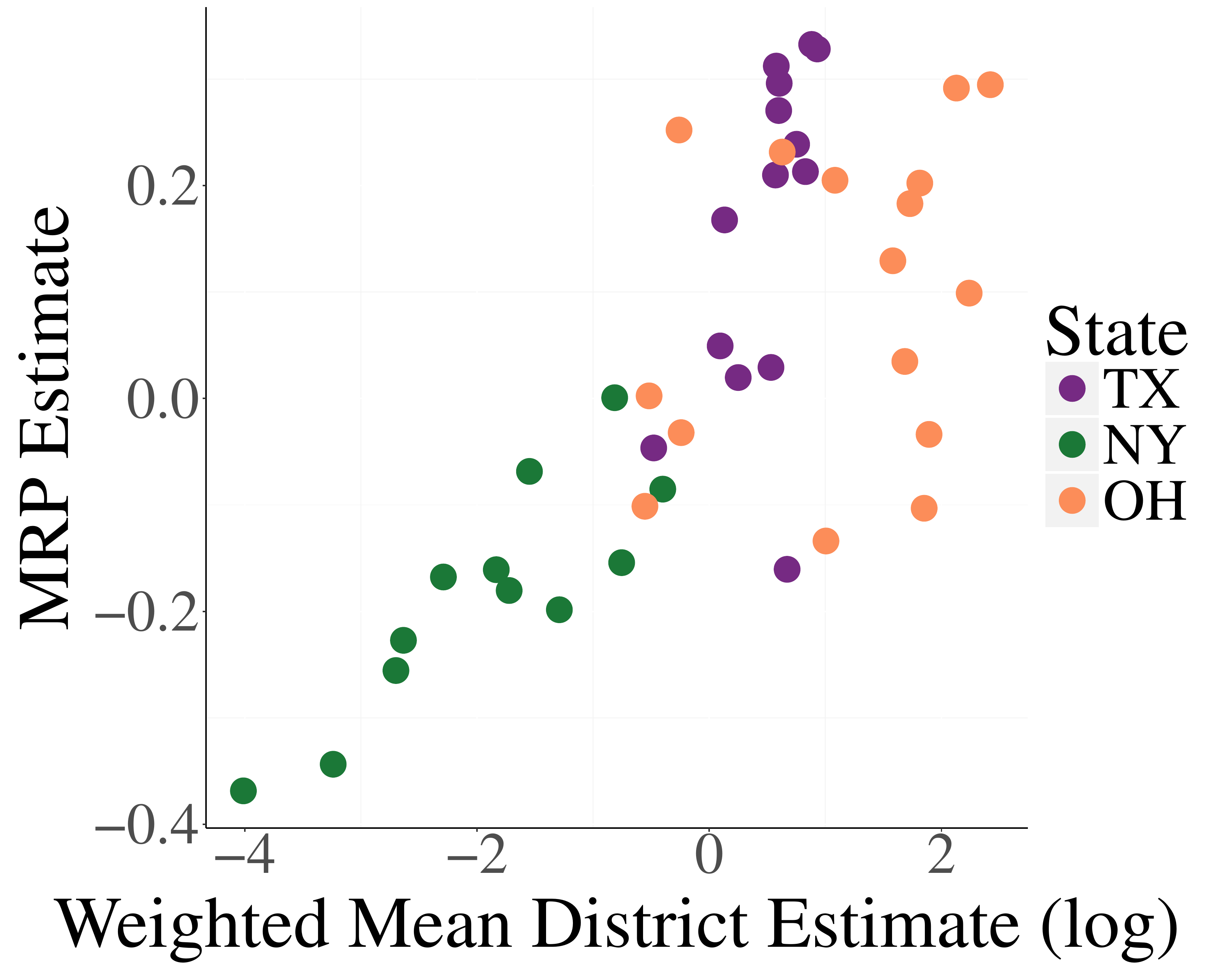}
        } \\
    \end{minipage}
    \par

\caption{In all of these cases, the inferred district-level voter preferences are based on the model varying its assumption of the underlying precinct distribution and are weighted mean district estimates transformed from $x$ to $sign(x) \, log( |x| + 1 )$. (Left) Inferences compared with CCES question of self-reported ideologies on a discrete scale. (Center) Compared with the CCES self-reported ideologies on a continuous scale from 0 to 100. (Right) District-level inferences of a full decade compared with MRP estimates \cite{warshaw}. }
\label{fig:precdist-cces-comp}
\end{figure}

Figure \ref{fig:precdist-cces-comp} visualizes the comparisons between the results derived of alternative component distributions and alternative data sources described in Sect. \ref{sec:dist-agg-desc}. We find significant positive correlations between our district-level point estimates and all of the alternative data sources. Assuming Laplace component distributions, our results have a correlation of 0.3216 with the responses selecting ideology given a discrete scale (left column in Fig. \ref{fig:precdist-cces-comp}), 0.2514 with the responses selecting ideology along a continuous scale (middle column), and 0.5323 with the MRP estimates (right column). All of these correlations were significant with p-values less than 0.01. Assuming Uniform component distributions, our results have a correlation of 0.3652 with the responses selecting ideology given a discrete scale, 0.2404 with the responses selecting ideology along a continuous scale, and 0.6331 with the MRP estimates. Again, all of these correlations were significant with p-values less than 0.01.

\subsection{Varying the Number of Clusters}
\begin{figure}
    \begin{minipage}{0.2cm}
        \rotatebox{90}{\textit{\textbf{K = 2}}}
    \end{minipage}
    \begin{minipage}{\dimexpr\linewidth-1cm\relax}%
        \raisebox{\dimexpr-.5\height-1em}{
            \includegraphics[width=0.35\linewidth]{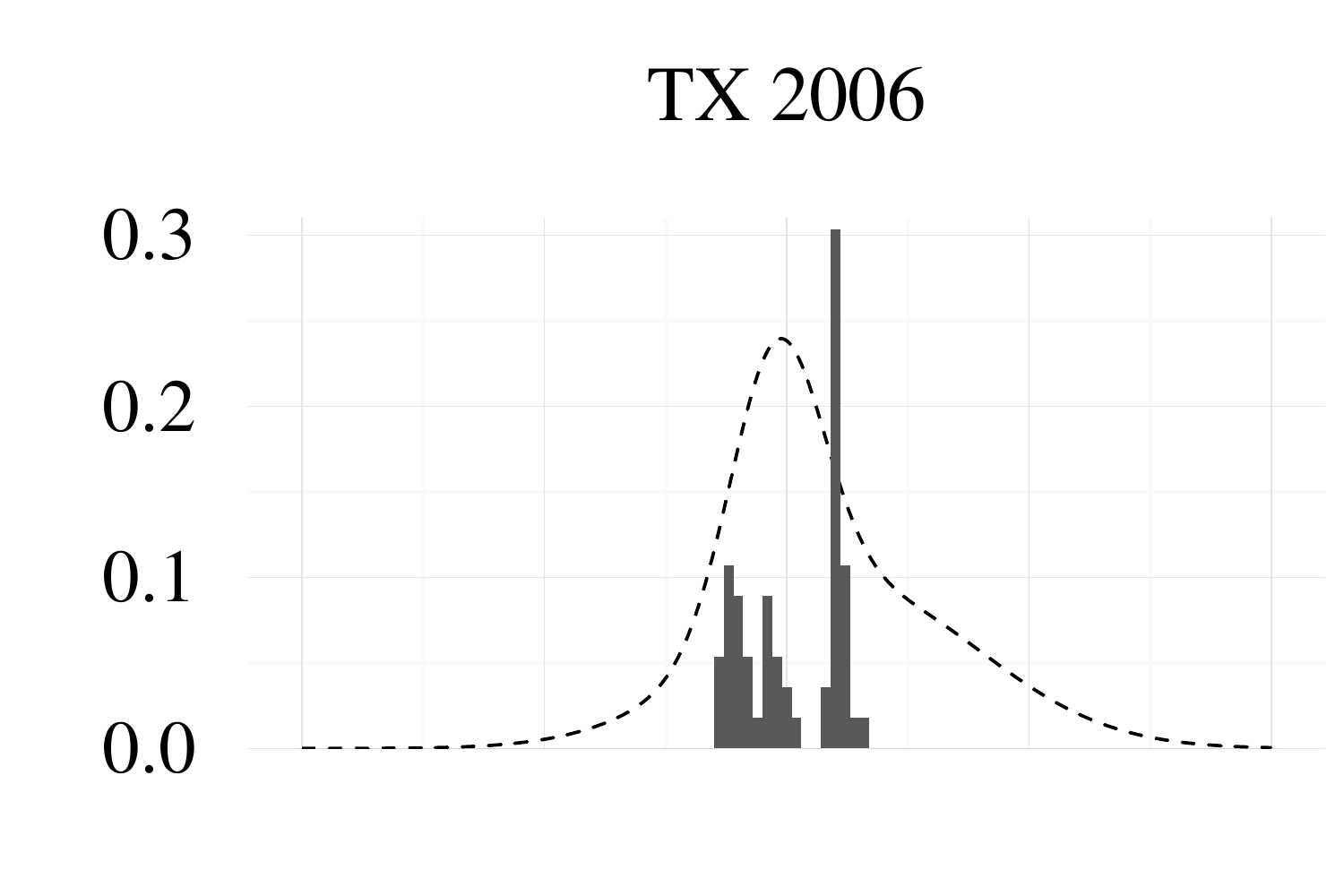}
            \hspace{-0.5cm}
            \includegraphics[width=0.35\linewidth]{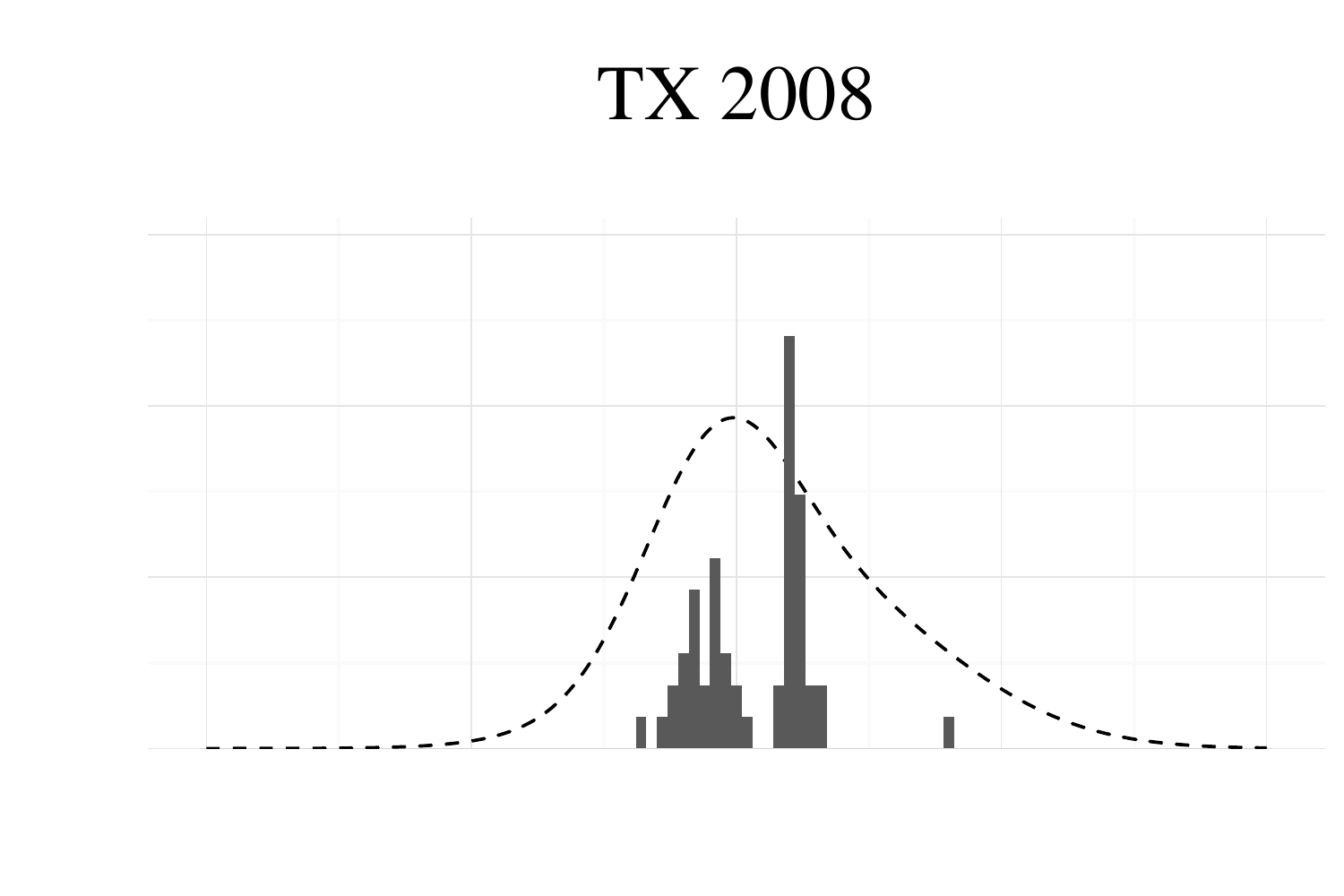}
            \hspace{-0.5cm}
            \includegraphics[width=0.35\linewidth]{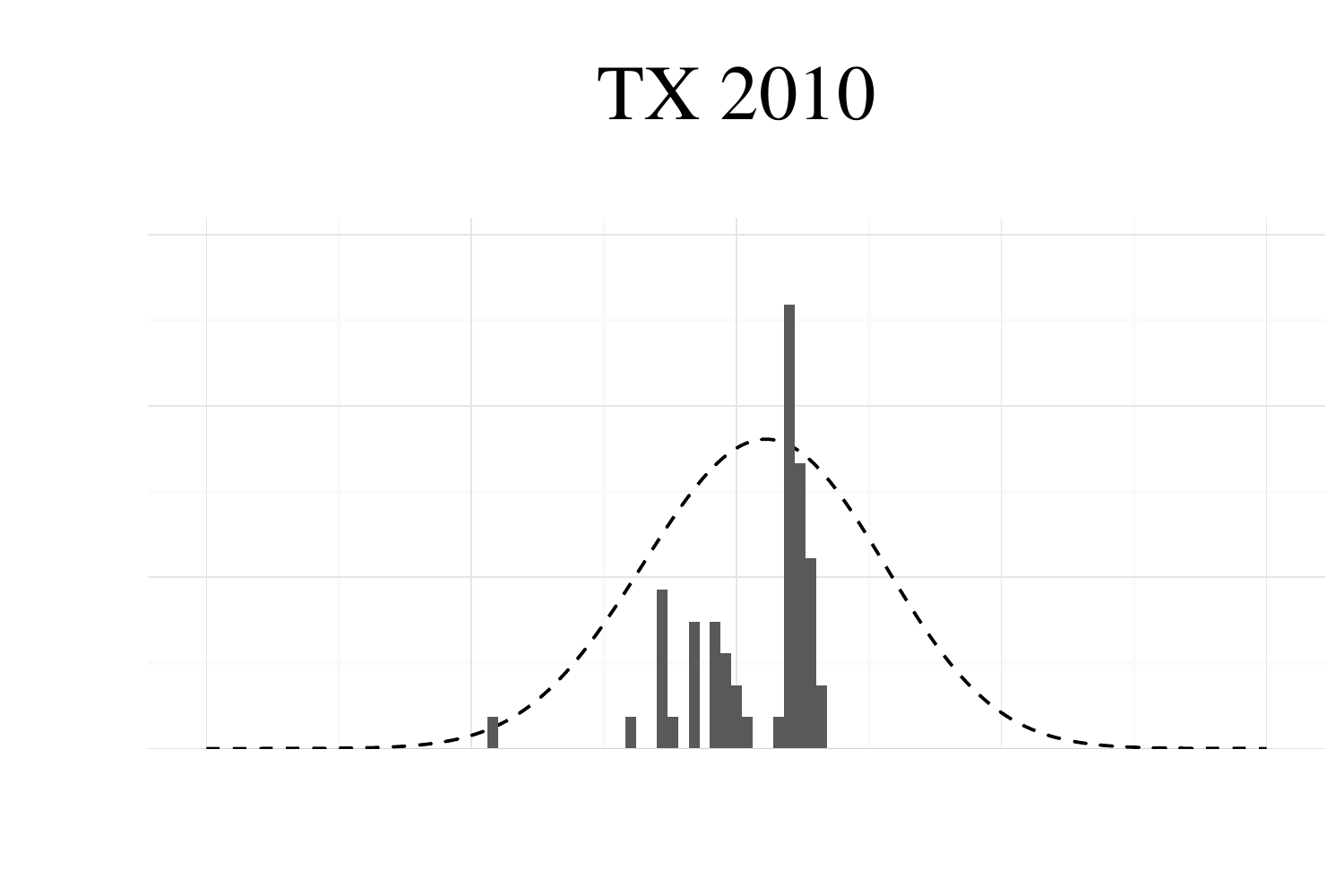}
            \par
        } \\[-15pt]
        \raisebox{\dimexpr-.5\height-1em}{
            \includegraphics[width=0.35\linewidth]{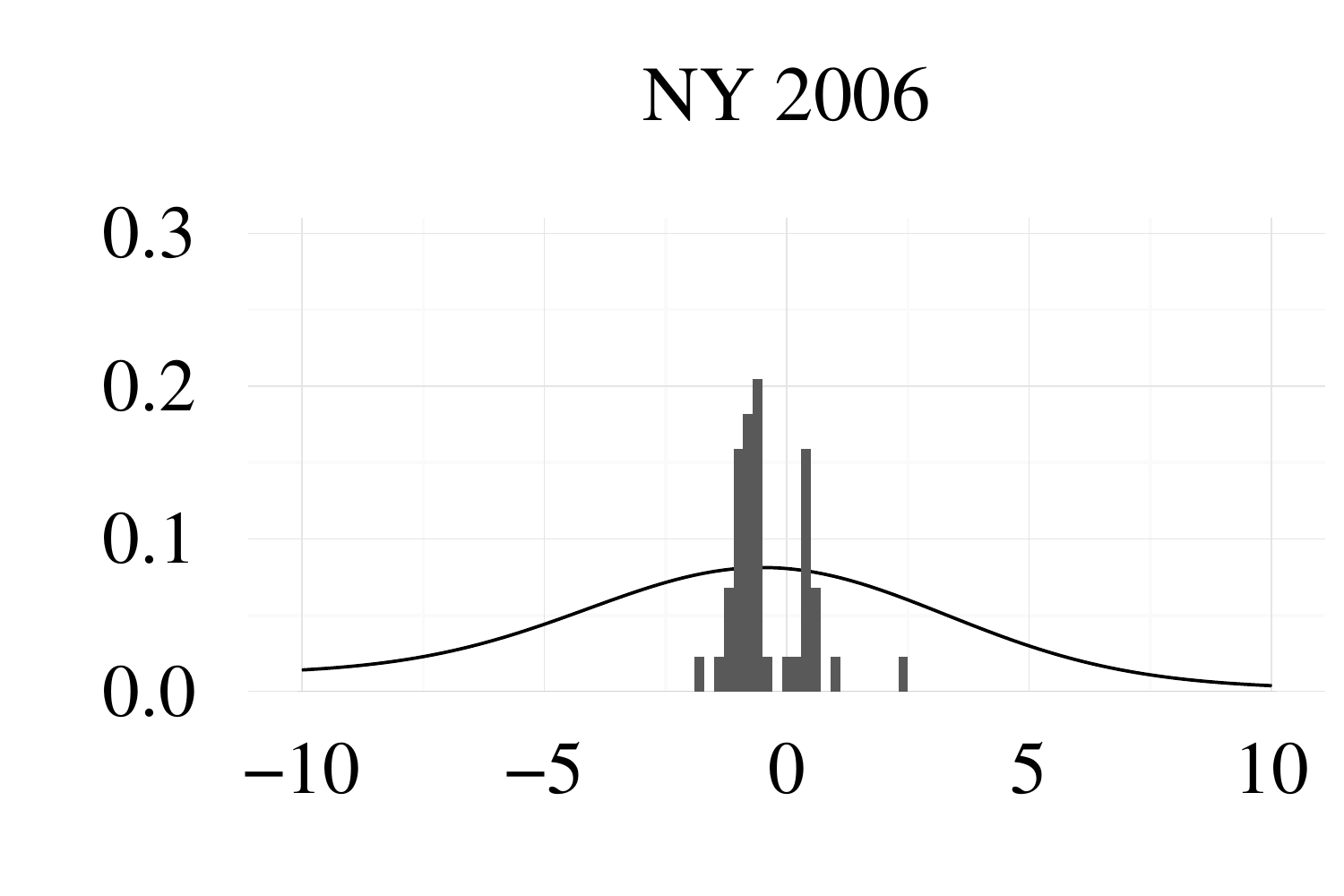}
            \hspace{-0.5cm}
            \includegraphics[width=0.35\linewidth]{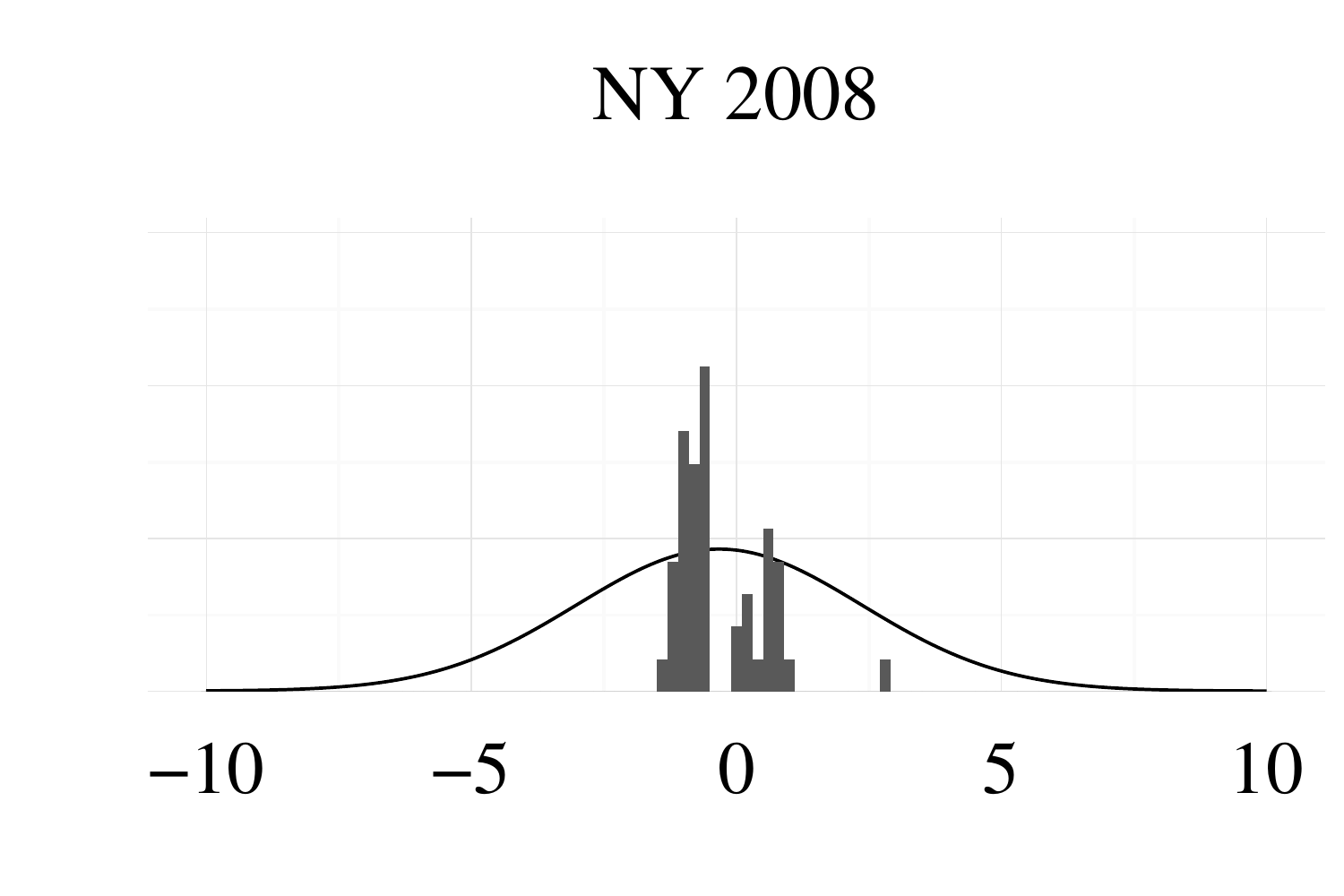}
            \hspace{-0.5cm}
            \includegraphics[width=0.35\linewidth]{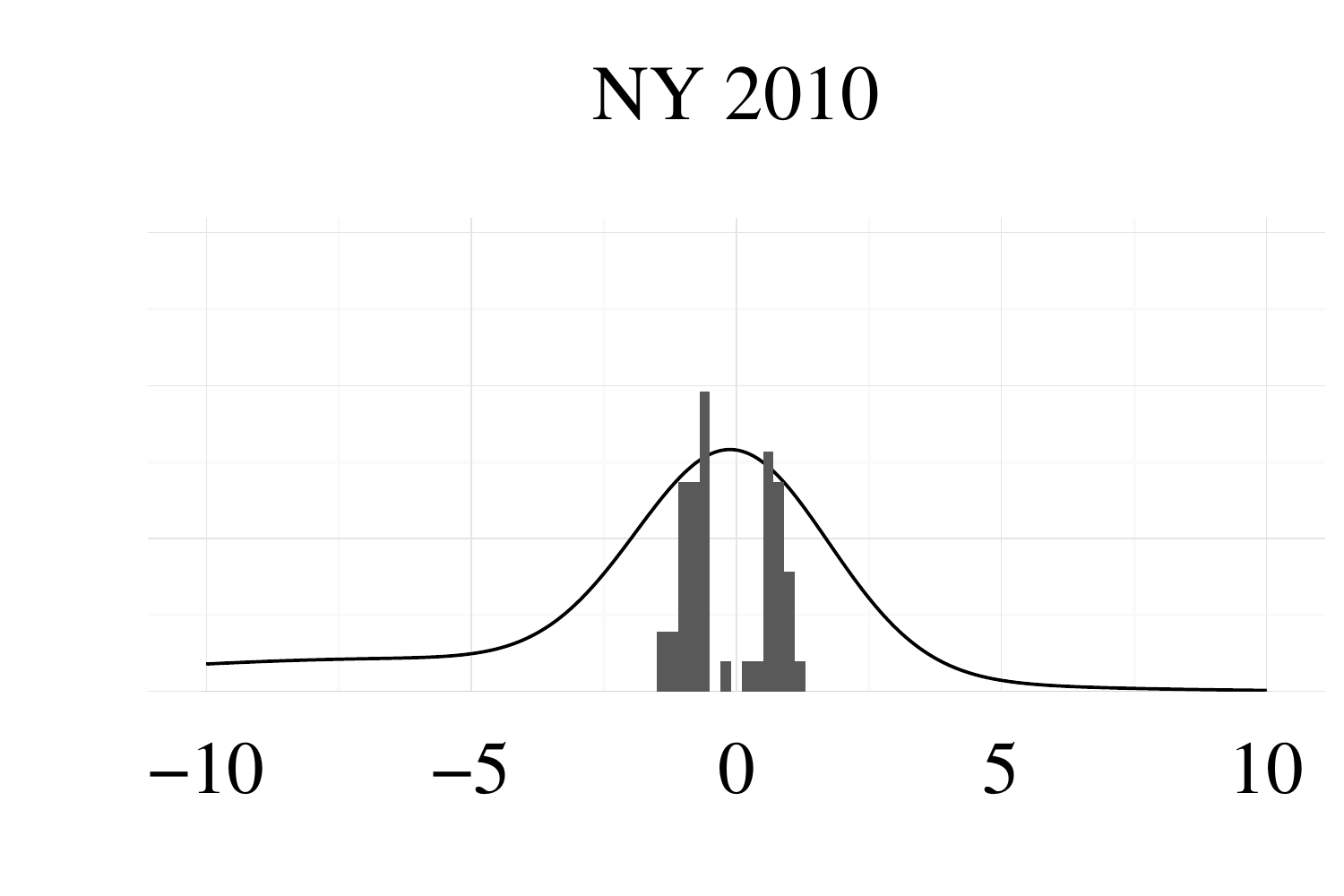}
            \par
        }
    \end{minipage}
    \par

    \begin{minipage}{0.2cm}
        \rotatebox{90}{\textit{\textbf{K = 8}}}
    \end{minipage}
    \begin{minipage}{\dimexpr\linewidth-1cm\relax}%
        \raisebox{\dimexpr-.5\height-1em}{
            \includegraphics[width=0.35\linewidth]{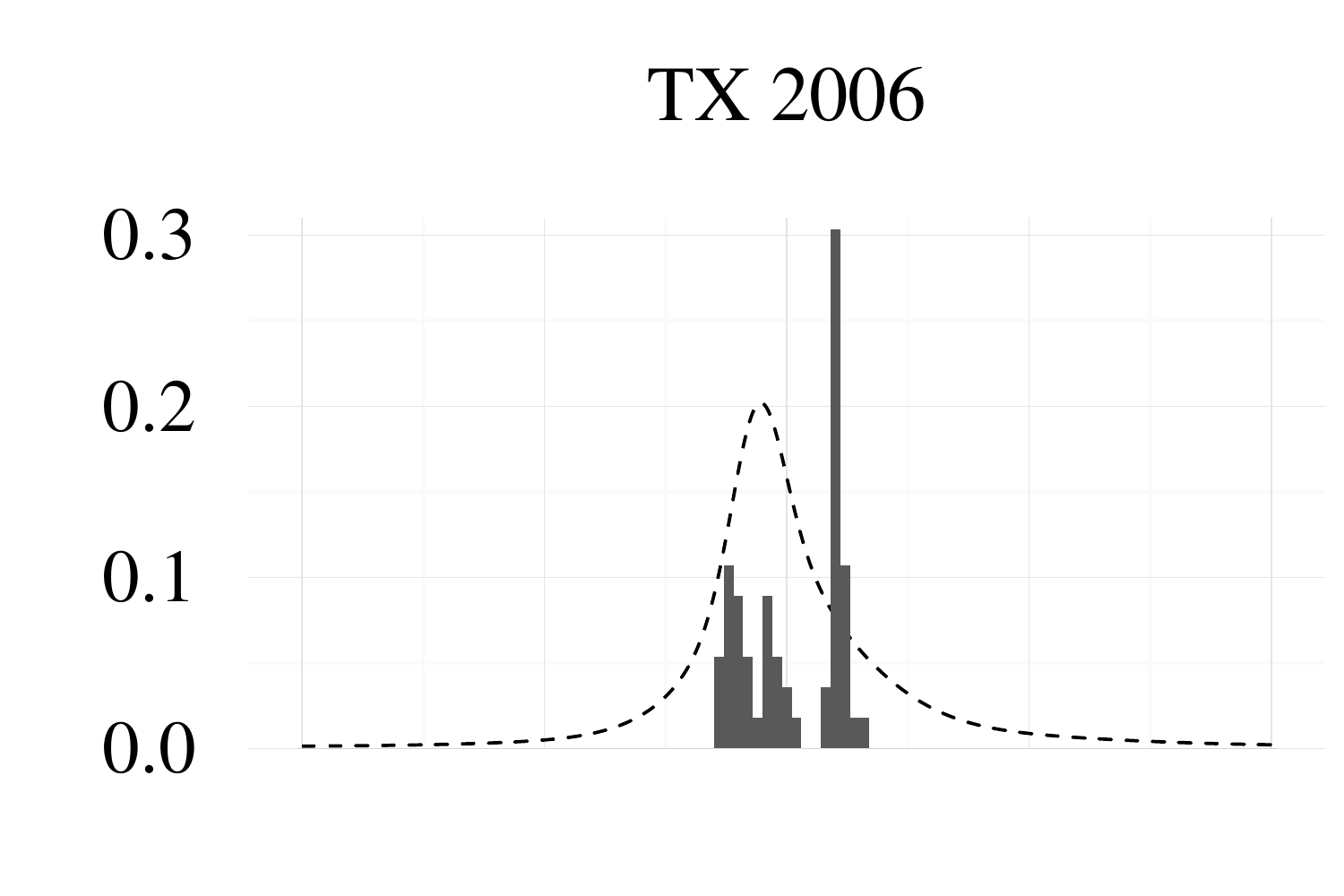}
            \hspace{-0.5cm}
            \includegraphics[width=0.35\linewidth]{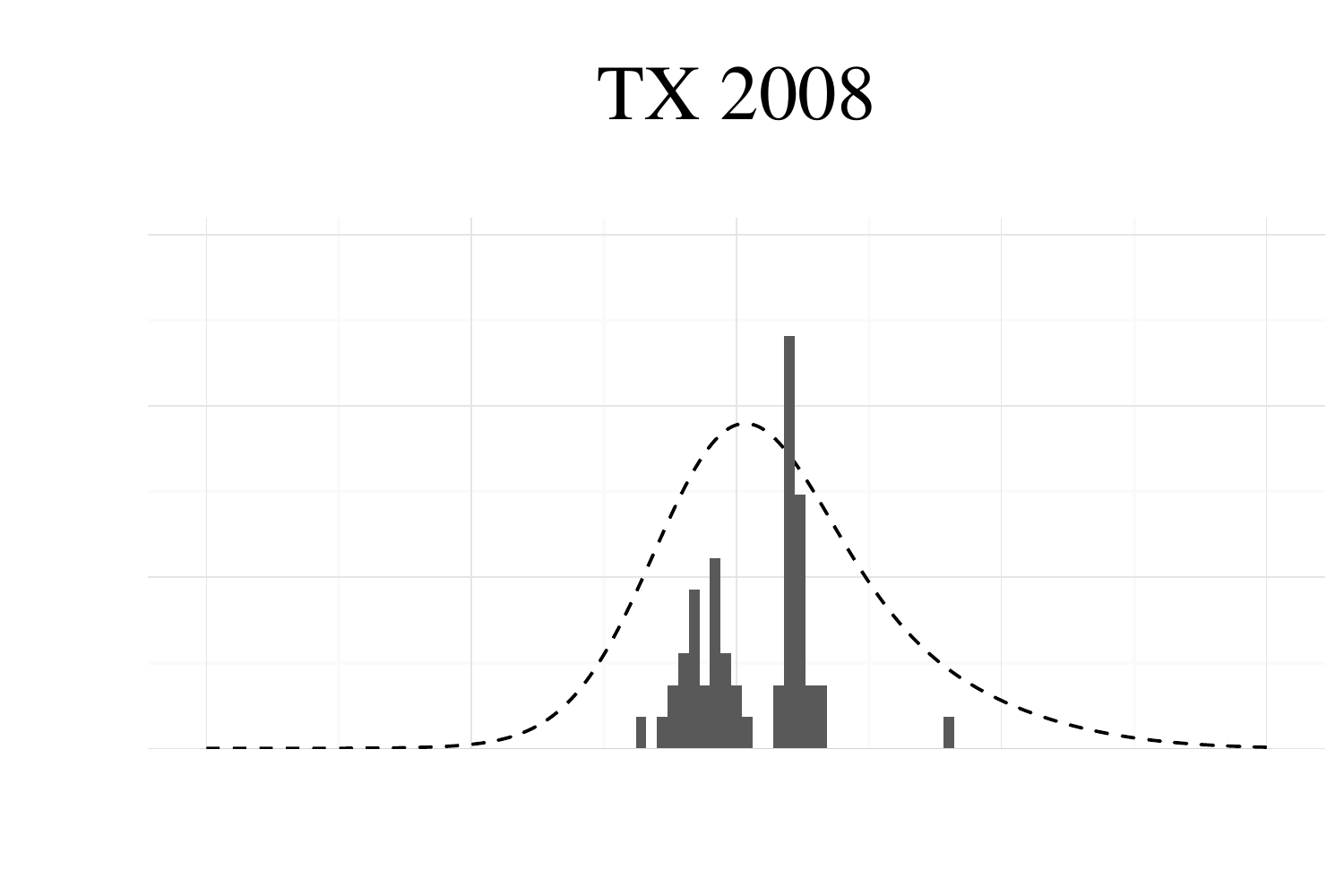}
            \hspace{-0.5cm}
            \includegraphics[width=0.35\linewidth]{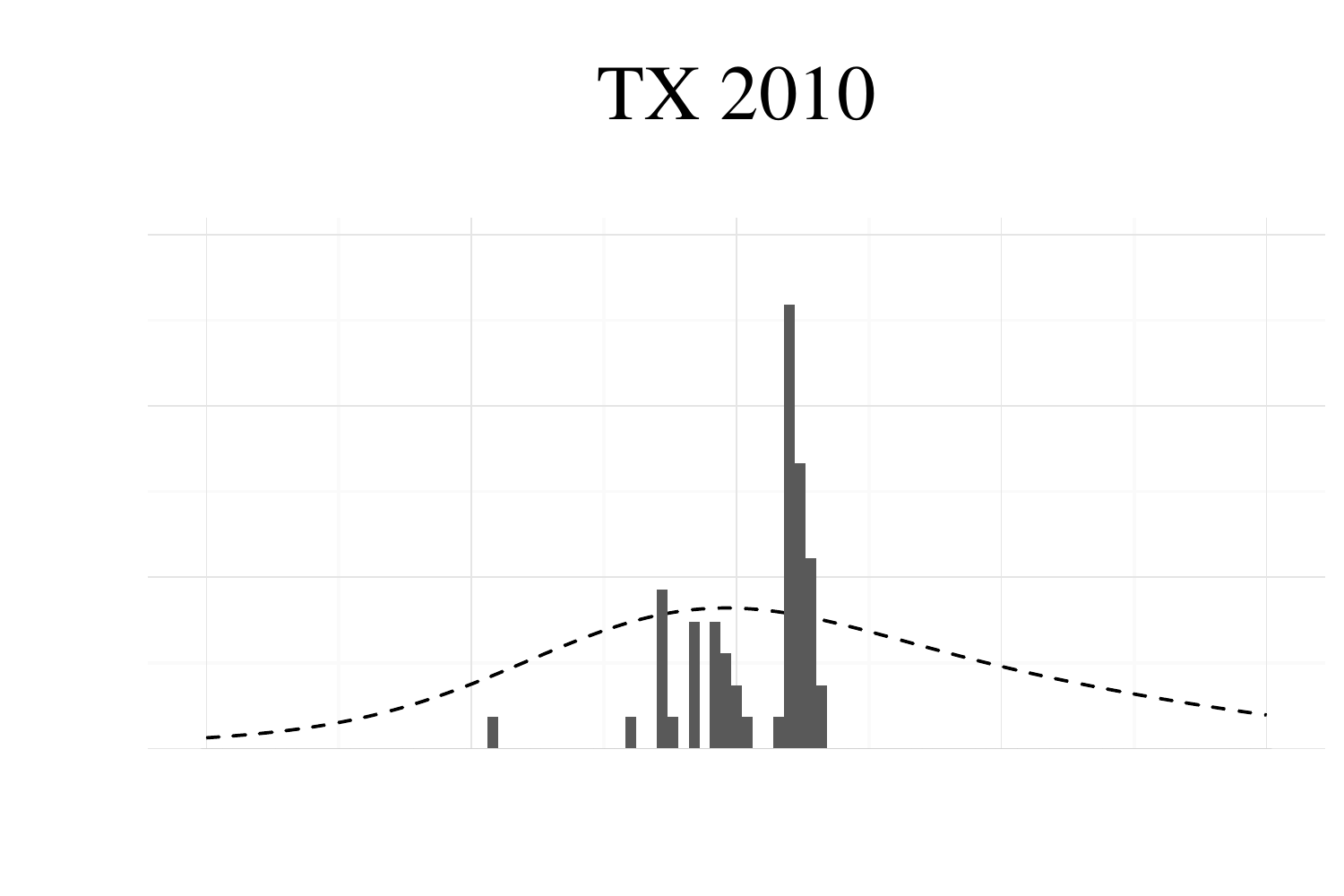}
            \par
        } \\[-15pt]
        \raisebox{\dimexpr-.5\height-1em}{
            \includegraphics[width=0.35\linewidth]{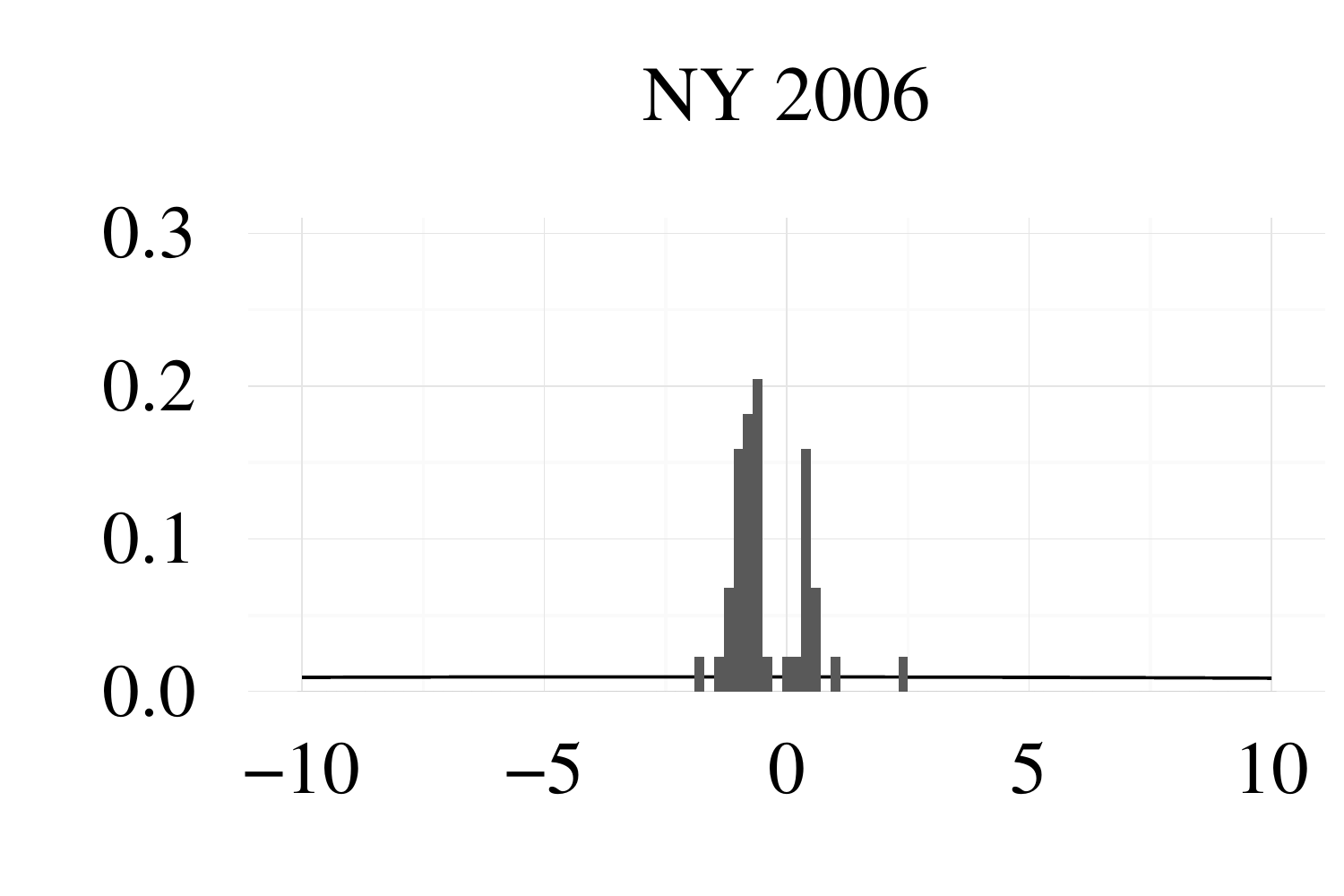}
            \hspace{-0.5cm}
            \includegraphics[width=0.35\linewidth]{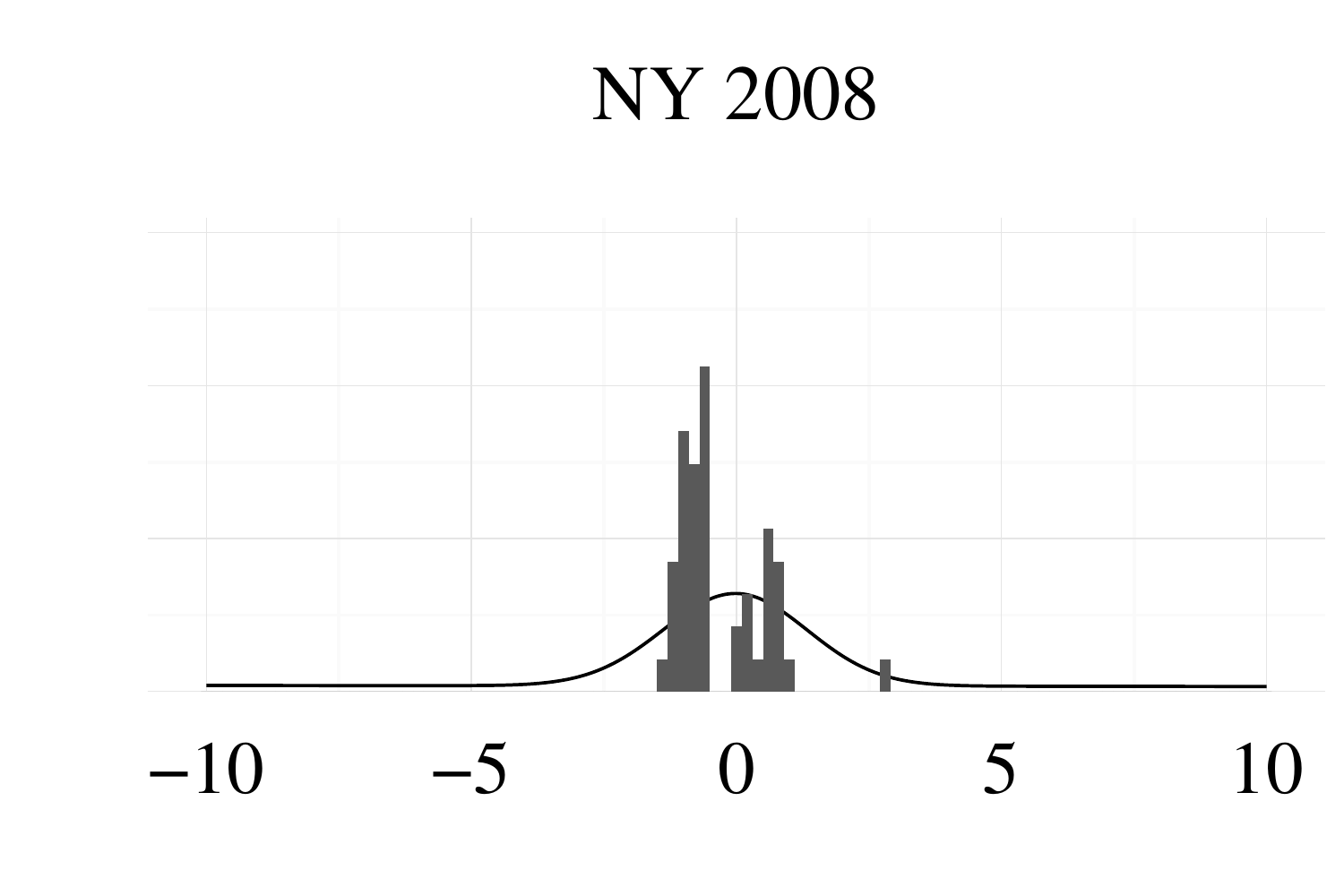}
            \hspace{-0.5cm}
            \includegraphics[width=0.35\linewidth]{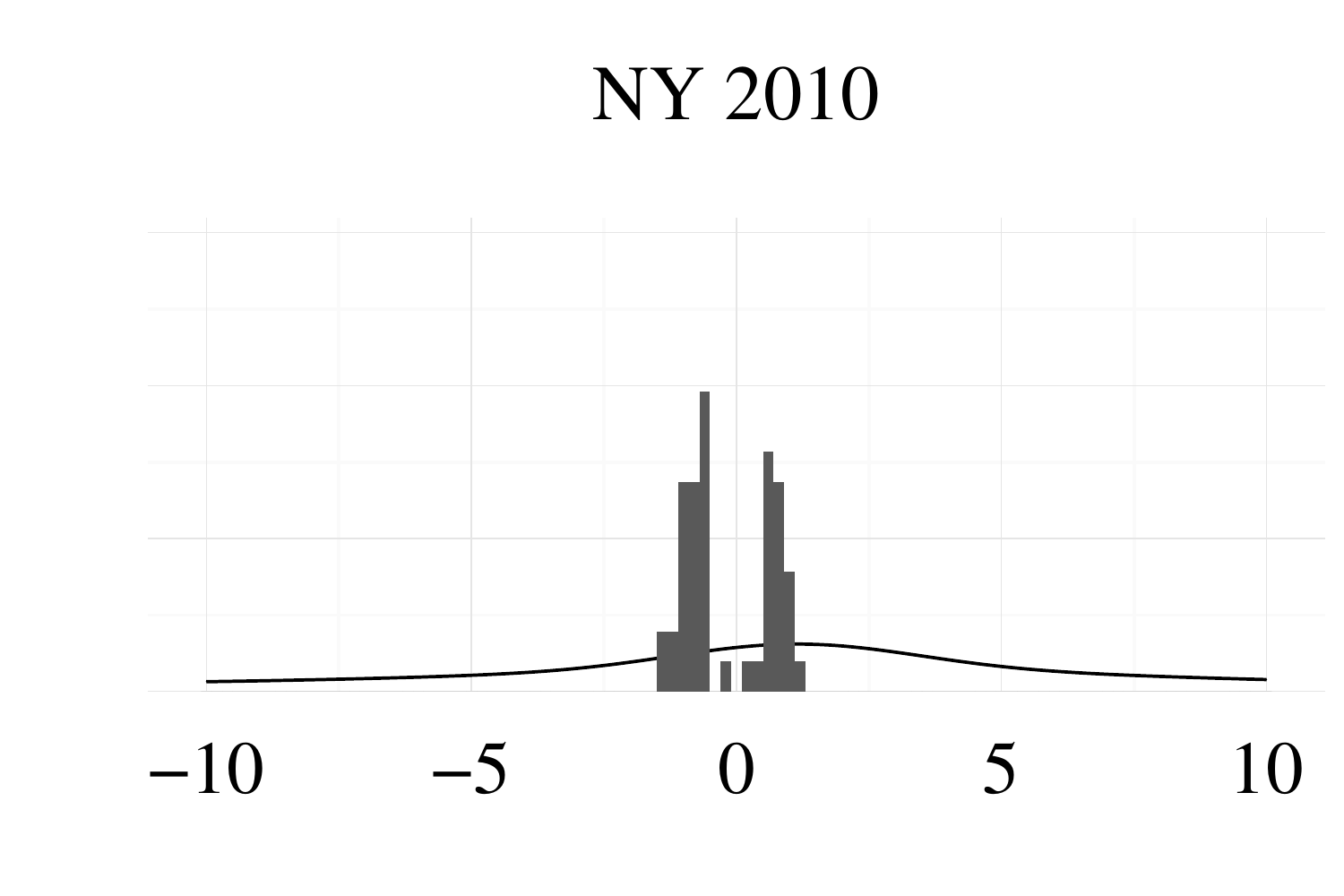}
            \par
        }
    \end{minipage}
    \par

  \caption{Distributions of inferred voter preferences (represented by lines) under alternative assumed mixture component distributions, and candidate preferences (histograms) based on data of the 2006, 2008 and 2010 congressional elections in Texas (dotted line) and New York (solid line).}
  \label{fig:clust-post-pred}
\end{figure}

\begin{figure}
    \begin{centering}
        \begin{minipage}{0.75cm}
            \rotatebox{90}{\textit{\textbf{K = 2}}}
        \end{minipage}
        \begin{minipage}{\dimexpr\linewidth-3cm\relax}%
            \centering
            \includegraphics[width=\linewidth]{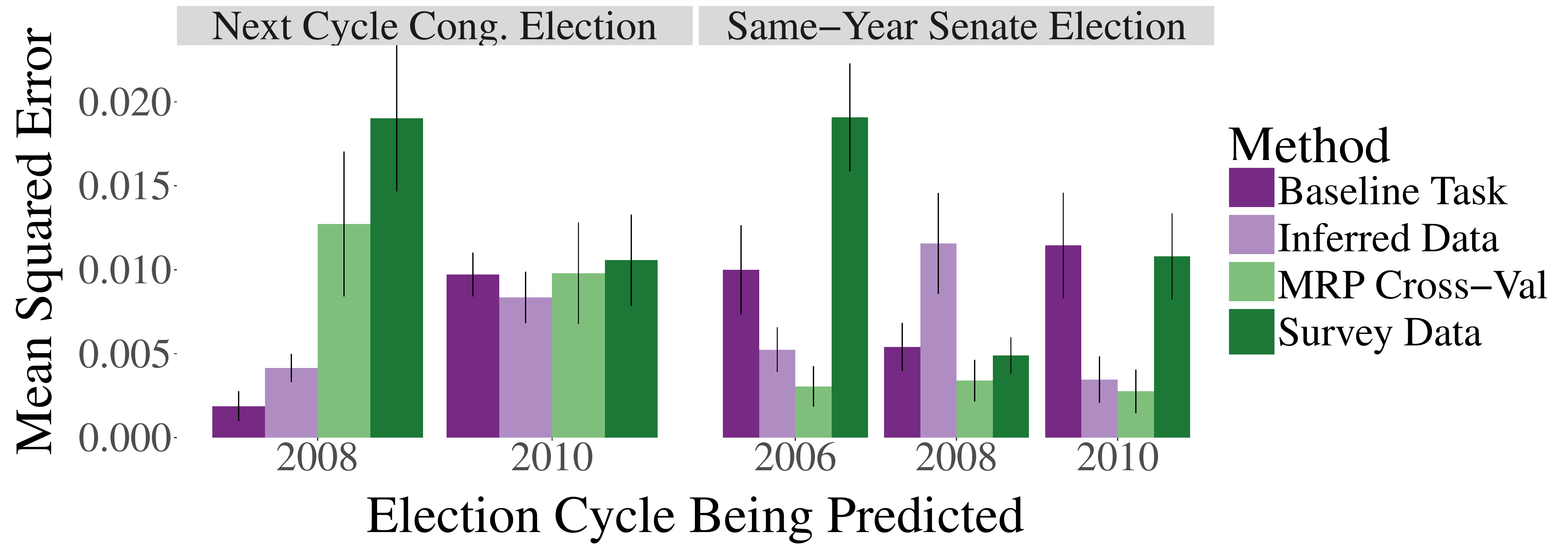}
        \end{minipage}
        \par \vspace{0.5cm}
    
        \begin{minipage}{0.75cm}
            \rotatebox{90}{\textit{\textbf{K = 8}}}
        \end{minipage}
        \begin{minipage}{\dimexpr\linewidth-3cm\relax}%
            \centering
            \includegraphics[width=\linewidth]{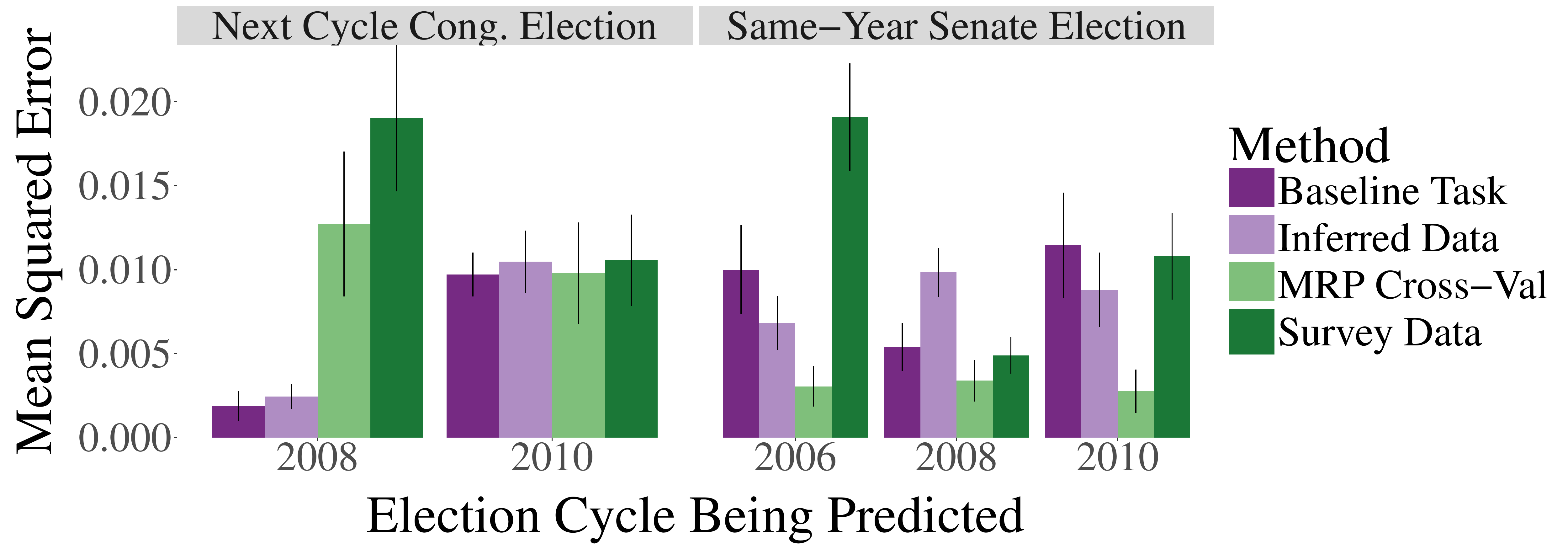}
        \end{minipage}
        \par
    \end{centering}
            
        \caption{Mean squared error of the actual and predicted vote share yielded by various prediction methods. Inferred data prediction method is based on our model assuming different numbers of clusters $(K)$.}
        \label{fig:clust-error-comp}
    \end{figure}

We also varied the number of clusters $(K)$ used in the model. The main results section in the paper presented results assuming $K=4$, but below we include the inferred distributions in Fig. \ref{fig:clust-post-pred}, derived polarization metrics in Table \ref{tbl:clust-pol-metrics}, and prediction comparisons in Fig. \ref{fig:clust-error-comp} for $K=2$ and $K=8$, assuming Normal underlying precinct distributions. Due to time constraints, we were only able to generate these results given the Texas and New York congressional elections.

\begin{figure}
    \begin{minipage}{0.2cm}
        \rotatebox{90}{\textit{\textbf{K = 2}}}
    \end{minipage}
    \begin{minipage}{\dimexpr\linewidth-1cm\relax}%
        \raisebox{\dimexpr-.5\height-1em}{
        \includegraphics[width=0.32\linewidth]{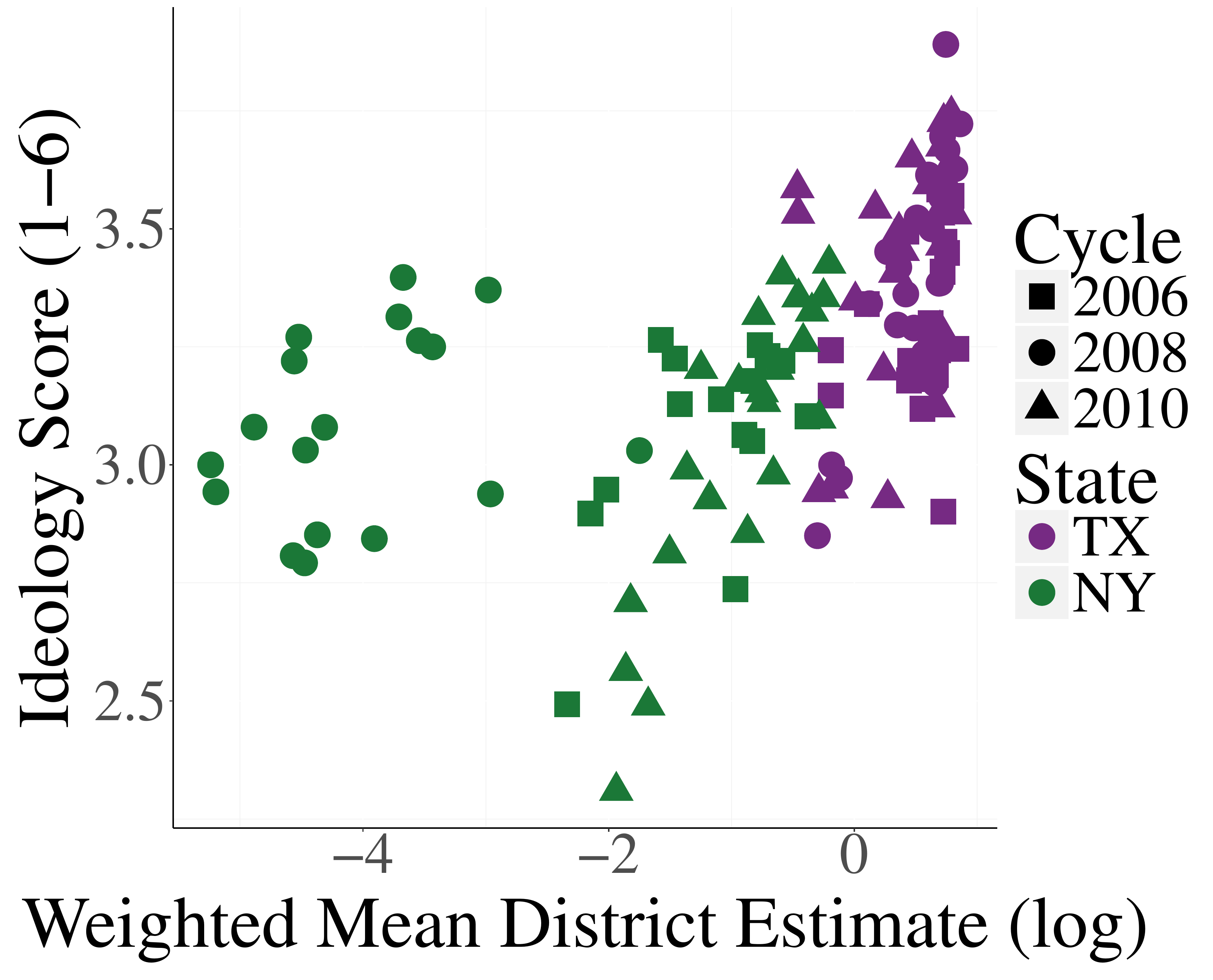}
        \includegraphics[width=0.32\linewidth]{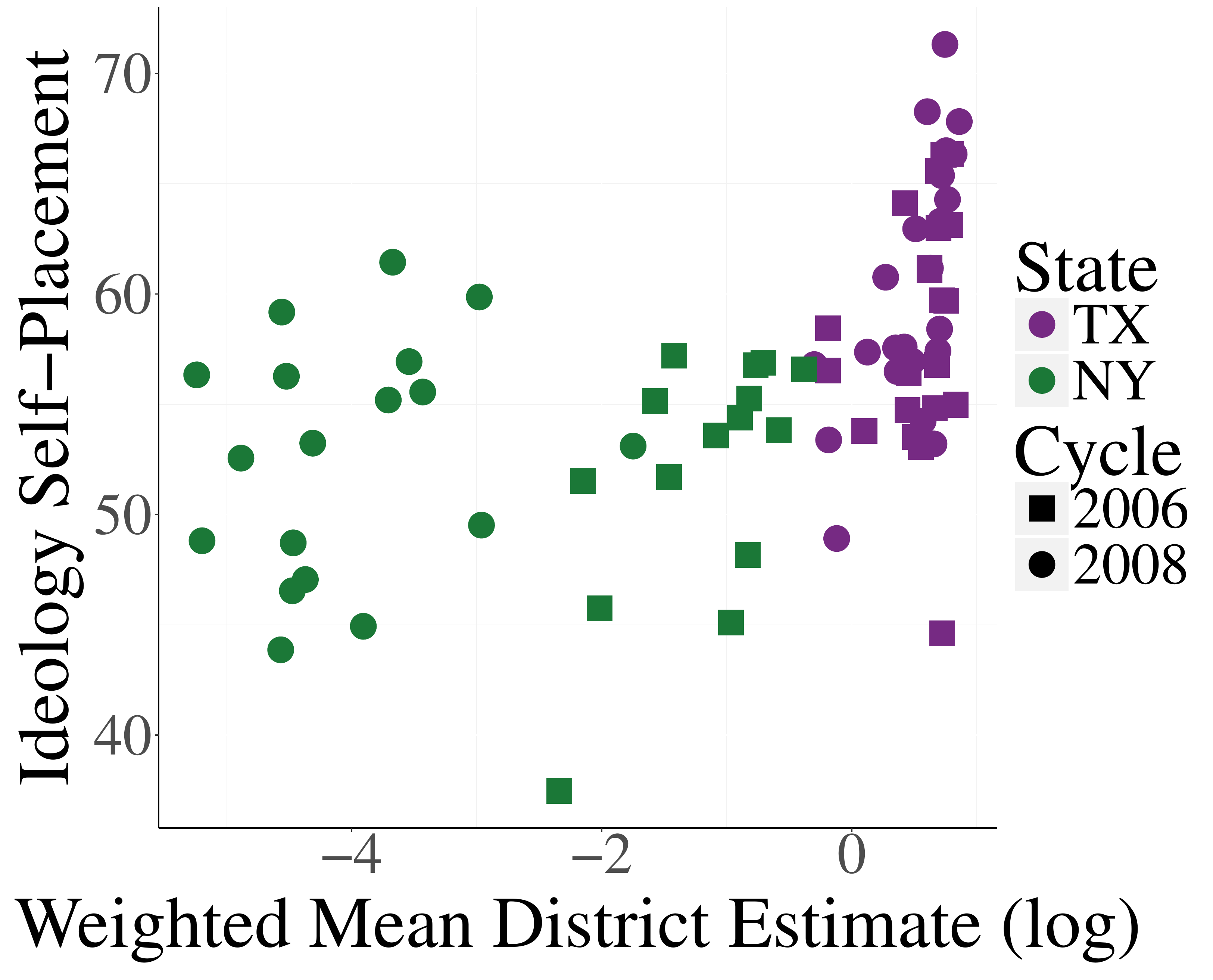}
        \includegraphics[width=0.32\linewidth]{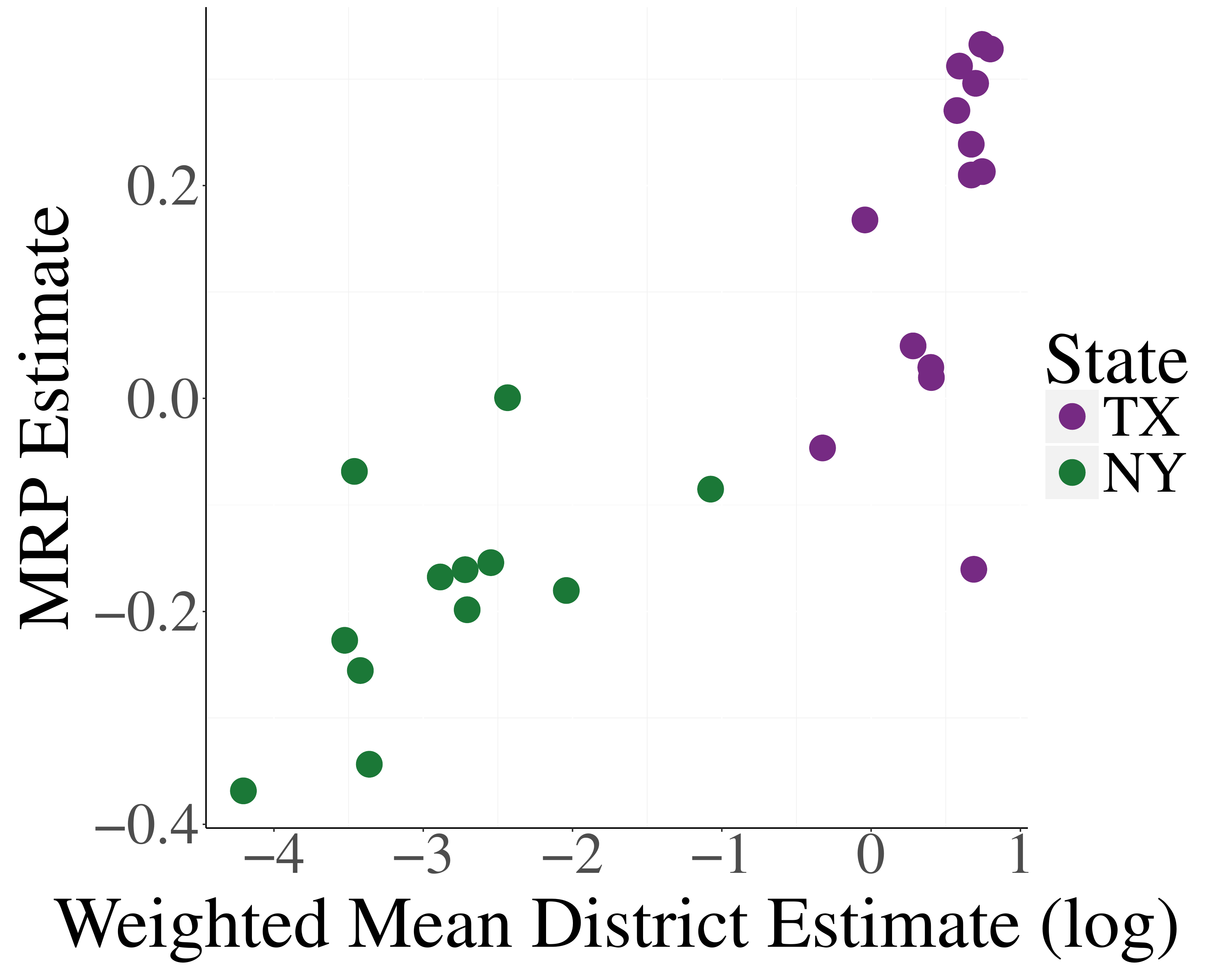}
            \par
        }
    \end{minipage}
    \par

    \begin{minipage}{0.2cm}
        \rotatebox{90}{\textit{\textbf{K = 8}}}
    \end{minipage}
    \begin{minipage}{\dimexpr\linewidth-1cm\relax}%
        \raisebox{\dimexpr-.5\height-1em}{
        \includegraphics[width=0.32\linewidth]{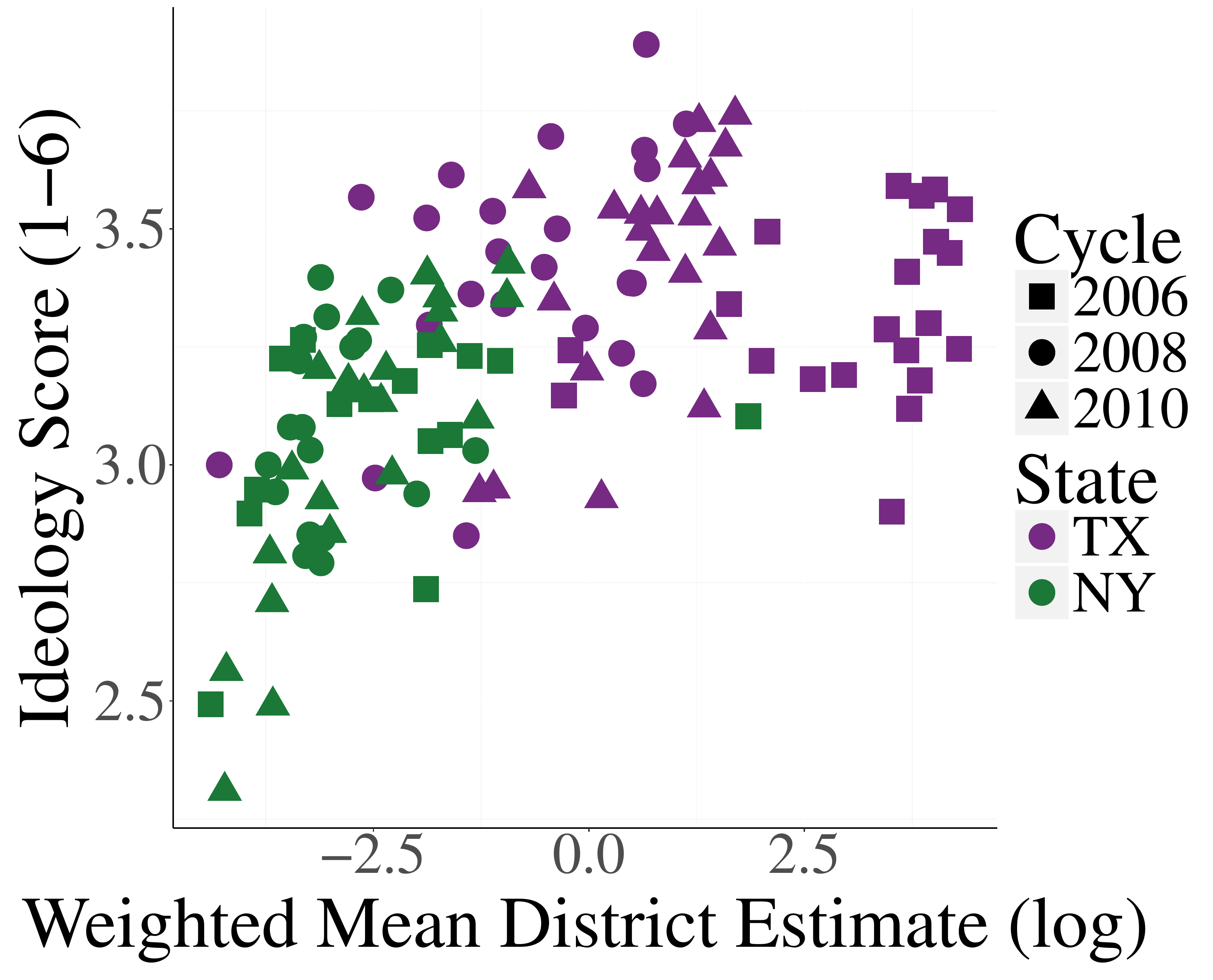}
        \includegraphics[width=0.32\linewidth]{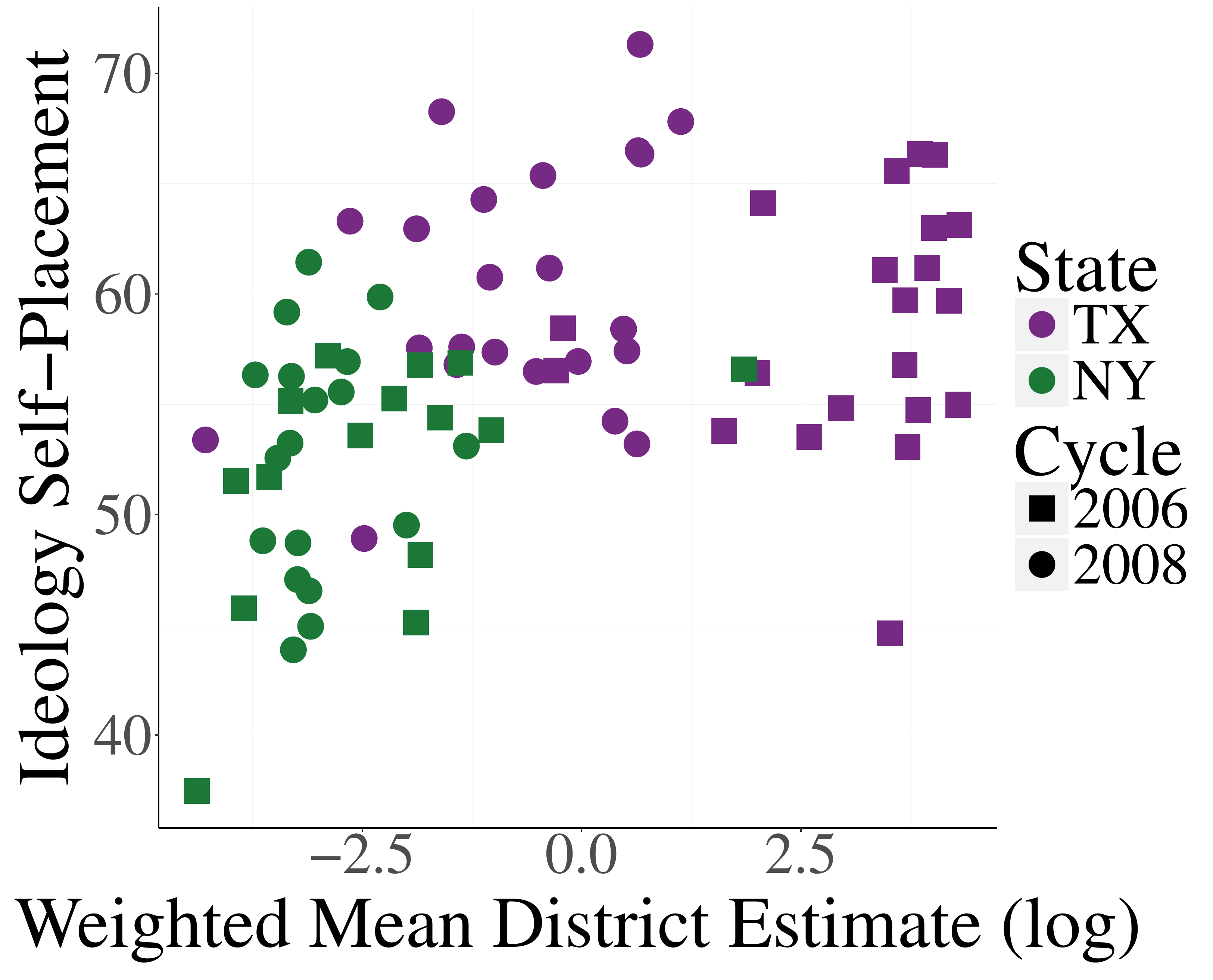}
        \includegraphics[width=0.32\linewidth]{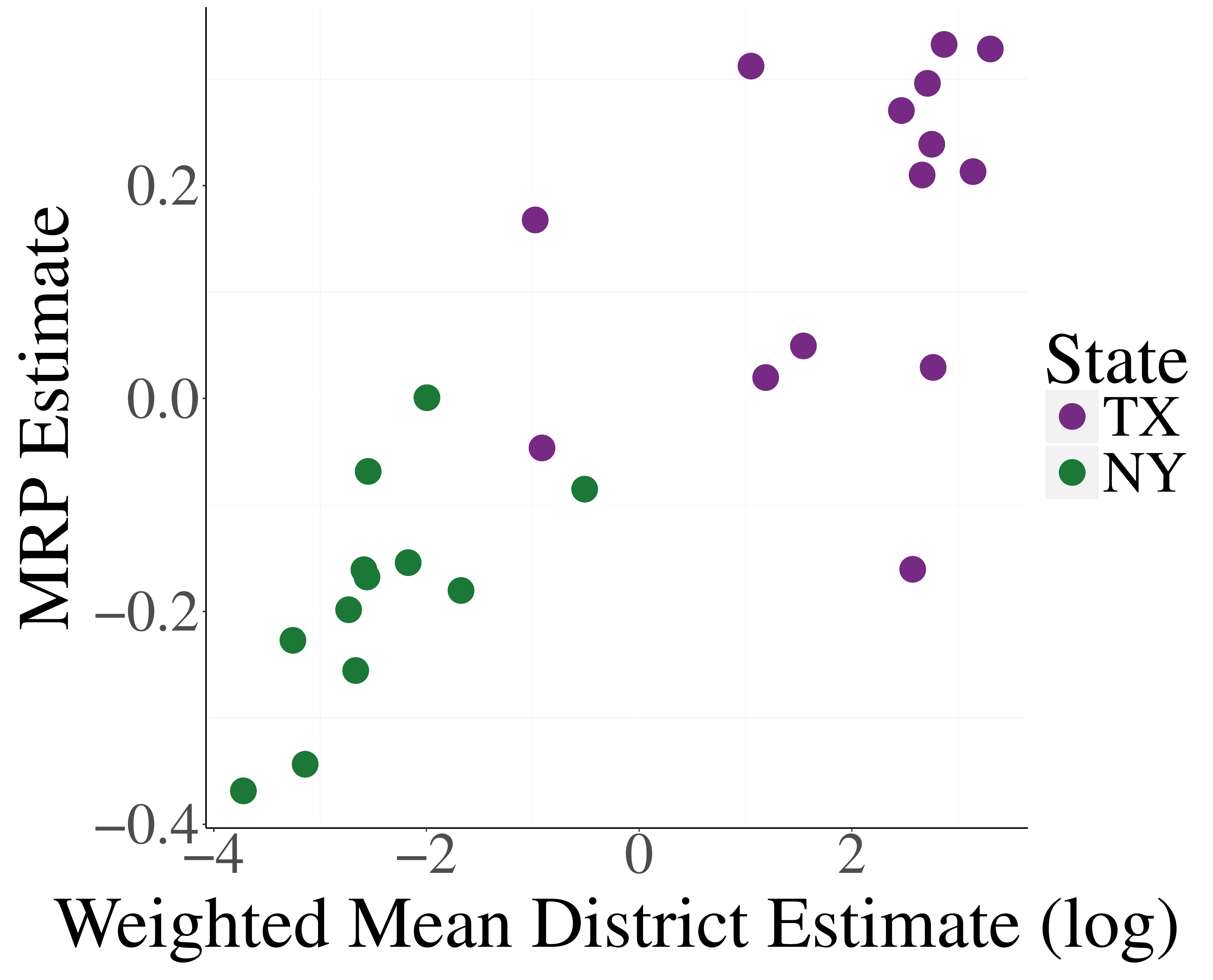}
        } \\
    \end{minipage}
    \par

\caption{In all of these cases, the inferred district-level voter preferences are based on the model varying its assumption of $K$ and are weighted mean district estimates transformed from $x$ to $sign(x) \, log( |x| + 1 )$. (Left) Inferences compared with CCES question of self-reported ideologies on a discrete scale. (Center) Compared with the CCES self-reported ideologies on a continuous scale from 0 to 100. (Right) District-level inferences of a full decade compared with MRP estimates \cite{warshaw}. }
         \label{fig:clust-cces-comp}
\end{figure}

Figure \ref{fig:clust-cces-comp} visualizes the comparisons between the results derived of alternative component distributions and alternative data sources described in Sec. \ref{sec:dist-agg-desc}. We find significant positive correlations between our district-level point estimates and all of the alternative data sources. When our model assumes 2 clusters rather than 4 clusters, the results of our model have a correlation of 0.2845 with the responses selecting ideology given a discrete scale (left column in Fig. \ref{fig:precdist-cces-comp}), 0.2646 with the responses selecting ideology along a continuous scale (middle column), and 0.7160 with the MRP estimates (right column). All of these correlations were significant with p-values less than 0.01. When our model assumes 8 clusters, the results of our model have a correlation of 0.2925 with the responses selecting ideology given a discrete scale, 0.1867 with the responses selecting ideology along a continuous scale, and 0.6861 with the MRP estimates (right column). Again, all of these correlations were significant with p-values less than 0.01.

    \begin{table}
        \centering
        \caption{Polarization metrics of the voters given inferences assuming different numbers of clusters $(K)$ and of the candidate CFscores (cand.) \cite{bonica}. The ``diff'' of each polarization metric is the difference between the voter and candidate metric.}
        \vspace{1mm}

        \def\arraystretch{1.1}%
        \setlength\tabcolsep{1mm}
        \begin{tabular}{|c|cl|ll|ll|ll|}
        \hline
        ~ & ~ & ~ & \multicolumn{2}{c|}{\centering{2006}} & \multicolumn{2}{c|}{2008} & \multicolumn{2}{c|}{2010} \\
        ~ & ~ & ~ & TX & NY & TX & NY & TX & NY \\ \hline
        \multirow{9}{*}{\rotatebox[origin=c]{90}{\textit{\textbf{K = 2}}}} & Difference- & Voters & 0.86 & 2.53 & 0.75 & 2.05 & 1.58 & 1.54 \\
        ~ & of-Means & Cand. & 1.87 & 1.58 & 1.83 & 0.37 & 0.41 & 1.433\\
        ~ & ~ & \textbf{Diff} & \textbf{-1.01} & \textbf{0.95} & \textbf{-1.08} & \textbf{1.68} & \textbf{1.17} & \textbf{0.11} \\ \cline{2-9}

        ~ &Standard & Voters &  2.40 & 7.75 & 2.36 & 188.98	& 2.15	& 4.56 \\
        ~ &Deviation & Cand. & 0.97 & 0.77 & 1.07 & 0.83 & 1.22 & 0.81 \\
        ~ &~ & \textbf{Diff} & \textbf{1.42} & \textbf{6.98} & \textbf{1.28} & \textbf{188.15} & \textbf{0.94} & \textbf{3.74}\\ \cline{2-9}

        ~ &~ & Voters & 0.49	& 3.13 & 0.27	& 4.07	& -0.15 & 3.24 \\
        ~ &Kurtosis & Cand.  & -1.62 & 1.67	& 0.49	& 1.79	& 3.24	& -1.61 \\
        ~ & ~ & \textbf{Diff} & \textbf{2.11} & \textbf{1.46} & \textbf{-0.23} & \textbf{2.27} & \textbf{-3.39} & \textbf{4.84}\\ \hline

        \multirow{9}{*}{\rotatebox[origin=c]{90}{\textit{\textbf{K = 8}}}} & Difference- & Voters & 9.06 & 9.30		& 3.39	& 6.80	& 1.80	& 1.81\\
        ~ & of-Means & Cand. & 1.74	& 0.54	& 	1.82 &	1.55 & 2.17 &	1.42 \\
        ~ & ~ & \textbf{Diff} & \textbf{7.32} & \textbf{8.76} & \textbf{1.56} & \textbf{5.25} & \textbf{-0.37} & \textbf{0.39} \\ \cline{2-9}

        ~ & Standard & Voters & 110.59	& 60.49 & 	38.55 &	85.00 & 		5.59 &	55.75\\
        ~ & Deviation & Cand. & 0.97	& 0.77	 &	1.07 &	0.83 &		1.22 &	0.81\\
        ~ & ~ & \textbf{Diff} & \textbf{109.61} & \textbf{	59.72} & \textbf{37.48} & \textbf{84.17} & \textbf{4.37} & \textbf{54.94}\\ \cline{2-9}
        ~ & ~ & Voters & 5.64	& 2.65	& 48.57 & 4.06 & -0.03 &	0.50 \\
        ~ & Kurtosis & Cand.  & -1.62	& 1.67	 &	0.49 &	1.79 &		3.24 &	-1.61\\
        ~ & ~ & \textbf{Diff} & \textbf{7.26} & \textbf{0.98} & \textbf{48.07} & \textbf{2.26} & \textbf{-3.27} & \textbf{2.11}\\ \hline

        \end{tabular}
        \label{tbl:clust-pol-metrics}
    \end{table}

\end{document}